\documentclass{article}

\usepackage{arxiv}
\newif\ifarxiv
\arxivtrue

\usepackage{subfig}
\usepackage{graphicx}%
\usepackage{multirow}%
\usepackage{amsmath,amssymb,amsfonts}%
\usepackage{amsthm}%
\usepackage{mathrsfs}%
\usepackage[title]{appendix}%
\usepackage{xcolor}%
\usepackage{textcomp}%
\usepackage{manyfoot}%
\usepackage{booktabs}%
\usepackage{listings}%
\usepackage{todonotes}
\usepackage{tikz,lipsum,lmodern}
\usepackage[most]{tcolorbox}

\usepackage{url}            
\usepackage{nicefrac}       
\usepackage{microtype}      

\usepackage{lipsum}
\usepackage{placeins}

\usepackage{svg}

\usepackage{natbib}
\usepackage{amssymb,amsmath}
\usepackage{graphicx}
\usepackage{url}
\usepackage{amsmath}
\usepackage{color}
\usepackage{todonotes}
\def\log{\,\mathrm{log}}

\usepackage{adjustbox}

\usepackage{subfig}

\def\<{\langle}
\def\>{\rangle}

\DeclareMathOperator*{\argmax}{arg\,max}
\DeclareMathOperator*{\argmin}{arg\,min}

\def\bbeta{\boldsymbol{\beta}}

\def\bmu{\boldsymbol{\mu}}

\def\blambda{\boldsymbol{\lambda}}

\def\btau{\boldsymbol{\tau}}

\def\bphi{\boldsymbol{\phi}}

\def\C{{\boldsymbol{C}}}
\def\c{{\boldsymbol{c}}}

\def\E{{\boldsymbol{E}}}

\def\f{{\boldsymbol{f}}}

\def\g{{\boldsymbol{g}}}

\def\h{{\boldsymbol{h}}}
\def\I{{\boldsymbol{I}}}

\def\K{{\boldsymbol{K}}}
\def\k{{\boldsymbol{k}}}

\def\M{{\boldsymbol{M}}}

\def\R{{\boldsymbol{R}}}

\def\S{{\boldsymbol{S}}}
\def\s{{\boldsymbol{s}}}

\def\X{{\boldsymbol{X}}}
\def\x{{\boldsymbol{x}}}

\def\y{{\boldsymbol{y}}}

\def\0{{\boldsymbol{0}}}
\def\1{{\boldsymbol{1}}}

\def\<{\, \left\langle \,}
\def\>{\, \right\rangle \,}

\def\max{\mathrm{max}}

\def\bbeta{\boldsymbol{\beta}}

\def\bepsilon{\boldsymbol{\epsilon}}

\def\btau{\boldsymbol{\tau}}

\def\DN{\mathcal{N}}

\def\Exp{\mathrm{Exponential}}
\def\Uni{\mathrm{Uniform}}

\def\ctilde{\kern -.04em\lower .7ex\hbox{\~{}}\kern .04em}

\usepackage{algorithm}
\usepackage{algorithmic}
\usepackage{tabularx}

\title{
Exploiting chemical shift variability enables recovery of overlapping metabolites from  \textsuperscript{1}H nuclear magnetic resonance spectra}

\author{
  Jesper L{\o}ve Hinrich\textsuperscript{1}\thanks{Corresponding author jehi@dtu.dk}, Pia Susan Mayer\textsuperscript{2}, Bekzod Khakimov\textsuperscript{2}, S{\o}ren Balling\textsuperscript{2}, Morten M{\o}rup\textsuperscript{1}\\ \\
  \textsuperscript{1} Department of Applied Mathematics and Computer Science, \\ Technical University of Denmark,\\ Richard Petersens Plads 321, Kgs. Lyngby, 2800, Denmark.
    \\ \\
    \textsuperscript{2} Department of Food Science, \\ University of Copenhagen, \\ Rolighedsvej 26, Frederiksberg, 1958, Denmark.
}

\begin{document}
\maketitle

\begin{abstract}
Overlapping peaks and sample-dependent chemical shift variability prevent reliable metabolite recovery from complex biological spectra. This problem is critical in 
    one-dimensional proton (1D \textsuperscript{1}H) NMR which has become the standard method providing fast acquisition and information-rich spectra in metabolomics and foodomics. 
    This study demonstrates how chemical shifts can be utilised as a strength in 1D \textsuperscript{1}H NMR, when suitably modeled through the proposed Bayesian Shift-Invariant Non-negative Matrix Factorization (BSI-NMF) procedure. 
    We find that BSI-NMF accurately recovers the underlying chemical signals in 1D \textsuperscript{1}H NMR spectra missed by existing analyses approaches across simulations, laboratory created datasets, and a large urine dataset obtained from 2439 people across Europe. 
    Our study highlights how shifts in the chemical signatures -- until now perceived as a nuisance -- can in fact when suitably modelled be instrumental for unique recovery of metabolites. This creates an opportunity to experimentally induce chemical shifts changes to facilitate unique recovery of spectra.
\end{abstract}
\FloatBarrier

\keywords{Nuclear Magnetic Resonance Spectroscopy  \and Proton NMR \and Multivariate Curve Resolution \and Non-negative Matrix Factorization \and Shift-invariant Non-negative Matrix Factorization \and Bayesian Inference \and Metabolomics \and Human Urine}

\section{Introduction}\label{sec:intro}
 Many biological measurement technologies contain structured variability across samples that if normalized away can destroy important information required for source separation and signal recovery.
The field of metabolomics focuses on the global metabolite fingerprint of complex biological mixtures such as tissue extracts and biofluids \cite{Beckonert2007-nz} including urine that has gained substantial attention as an abundant and non-invasive biofluid providing rich information on the health status and diet of an individual \cite{Bezabeh2019-cm}.
One of the most versatile analytical metabolomics instruments recognised as one of the pillars in life science research \cite{moco2022} is 
nuclear magnetic resonance (NMR) spectroscopy. NMR offers fast and robust data acquisition, high reproducibility, simple sample preparation, non-destructiveness of the sample, structural information of the metabolite and the possibility of absolute quantification \cite{Giraudeau2023, Gowda2019}. The most commonly applied method in NMR-based metabolomics is one-dimensional (1D) proton (\textsuperscript{1}H) NMR, where each proton nuclei environment gives rise to Lorentzian-shaped peaks which vary in multiplicity, intensity and chemical shift, giving unique information about the metabolite. This information is critical for accurate annotation and quantification, but chemical shift changes of peaks between samples are currently treated as a nuisance, hampering the recovery of metabolites when peaks in crowded regions overlap.

Chemical shift changes can arise from differences in pH, ionic strength, temperature, matrix effects or instrumental factors \cite{Savorani2010-dz,Bezabeh2019-cm}. While some of these factors can be minimised with rigorous standard operating procedures for sample preparation and spectral acquisition, inter-individual biological samples, e.g. urine, are prone to chemical shift changes. Peak overlap occurs in crowded regions of the chemical shift axis for \textsuperscript{1}H nuclei, when the nuclei experience similar chemical environments (such as structurally similar metabolites). The limitations from chemical shift changes and peak overlaps are recognized as critical for the downstream data processing of large metabolomics data sets. The problem can be partially solved by peak alignment algorithms (for chemical shift)\cite{Savorani2010-dz,NIELSEN1998} and by use of decomposition techniques (for peak overlap) \cite{Lee1999, Puig-Castellv2017} when the spectra can be subdivided into individual multiplets. However, metabolomic studies can consist of hundreds of samples of complex mixtures made up of different classes of metabolites and varying concentrations that present multiple regions of overlapped peaks with 
different chemical shift changes. These regions and peaks are often not baseline separated and thus cannot be further subdivided, which limits existing methods to resolve the individual underlying peaks.

Existing NMR metabolomics workflows typically rely on simultaneous analysis of spectral ensembles (2D data analysis from stacked 1D spectra) and involve a two-step procedure that applies: i) \emph{Peak alignment methods} which includes approaches such as 
 non-linear correlation optimized warping (COW) and dynamic time warping (DTW) \cite{NIELSEN1998, Tomasi2004}, or linear methods such as Interval Correlation Shifting (\textit{i}coshift)\cite{Savorani2010-dz}.
ii) \emph{2D peak deconvolution methods} to the aligned data \cite{Vu2013-iy} which includes multivariate peak fitting \cite{Li2023,Smith2017} and multivariate curve resolution (MCR) \cite{Lawton1971, Juan2014, Engelsen2013} typically imposing positivity constraints on the extracted matrices using non-negative matrix factorization (NMF) techniques \cite{paatero1994positive,Lee1999}.

Notably, NMF can potentially uniquely recover the underlying components if the data sufficiently span the positive orthant\cite{donoho2003does,laurberg2008theorems,gillis2012sparse,huang2013nmf} . Often, this requirement is not satisfied, but inclusion of a baseline in the modeling \cite{laurberg2007affine,laurberg2008theorems} as well as regularization of the solution to favor sparse representations \cite{laurberg2008theorems,gillis2012sparse,huang2013nmf} can help alleviate model redundancies. NMF and the baseline corrected and sparse extensions all assume that the chemical signals for the same metabolite only varies in concentration across samples, whereas resolving issues of uniqueness require imposing additional constraints that may not comply with the underlying structure of the signals of interest.

\begin{figure}[tbp]
    \centering
\begin{tcolorbox}[arc=0pt, colback=gray!50, boxrule=0pt,height=0.5cm]
   \centering\vspace{-0.25cm}
    \textbf{BSI-NMF Assumed Generative Process for \textsuperscript{1}H NMR data}
    \end{tcolorbox}
    \vspace{-0.5cm}
     \begin{tcolorbox}[arc=0pt, colback=gray!10, boxrule=0pt]
   \centering    
\vspace{-0.5cm}
\begin{align*}
 & \underbrace{\beta_n \sim\Uni(0,\infty)}_{\text{Baseline level}},\quad 
    \underbrace{\lambda_d \sim \Exp(\eta)}_{\text{Component relevance}},\quad 
    \underbrace{\tau_{n,d}\sim\Uni(\tfrac{T}{2},\tfrac{T}{2})}_{\text{Component shifts}},
    \\
   &\qquad \qquad \underbrace{c_{n,d} \sim \Exp(\lambda_d)}_{\text{Sample concentration}},\quad    
\underbrace{s_{t,d}\sim \Exp(\lambda_d)}_{\text{Component spectra}},
\\
&\qquad \qquad \qquad \qquad \underbrace{x_{n,t} \sim  \DN\left(\beta_n + \sum_{d=1}^D c_{n,d} s_{t-\tau_{n,d},d}, \sigma^2\right)}_{\text{The observed data at noise level }\sigma^2}
\end{align*}
\end{tcolorbox}
\vspace{-0.5cm}
\begin{tcolorbox}[arc=0pt, colback=gray!50, boxrule=0pt,height=0.5cm]
   \centering\vspace{-0.2cm}
    \textbf{Computer simulated dataset}
    \end{tcolorbox}        
     \includegraphics[width=0.98\linewidth,trim={0 0 0 1cm},clip]{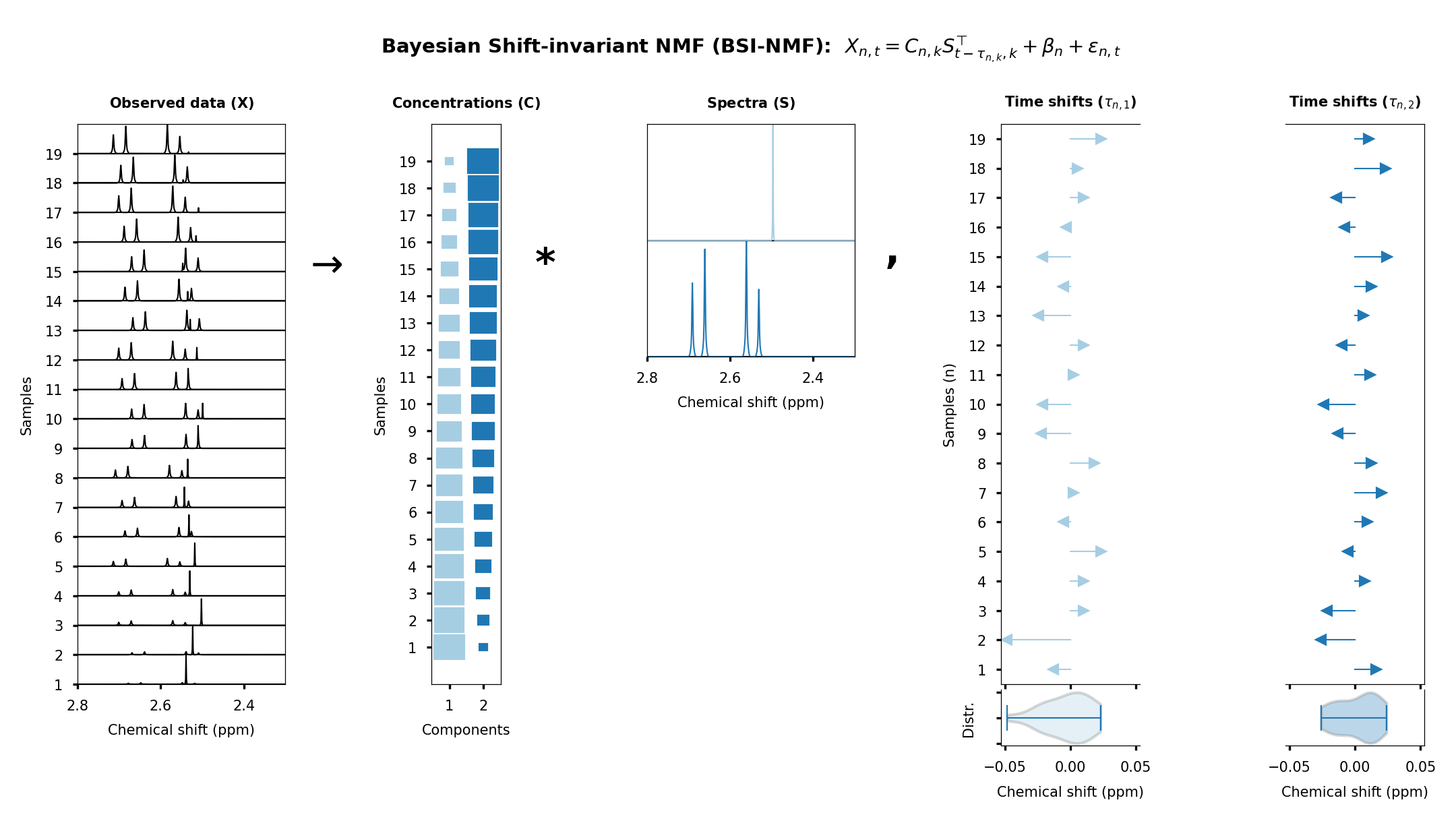}
         \\
         \vspace{-0.3cm}
\begin{tcolorbox}[arc=0pt, colback=gray!50, boxrule=0pt,height=0.5cm]
   \centering\vspace{-0.25cm}
    \textbf{\textsuperscript{1}H NMR spectra of human urine (N=2439, 2.5-2.75 ppm)
    }
    \end{tcolorbox}         
    \includegraphics[width=0.98\linewidth]{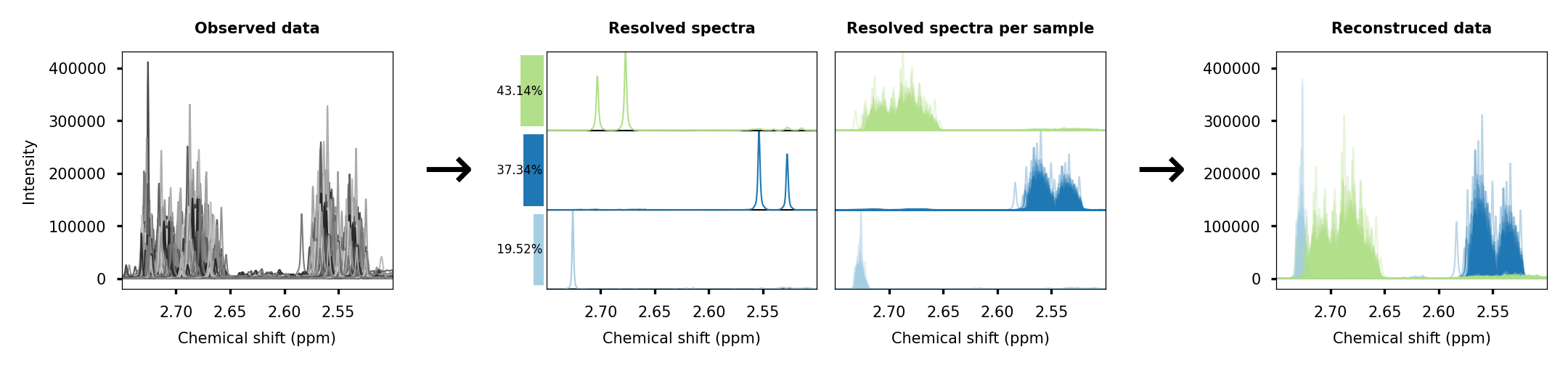} 
        
        \caption{
        Top panel: The proposed generative process of \textsuperscript{1}H NMR data forming the Bayesian Shift-invariant Non-negative Matrix Factorization (BSI-NMF) procedure. $p(\cdot)$ are the priors placed on the parameters and $\mathcal{L}(\cdot)$ is the likelihood function. The Uniform, Exponential and Normal distribution are denoted by $\Uni(\cdot)$, $\Exp(\cdot)$ and $\DN(\cdot)$, respectively. The  relevance of each component is modelled by $\blambda$ with hyperparameter $\eta$
        . 
        The concentrations are $\C\in \mathbb{R}_{\geq 0}^{N\times D}$ and spectra are $\S\in \mathbb{R}_{\geq 0}^{T\times D}$. The sample and component wise shifts $\tau_{n,d}$ shifts the associated spectra $\s_{d}$ by $\tau_{n,d}$ indices with positive values shifting the spectrum to higher ppm values.
        Middle and bottom panels: The BSI-NMF procedure is applied to a computer simulated (N=19) and a \textsuperscript{1}H NMR human urine dataset (N=2439) respectively. BSI-NMF decomposes the data $\X$ into concentrations $\C$ and spectra $\S$ with and additive offset component per sample (not shown) and with additive normal distributed noise $\bepsilon$. In this example, each component $d$ represents a metabolite \textsuperscript{1}H NMR spectrum $\s_d$, concentration $\c_d$, and shift in each sample $\tau_{n,d}$. The metabolites are dimethylamine ($d=1$) and citric acid ($d=2$). The real data example (c) show BSI-NMF is able to automatically recover the overlapping and shifting spectra and will split a metabolite spectrum (citric acid) into two spectra (dark blue and light green) when supported by the data.}
        \label{fig:1}
\end{figure}

Importantly, the chemical shift variability is in the existing approaches treated as a nuisance to be removed. \emph{In this work, we hypothesize that chemical shifts are in fact an important source of information providing structured variability across samples that can enhance metabolite recovery}. \emph{The aim of the study is therefore to exploit this variability by developing a modeling framework that jointly account for chemical shift changes and spectral overlap in \textsuperscript{1}H NMR spectra thereby making chemical shift variability informative of metabolites.}

In the literature, NMF has been extended to shift-invariant non-negative matrix factorization (SI-NMF) \cite{Morup2007,morup2008shiftcp}, which learns sample and component specific shifts and can lead to unique recovery \cite{harshman2003shifted}. NMF decompositions have been further advanced with automatic relevance determination (ARD)\cite{Mackay1995,morup2009automatic,Hinrich2018}, which acts as a regularization and an efficient framework to learn the number of components at assumed signal-to-noise (SNR) levels \cite{morup2009automatic}.  These advancements are presently explored for the modeling of \textsuperscript{1}H NMR spectral ensembles by developing the Bayesian Shift-invariant NMF (BSI-NMF). BSI-NMF simultaneously learns component level chemical shift changes, the associated pure spectra, and sample concentrations as well as a baseline spectrum, while quantifying the number of metabolites supported by the data at given levels of SNR using ARD that promotes sparse representations. The BSI-NMF thereby aims to optimally leverage all relevant information and model-structure known to enable unique recovery of metabolites. The developed BSI-NMF model with $D$ initial components can be formulated in terms of the imposed generative process of \textsuperscript{1}H NMR data as described in the top panel of Figure~\ref{fig:1} whereas the middle and bottom panels demonstrate the principle of BSI-NMF when applied to an ensemble of \textsuperscript{1}H NMR spectra. Given $N$ \textsuperscript{1}H NMR samples acquired across $T$ measured frequencies (given in ppm) results in a data matrix $\X \in \mathbb{R}^{N \times T}$.  The noise variance, $\sigma^2$, is considered known, but in practice it is initialized at a low noise variance (assuming all data is signal) and annealed to a high noise variance (i.e., assuming all data is noise). This gradually places more weight on the learned component wise length scales $\lambda_d$ used for automatic relevance determination and consequently provides a  pruning path that can be traced in terms of the supported numbers of components and reconstruction quality as measured by the explained variance $R^2$. 
Full details on the methods, optimization, and experimental settings are provided in the Online Methods, Section \ref{sec:methods}.

The BSI-NMF performance is systematically investigated on computer simulated spectra and spectra recorded on artificial mixture samples with known concentration levels of metabolites. For external validation, the methodology is applied to human urine samples from a large cohort (2439 spectra) \cite{Trimigno2019}. BSI-NMF performance is compared to conventional NMF considering unaligned and \textit{i}coshift ($n=1$) aligned data. Then BSI-NMF is compared to commonly applied multivariate peak fitting and peak alignment and \textit{i}coshift coupled with NMF decomposition. The study highlights how BSI-NMF uncovers chemically meaningful spectral signatures that are difficult to access by existing analysis approaches. Importantly, we establish that the chemical shift changes, previously perceived as a nuisance in \textsuperscript{1}H NMR data, are instrumental for the successful unique recovery of metabolites.

\section{Results}\label{sec:results}
The BSI-NMF model reduces to existing NMF based modeling procedures for \textsuperscript{1}H NMR data as follows i) the standard NMF model \cite{paatero1994positive,Lee1999,deJuan2021}  by turning off the regularization (i.e, fixing $\blambda=\mathbf{0}$) and shift modelling (i.e., fixing $\boldsymbol{\tau}=\mathbf{0}$) while including also a baseline component which has been found to promote unique component recovery \cite{laurberg2007affine,laurberg2008theorems}, ii) a Bayesian NMF (B-NMF) model with relevance determination \cite{Hinrich2018} by turning off the shift modelling (i.e, fixing $\boldsymbol{\tau}=\mathbf{0}$) where the relevance determination also acts as a sparsity promoting (i.e., $\ell_1$-norm) regularization previously explored for sparse NMF in order to also enhance model uniqueness \cite{laurberg2008theorems,wang2012nonnegative}, and iii) what is here termed Bayesian (B-)CoShiftNMF defined by sample specific shift-values shared across the components, e.g. $\tau_{n,1} =\ldots= \tau_{n,D}$. CoShiftNMF corresponds to an integrated procedure simultaneously performing CoShift based alignment of samples and NMF. This notably avoids the necessity of specifying an optimal reference spectrum which can sometimes be challenging when applying the CoShift procedure \cite{Vu2011}. 

B-NMF, B-CoShiftNMF and BSI-NMF were systematically investigated and compared by initializing at the maximum component order ($D=10$), while SNR was annealed from 50 to 0 such that the ARD optimized $\blambda$ and consequently pruned components deemed unnecessary at the specific noise level, while NMF is fitted for models specified using 1 to 10 components. To improve recovery of small peaks all samples were standardized by their Frobenius norm and the standardization was corrected to not over-amplify noise. For further preprocessing and modelling details see the Online Methods, Section~\ref{sec:methods:experimentaldetails}.

Computer simulated samples were used to explore the possibilities and limitations of BSI-NMF prior to its application on acquired data, which included metabolite mixtures (Urine Metabolites A: 20 metabolites, Urine Metabolites B: 18 metabolites) with known concentration levels in a triangular Design of Experiment (with $N=42$) and a dataset of human urine ($N=2439$) \cite{Trimigno2019}. The simulation and experimental details are reported in the Online Methods (Section~\ref{sec:methods:simstudy} and Section~\ref{sec:methods:experimentaldetails}). Three cases that pose different challenges in urine metabolomics were chosen for the evaluation of the proposed algorithm. A comprehensive comparison is shown exclusively for the computer simulated data, while for the remaining cases, only BSI-NMF is presented for clarity. Full comparisons of all methods across all ten model orders are available in the Appendix Section~\ref{sec:app:allcomponents}. 
\begin{figure}[H]
    \centering
        \includegraphics[width=0.98\linewidth]{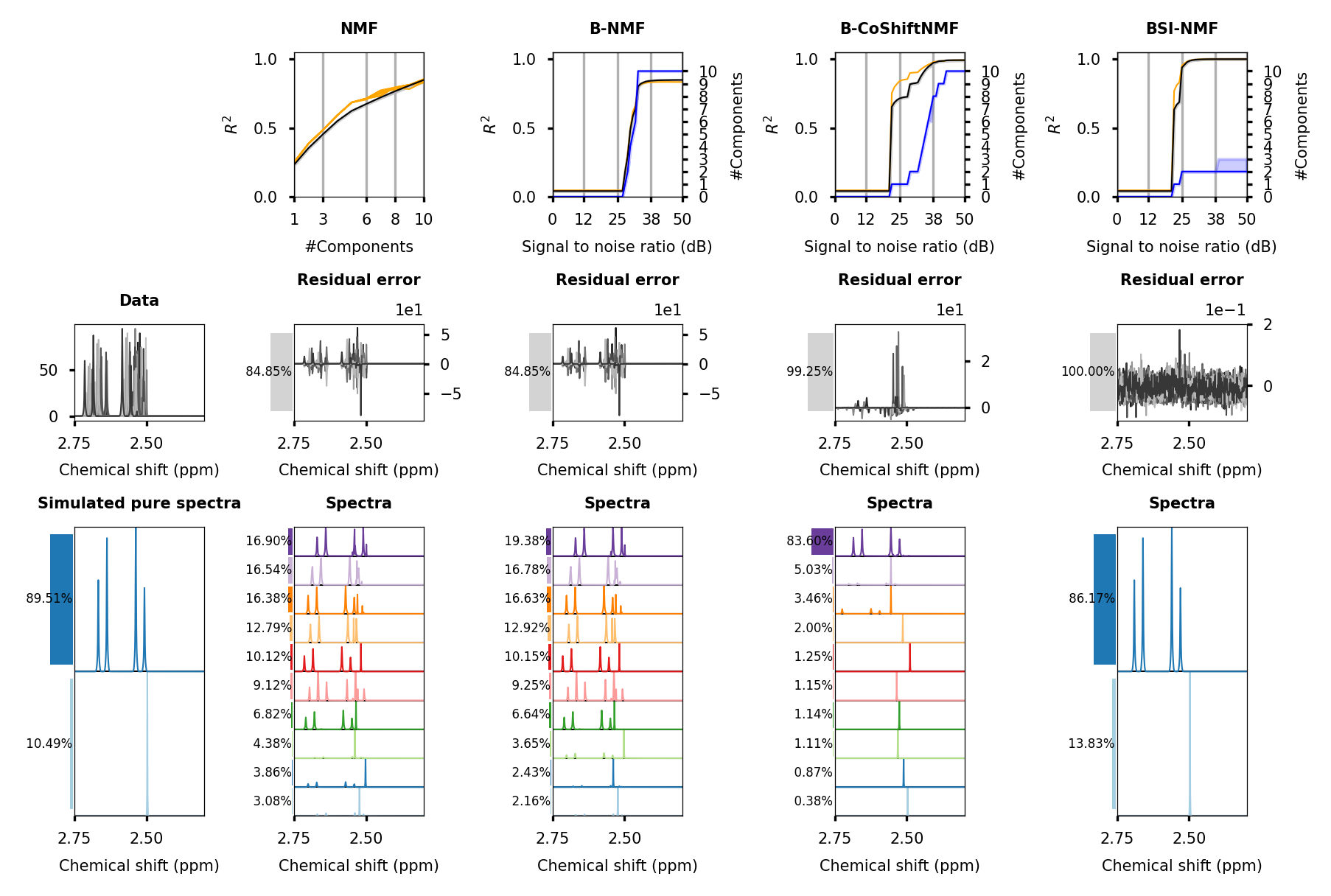}    
    \caption{
    Simulated data of citric acid and dimethylamine with compound specific shifts ($N=19, T=1801, D=2$): First column shows data and the pure simulated spectra. The results of applying NMF, B-NMF, B-CoShiftNMF, and BSI-NMF (column 2 - 5) show the model performance across the chosen SNR range 50 to 0 dB (no. components for NMF) as coefficient of regression $R^2$ with (black line) and without (dotted orange) the baseline component. The blue line indicates the number of components identified by the automatic relevance determination at each SNR (or number of component), while shaded gray and blue area denote the range of values across repeats. In the second row, the residual error from each model are shown with a bar of explained variance (\%). The third row, depicts the simulated spectra (column 1) and the latent spectra of the best model and bars for the contribution (\%) to the data reconstruction.
}
    \label{fig:simulated}
\end{figure}

\subsection{Case 1: citric acid and dimethylamine}
\label{sec:results:citric}
This case considers citric acid, which is particularly sensitive to pH \cite{Moore1994,Madurga2017} and often shows strong chemical shift changes in urine samples, making it difficult to align with existing tools.

\paragraph{Computer Simulated Data}
\label{sec:simulated}
Computer simulations of the spectra of citric acid and dimethylamine were used to investigate the strengths and limitations of the methods. Four simulation scenarios were evaluated: 1) No shifts and no overlap between the spectra; 2) No shifts; 3) Sample specific shifts only, e.g. the two components shift by the same amount, equivalent to shifting the entire sample spectrum; 4) Metabolite specific shifts, e.g. the two metabolites have different chemical shift changes within and across samples. The simulation settings are detailed in the Online Methods, Section~\ref{sec:methods:simstudy}. 

The simpler simulation scenarios (1-3) are shown in the Appendix~\ref{fig:app:simulated}. In short, for scenarios 1 (no shift, no overlap) and 2 (no shift) all methods worked and identified the correct number of components, whereas in scenario 3 (sample specific shifts) only B-CoShiftNMF and BSI-NMF identified the correct number of components. However, in all three scenarios, all methods suffer from non-unique solutions. 

The most realistic scenario (4) of metabolite specific shifts 
is shown in Figure~\ref{fig:simulated}. For BSI-NMF the first drop in R\textsuperscript{2} from the ARD pointed to the best number of components (2). Indeed, the two component model explained almost 100\% of the data and recovered two chemically meaningful latent spectra of citric acid and dimethylamine. In contrast, NMF, B-NMF or B-CoShiftNMF provided no clear indication of the right number of components based on R\textsuperscript{2} and even the ten component model had structured residuals left. The latent components of NMF and B-NMF duplicated the spectrum of citric acid and dimethylamine at different chemical shifts to describe the data. B-CoShiftNMF captured the citric acid profile in its first latent spectrum - as it described most of the signal - but this entailed that the shifts of dimethylamine were not modeled, and instead its spectrum was duplicated at different chemical shifts.

\paragraph{\textsuperscript{1}H NMR Data}
For the real datasets, citric acid and dimethylamine were investigated in the subinterval of the spectra from 2.5 to 2.75 ppm. For Urine Metabolites A, the chemical shift was negligible and all methods identified the correct number of metabolites $D=2$ and their unique underlying metabolite spectra. The results for BSI-NMF are shown in Figure~\ref{fig:real:interval:citric} and the results of the other methods are shown in the Appendix, Figure~\ref{fig:app:allcomp:urineA:citric}. 

For Urine Metabolites B, the BSI-NMF model described 93\% of the variance using only two components $D=2$ (Figure~\ref{fig:real:interval:citric}). NMF and B-NMF failed to adequately model the data due to the presence of small chemical shift changes - neither correctly identified the number of components or their corresponding spectra (Figure ~\ref{fig:app:allcomp:urineB:citric}). B-CoShiftNMF achieved good performance, but did not clearly identify the correct number of components. Mostly, it aligned the citric acid spectrum with one component and used additional components to capture the dimethylamine spectrum at different chemical shifts.

For the Human Urine dataset, there were multiple components with shifts, which violated the assumptions of NMF, B-NMF, and B-CoShiftNMF, making them unable to correctly identify the number of components. In contrast, BSI-NMF pointed to a three component ($D=3$) solution and recovered the spectra uniquely, see Figure~\ref{fig:real:interval:citric}. It did not recover citric acid as one component with two doublets, but as two components with one doublet each due to the varying degree of shift between them, which becomes especially clear in the Appendix, Figure~\ref{sec:app:allcomp:human}.

\begin{figure}[H]
    \centering
    \includegraphics[width=.98\linewidth]{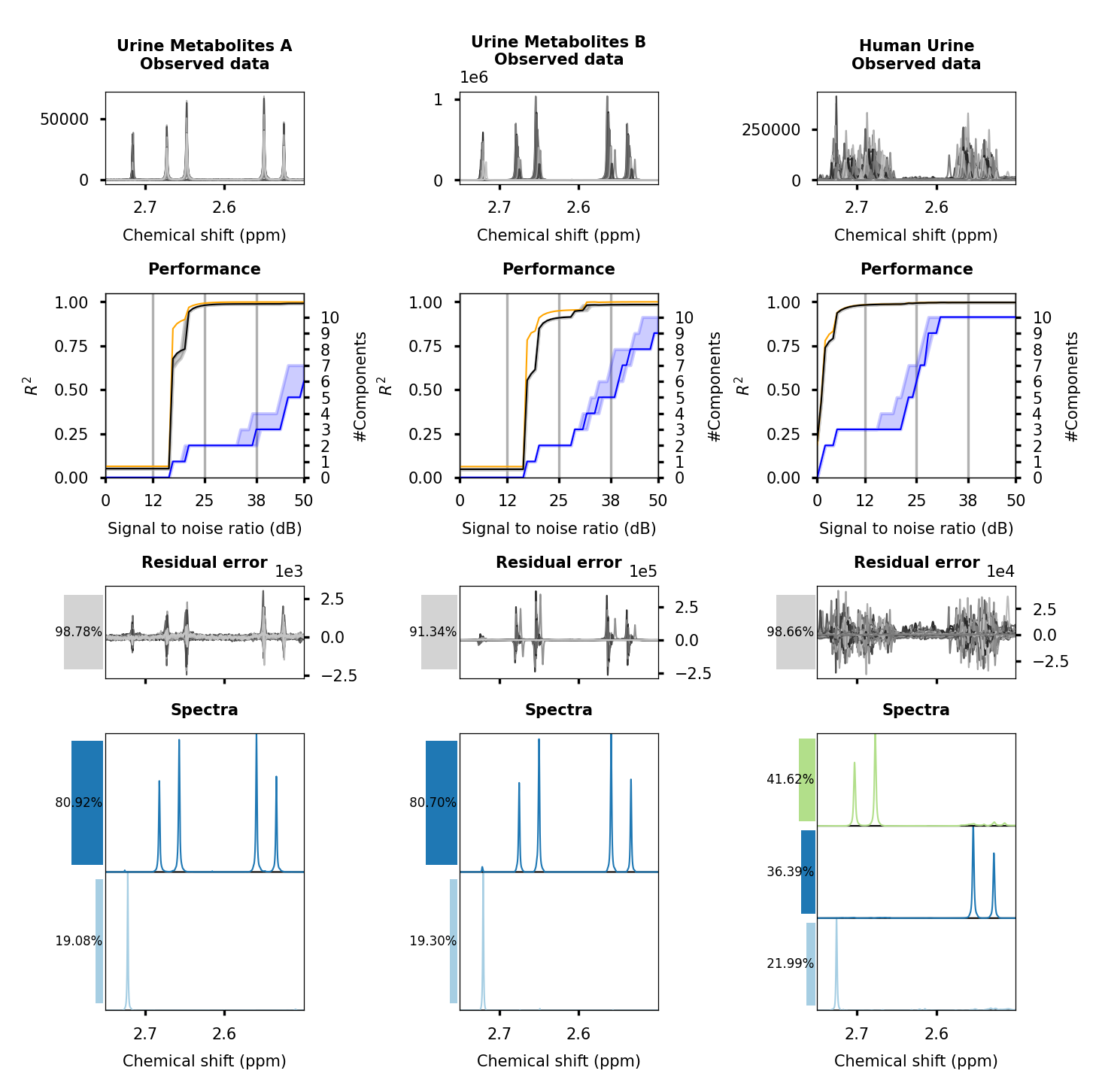}    
    \caption{Case 1: BSI-NMF applied to interval 2.5 to 2.75 ppm in Urine Metabolites (A (1st column) \& B (2nd column)) and Human Urine (3rd column). The first row shows the observed data of each dataset, second row highlights the performance across the chosen SNR range 50 to 0 dB as coefficient of regression $R^2$ with (black line) and without (dotted orange) the baseline component. The blue line indicates the number of components identified by the automatic relevance determination at each SNR. The shaded gray and blue area denotes the range of values across repeats. Third row indicates the residual error for each model, while the bar gives explained percent. The last row, the latent spectra for the best model are shown with their respective percentage contribution of the reconstruction as bars next to the spectrum.
}
    \label{fig:real:interval:citric}
\end{figure}

\subsection{Case 2: Creatine and Creatinine}
\label{sec:results:creatine}
\begin{figure}[tbp]
    \centering
    \includegraphics[width=.98\linewidth]{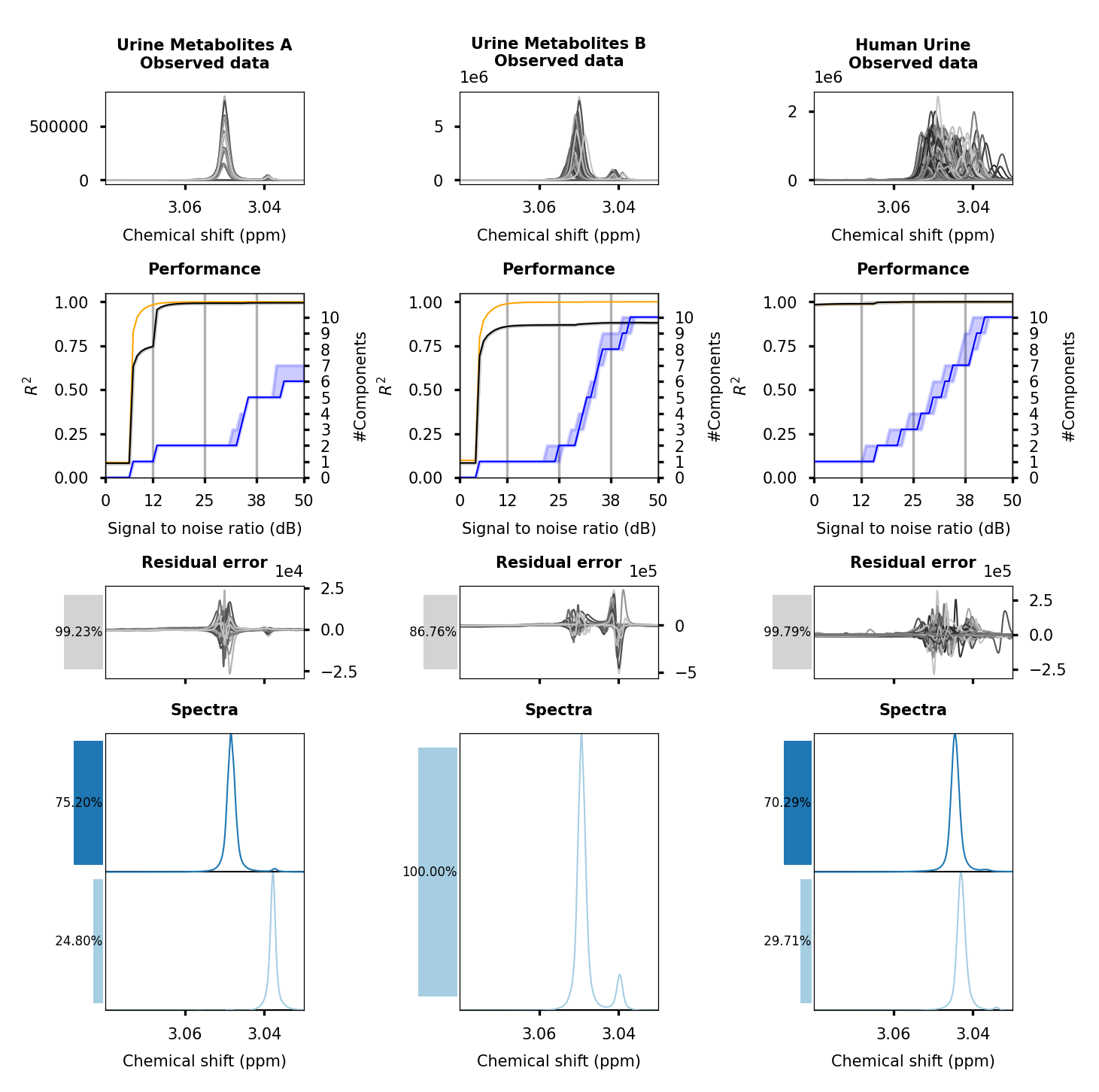}    
    \caption{Case 2: BSI-NMF applied to interval 3.03 to 3.08 ppm in Urine Metabolites (A \& B) and Human Urine. For explanation of the individual plots, see caption of Figure~\ref{fig:real:interval:citric}.}
    \label{fig:real:interval:creatine}
\end{figure}

This case considered the subinterval of 3.03 to 3.08 ppm containing creatine and creatinine signals, which are important bulk metabolites in urine and creatinine can be used to determine the dilution of urine \cite{Vought1963,Le2020}.
For Urine Metabolites A, the BSI-NMF model identified a two component solution, which can be observed from the distinct drop in R\textsuperscript{2} going from $100\%$ to $75\%$  (Figure~\ref{fig:real:interval:creatine}). A similar clear recommendation of the correct number of components was observed in Urine Metabolites B, where the drop in R\textsuperscript{2} indicated that only one component was needed to describe the data. This result was due to the experimental design in which creatinine and creatine were placed in the same stock solution and, therefore, co-varied in concentration and chemical shifts. 

For the Human Urine dataset, two components were well supported by BSI-NMF, resulting in creatine and creatinine latent spectra. The $R^2$ measure for selecting the optimal number of components decreased only slightly when creatine was not included in the model, due to its small contribution to the overall signal. 

\subsection{Case 3: Urea and Cis-aconitic Acid}
\label{sec:results:urea}
This case considered an interval from 5.66 to 5.987 ppm containing urea, and it presented a difficult challenge due to the broad urea signal and the large changes in chemical shift of cis-aconitic acid. Here, only the Human Urine dataset was used, as these metabolites were not included in the Urinary metabolite mixtures.

\begin{figure}[tbp]
    \centering
    \includegraphics[width=0.95\linewidth]{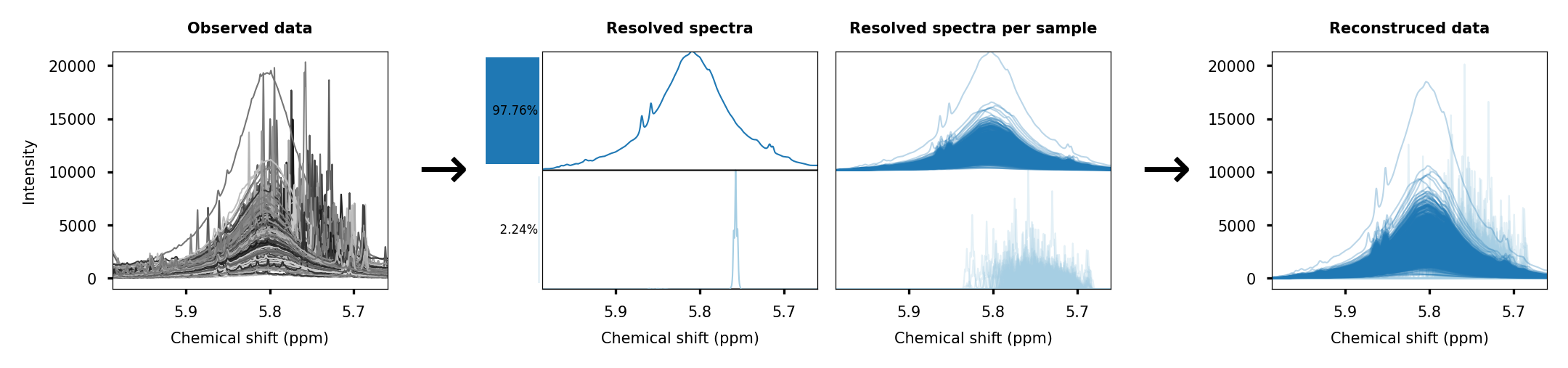}
    \caption{Case 3: BSI-NMF applied to interval 5.66 to 5.987 ppm on Human Urine. Observed \textsuperscript{1}H data, followed by the resolved spectra from a two component model (R\textsuperscript{2} $99.56\%$)as well as the resolved spectra with sample specific shift and concentration. Lastly, the fully reconstructed data.}
    \label{fig:interval:urea:1B}
\end{figure}

The BSI-NMF method identified a two component model as the right model order (Appendix, Figure~\ref{fig:app:allcomp:human:urea}). In Figure~\ref{fig:interval:urea:1B}, the results of the two component BSI-NMF model is shown including the observed \textsuperscript{1}H NMR spectra, the two resolved spectra, the two reconstructed spectra individually as they shift and change in concentration, and the reconstructed data. The two resolved spectra show the broad urea peak together with other co-varying peaks and the triplet of cis-aconitic acid. The co-varying peaks together with the urea peak had small magnitudes and the reliance on $R^2$ to assess model fit limits the discovery of metabolites with low concentrations (relative to urea). Increasing the number of components split urea from the co-varying peaks, see Appendix, Figure~\ref{fig:app:allcomp:human:urea}.

The component representing urea (with co-varying peaks) can also be recovered with a one component model for NMF, B-NMF, and B-CoShiftNMF, see Appendix, Figure~\ref{fig:app:allcomp:human:urea}. However, the signals of the triplet cis-aconitic compound were not resolved.

\subsection{Comparison to peak fitting and \textit{i}coshift}
\label{sec:results:comparison}
The results of BSI-NMF for the different cases and datasets were validated against two existing methodologies for quantification, namely peak picking and fitting, as well as \textit{i}coshift with NMF. The full details on the comparison are available in Section~\ref{sec:methods:comparision}.

For Urine Metabolites A and B, the estimated concentrations for each method were compared to the known concentration levels (in percentages 0-20-40-60-80). For the Human Urine dataset, the true concentrations were unknown and the estimated concentrations of \textit{i}coshift with NMF and Peak fitting were compared to BSI-NMF. Scatter plots and comparison were based on $R^2$ coefficients between the concentrations and are given in Appendix~\ref{sec:app:comparision}, Table~\ref{tab:comparison} for case 1 and case 2.  

For Urine Metabolites A and B, all the methods performed equally well in determining the level of citric acid, dimethylamine (DMA), and creatinine in the samples. However, peak fitting was not able to fully resolve creatine ($R^2=0.81$) and ($R^2=0.96$) in Urine Metabolites A and B, respectively.  In contrast, BSI-NMF perfectly recovered creatine.

For human urine, all the methods performed equally well in determining the level of citric acid, but DMA was only correctly estimated by peak fitting and BSI-NMF. Since \textit{i}coshift with NMF and BSI-NMF should theoretically be able to find the same solution, three hundred NMR spectra were manually examined and aligned in smaller batches, which is denoted DMA (man.) in the table. This resulted in a perfect correlation with BSI-NMF, but the peak fitting (not further optimized) then exhibited poorer performance $R^2=0.86$. Creatinine, which is the signal with higher intensity, is estimated for BSI-NMF and peak fitting almost equally with $R^2=0.93$, but for creatine a larger difference between BSI-NMF and peak fitting was observed with $R^2=0.47$. A manual examination of one hundred spectra for correct peak picking and fitting of creatinine and creatine (and not some side shoulders or other peaks) improved the correlation of BSI-NMF and peak fitting to almost $R^2=0.98$.

\section{Discussion}\label{sec:discussion}

Computer simulated data has shown that only BSI-NMF was able to successfully recover the underlying metabolite information despite individual sample and metabolite specific chemical shift changes. The NMF, B-NMF and B-CoshiftNMF methods failed as they do not account for the metabolite level shifts, and accordingly, they were unable to explore this crucial source of information to uniquely recover the metabolites. This is a well known issue of NMF when the data does not sufficiently span the positive orthant \cite{paatero1994positive,donoho2003does,laurberg2008theorems,huang2013nmf}. The lack of unique recovery for models imposing sample level shifts (i.e., the existing (I)CoShift procedure and our integrated B-CoShiftNMF) has previously been discussed \cite{harshman2003shifted} and followed by the observation that sample level shifts were equivalent to shifting each sample before analysis, i.e., forming an optimal CoShift followed by solving a standard NMF problem with its associated conditions established for uniqueness.  Importantly, BSI-NMF was able to \emph{uniquely} recover the true underlying spectra by exploring the chemical peak shifts unique to the metabolites. Uniqueness properties in shift-invariant decompositions have previously been discussed  \cite{harshman2003shifted, Morup2007} and it has been concluded that component-wise shifts strengthen model uniqueness. For detailed uniqueness discussion, see supplementary section~\ref{sec:methods:shiftnmf:unique},

While B-CoShiftNMF failed to identify the correct number of components under metabolite level shifts, it was observed that B-CoShiftNMF (in the two component case) correctly found and aligned the signal to the one component explaining most of the data and then added one component for each unique shift value (at most $N$ unique shift values). 

From the Urinary Metabolite mixtures (A and B) and the Human urine, it was demonstrated that the BSI-NMF was able to uniquely recover metabolites when other approaches failed. Furthermore, it was observed that both B-CoShiftNMF and BSI-NMF produced unique solutions in the Urine Metabolites B dataset well reflecting the underlying metabolites whereas standard NMF and B-NMF failed. This indicates that only shifts at the sample level and not metabolite level were present in this dataset and the presence of pure samples are sufficient for uniqueness.

Interestingly, BSI-NMF split the citric acid doublets into two components for the human urine data set. This was attributed to the sensitivity of citric acid to pH changes, which is known to influence the chemical shift, J-coupling and the chemical shift distance between the two doublets \cite{Moore1994,Madurga2017}. A combination of different levels of chemical shift changes and an increased distance between the doublets resulted in different shift patterns leading to a three component model using BSI-NMF. 

A key advantage of the proposed BSI-NMF method is the ability to separate multiple overlapping spectral signatures with individual chemical shift changes. In addition, if the data was sufficiently shifted, BSI-NMF returned unique components, whereas NMF, B-NMF and B-CoShiftNMF failed to do so. This was empirically validated on computer simulated datasets $N=19$, two laboratory made Urine Metabolite mixture datasets (A \& B) ($N=42$), and human urine data obtained from $N=2439$ individuals. Here, it was shown that when sufficient shift diversity across metabolites was present, a unique solution could be found. However, if shifts were not diverse, the problem remained like an NMF problem and required the uniqueness conditions of NMF for correct recovery. 

This study casts BSI-NMF in the context of Bayesian inference to exploit tools for automatic relevance determination (ARD) by use of component wise priors whose length scales were learned. While useful to prune superfluous components due to ambiguities or noise, the ARD methodology was highly sensitive to the assumed signal-to-noise ratio (or equiv. noise variance). Learning the noise variance as part of model fitting was problematic both because it was difficult to determine, but more importantly, the definition of noise is unclear. What is noise to a modelling framework may not be considered noise by a trained chemist and vice versa. Instead, a pruning path framework was proposed, based on an initial assumed noise level (practically zero noise), which was gradually increased (to practically everything is noise). Looking at a combination of model fit (R\textsuperscript{2}) and number of components, identification of the correct number of components was possible, when a minimal number of components was chosen before substantial reductions in the fit quality was observed.

The BSI-NMF model is a promising methodology in the context of \textsuperscript{1}H NMR for uncovering metabolites in complex biological mixtures. In contrast to existing work, it solves the chemical shift change problem in the latent space ($\C$) instead of the data space ($\X$) which is essential when having multiple overlapping metabolites and does not introduce artifacts (such as horizontal lines or deformed peaks) in the resolved spectra. BSI-NMF can be reduced to sample level shifts shared across components (B-CoShiftNMF) which obviates the need to define a reference spectrum for the alignment and which is proposed as an alternative to the currently used two step procedure applying \textit{i}coshift to the raw data and decomposing the spectra by NMF.

BSI-NMF uncovered metabolites in complex mixtures where existing approaches were limited or failed. This unique recovery relied on sufficient component level shifts. This study thereby established that the chemical shift changes - previously perceived as a nuisance in \textsuperscript{1}H NMR data - can be instrumental for the successful unique recovery of metabolites and paradoxically controlling for chemical shift changes may in fact limit the ability to recover the underlying metabolites. It is therefore, recommended to avoid experimentally controlling for chemical shift changes and instead use this important source of information actively in the modeling and further utilise the information for metabolite annotation and biological interpretation. In principle, it is desirable to ensure sufficient shift by making experiments that induce chemical peak changes, which is potentially powerful when combined with suitable modeling procedures such as the proposed BSI-NMF. If sufficient chemical shifts are absent or multiplets of the same order, such as singlets, cross each other on the ppm axis (crossing peaks), we expect more advanced acquisition procedures and modeling approaches are needed for unique recovery, such as 2D NMR methods combined with higher-order arrays whose decomposition has strong uniqueness conditions \cite{kruskal1977three}. Notably, the proposed shift-invariant modeling is generic and can be applied to other data domains beyond the considered including chromatography and time resolved spectroscopy. As such, we expect the approach if suitably extended to also enable enhanced component recovery in 2D NMR data. Future work should thus explore the use of the BSI-NMF framework to other domains.

\subsection*{Acknowledgements}
All authors were supported by the Independent Research Fund Denmark, grant no. 10.46540/2035-00294B.

\newpage
\FloatBarrier
\section{Methods}\label{sec:methods}


\subsection{Datasets Simulated, Artificial, and Biological}
\label{sec:methods:datasets}
This section describes how the different datasets where generated or acquired, e.g. the datasets for the simulation study (Section~\ref{sec:methods:simstudy}), the laboratory made Human Metabolites A and B datasets (Section~\ref{sec:methods:data:urineA} and \ref{sec:methods:data:urineB}), and the Human Urine dataset (Section~\ref{sec:methods:data:human}).

\subsubsection{Simulation Study Design}
\label{sec:methods:simstudy}
Computer simulated samples are used to investigate the strengths and limitations of methods when the data generation and noise is exactly controlled. This highlights the difference in the following scenarios; 1) No shift and no spectra overlap. 2) No shift. 3) Sample specific shift only (global shift). 4) Compound specific shift. 

Two compounds are simulated in the interval 2.3 to 2.75 ppm based on their peak assignments in \textsuperscript{1}H NMR (500 MHz, H\textsubscript{2}O) from the Human Metabolite Database, specifically citric acid (\url{https://hmdb.ca/metabolites/HMDB0000094}) and dimethylamine (\url{https://hmdb.ca/spectra/nmr_one_d/1077}). Citric acid consists of two doublets and has intensities $[0.6328, 0.9238, 1, 0.5787]$ at $[2.67, 2.64, 2.54, 2.51]$ ppm, respectively. Dimethylamine is a singlet and has intensity $1$ at $2.5$ ppm. The peak assignment for the compounds are then convoluted with an approximate Lorentzian peak shape model with $(\sigma_1, \sigma_2)=(0.0031, 0.001)$ and $(\sigma_1, \sigma_2)=(0.001, 0.0003)$ to generate the spectrum for Citric Acid and Dimethylamine, respectively. The approximate Lorentzian peak shape model is given by the sum of two Gaussian's, i.e. $Lorent(\sigma_1, \sigma_2) = \DN(0,\sigma_1) + \DN(0,\sigma_2)$. 

The ppm axis is simulated with a resolution of 0.001 ppm between 2.3 and 2.75 ppm, such that $T=1801$. Concentrations are simulated from 5 to 95 in steps of 5 for one compound and from 95 to 5 for the other compound such that the concentrations over compounds sums to 100 for each sample - thus $N=19$. Sample $\tau_n$ and compound $\tau_{n,d}$ shifts are drawn from a $\texttt{Uniform}(0,0.05)$ distribution thus allowing a maximum shift of 0.05 ppm. For visualization purposes, we only simulate positive shift (toward higher ppm). Lastly, homoscedastic white noise is added such that the signal to noise ratio is 50 dB. The resulting data $\X \in \mathbb{R}^{N\times T}$ and ground truth spectra are shown to the left in Figure~\ref{fig:simulated} and similarly in Figure~\ref{fig:app:simulated}. This covers the simulation for when either sample shifts, scenario 3), or compound shifts, scenario 4) are used. For no shifts, scenario 2), the shift is set to zero $\tau_{n,d}=0 \forall_{n,d}$. Similarly, for no shift and no overlap, $\tau_{n,1}=0$ for Citric Acid and $\tau_{n,2}=40$ for Dimethylamine which causes the two compounds to have zero spectral overlap. Notably, there are no pure samples as this often the case in the real setting and having $D-1$ pure sample will ensure the components span the non-negative orthant in the sample mode ($\C$ matrix) and thus have unique concentration and spectra recovery by design.  


\subsubsection{Artificial Mixture: Urine Metabolites A}
\label{sec:methods:data:urineA}
20 metabolites commonly found in human urine were chosen (betaine, creatine, creatinine, alanine, phenylalanine, acetic acid, citric acid, cystein, threonine, glutamine, tryptophan, tartaric acid, histidine, hippuric acid, succinic acid, dimethylamine, ethanolamine, methanol, trimethylamine, and trimethylamine N oxide) and divided into three groups representing one stock solution each.

\textbf{Stock solution A:} creatinine ($\geq$98 \%, Sigma-Aldrich, Germany), acetic acid ($\geq$99.7 \%, Sigma-Aldrich, Germany), citric acid ($\geq$98.5 \%, Sigma-Aldrich, Germany), L-alanine ($\geq$99.5 \%, Fluka Chemie, Switzerland), hippuric acid ($\geq$98 \%, Sigma-Aldrich, Germany), trimethylamine N oxide (95 \%, Sigma-Aldrich, Germany), methanol (HPLC grade, VWR, Germany), L-phenylalanine ($\geq$99 \%, Acros Organics)

\textbf{Stock solution B:} betaine ($\geq$99 \%, Sigma-Aldrich, Germany), dimethylamine (40wt \% in H$_2$O, Sigma-Aldrich, Germany), L-cysteine (97 \%, Sigma-Aldrich, Germany), succinic acid ($\geq$99 \%, Sigma-Aldrich, Germany)

\textbf{Stock solution C:} creatine (anhydrous, Sigma-Aldrich, Germany), ethanolamine ($\geq$99 \%, Sigma-Aldrich, Germany), L-threonine ($\geq$99.5 \%, Fluka Chemie, Switzerland), L-glutamine ($\geq$99 \%, Sigma-Aldrich, Germany), trimethylamine ($\sim$45 wt \% in H$_2$O, Sigma-Aldrich, Germany), L-tryptophan ($\geq$98 \%, Sigma-Aldrich, Germany), DL-tartaric acid ($\geq$99 \%, Sigma-Aldrich, Germany) and L-histidine ($\geq$99 \%, Sigma-Aldrich, Germany).

The stock solutions were prepared 12.5-times higher than the concentration found in real urine samples, while the ratio of the metabolites represents the ratio of real concentration measured in a previous study \cite{Bouatra2013}. All stock solutions were prepared in Milli-Q water and left to equilibrate at room temperature for 24 h. A triangular design of experiment with six levels (0 \%, 20 \%, 40 \%, 60 \%, 80 \%, 100 \%) was chosen for this synthetic data with the corner points of the triangle corresponding to 100 \% of one stock solution (e.g. stock solution A: 100 \%, B: 0 \%, C: 0 \%). This led to a total of 21 stock mixtures.

For preparation of the samples 750 µL Milli-Q (Merck Millipore, Germany)  water were transferred in a 1,5 mL Eppendorf tube with 150 µL of one of the stock mixtures and 100 µL phosphate buffer (1.5 M KH$_2$PO$_4$ ($\geq$98\%, Sigma-Aldrich, Germany) and K$_2$HPO$_4$ ($\geq$98\%, Sigma-Aldrich, Germany), pH 7.4 in D$_2$O (99.9\% atom D, Euroisotop Laboratories, France) with 0.1 \% TSP (3-(Trimethylsilyl)propinic acid-d$-4$, sodium salt, $\geq$98 \%, Sigma-Aldrich, Germany) and 2 mM sodium azide (NaN$_3$ $\geq$99.5 \%, Sigma-Aldrich, Germany) \cite{Khakimov2020}), adding up to 1 mL. The sample were then stored at -60º C until acquisition. For acquisition samples were thawed for 30 min at room temperature, vigorously vortexed, checked for any precipitation and 600 µL transferred into 5 mm NMR tubes (4’’ tubes from LabScape™) suitable for the SampleJet. Samples were prepared in duplicates (n = 42). For acquisition samples were randomized with additional QC sample (n = 5) to ensure spectrometer stability.

Data acquisition was performed on an Avance III 600 MHz (operating at 600.13 MHz) with a 5 mm broadband inverse (BBI) probe and a SampJet (Bruker BioSpin, Ettlingen, Germany) running with TopSpin Version 3.6 and IconNMR. Samples were stored at 277 K in SampleJet, then heated for 5 min at 297 K in the heating unit, before transfer into the magnet for another 5 min equilibration time inside the magnet \( \mathrm{T} = 297 \pm 0.1~\mathrm{K} \).  Afterwards, automatic locking, automated tuning and matching as well as topshim routine were performed within IconNMR. As for acquisition method, one dimensional proton spectra were recorded. Noesypr1d (Bruker library) was chosen for water suppression with the following acquisition parameters: number of dummy scans was 4, number of scans was 16, 60k data points were acquired during the acquisition time of 5 s for a spectral width of 10 ppm. Recycling delay (d1) was set to 35 s to ensure relaxation of spin, which was previously checked with T1 relaxation measurements. Mixing time of the noesy block was set to 10 ms. The receiver gain was optimised on the most concentrated sample, while the 90º hard pulse was optimised on the quality control (QC) sample and both parameters were kept constant for all other samples. 

Data processing was done in TopSpin Version 4.3.0 in batch mode using qumulti. Zero-filling was performed to 131k data points, an exponential window function and a line-broadening of 0.3 Hz were applied before Fourier transformation. Afterwards automatic phasing (apk) and automatic baseline correction (abs) were applied. Then chemical referencing was performed after checking the spectra manually for correct phasing or baseline. Data was then transferred to Matlab.

\subsubsection{Artificial Mixture: Urine Metabolites B}
\label{sec:methods:data:urineB}
Similar to Artificial mixture A, the 2nd artificial mixture contains commonly found human urine metabolites. Metabolites were slightly adapted and 18 metabolites where chosen and have again been divided into three stock solutions:

\textbf{Stock solution A:} betaine, creatine, creatinine, L-cysteine, L-alanin, DL-serine, DL-phenylalanine

\textbf{Stock solution B:} acetic acid, citric acid, hippuric acid, lactic acid ($\geq$98.5 \%, Sigma-Aldrich, Germany), succinic acid, DL-tartaric acid

\textbf{Stock solution C:} dimethylamine, ethanolamine, methanol, trimethylamine and trimethylamine N oxide

Preparation followed the same procedure as for artificial mixture A, however, the concentration was slightly higher so that the initial concentration of each metabolite was 20 times the average of the concentration found in human urine studies.

\subsubsection{Human Urine Dataset}
\label{sec:methods:data:human}
Data was taken from CHANCE project \cite{Trimigno2019} and exact sample preparation is noted there. In short, samples were frozen after collection and shipped from recruitment centers in five countries to two NMR laboratories. Italy, Serbia, and the UK were analyzed in Italy (CERM, Florence), while Lithuania and Finnland samples were shipped to Denmark (University of Copenhagen, Copenhagen). 630 µL of urine were centrifuged (14000 rcf) for 5 min and 540 µL supernatant was mixed with 60 µL phosphate buffer (1.5 M KH\textsubscript{2}PO\textsubscript{4} and K\textsubscript{2}HPO\textsubscript{4}, pH 7.4 in D\textsubscript{2}O with 0.1 \% TSP and 2 mM sodium azide). A final volume of 450 µL of sample were transferred into a 4.25 mm NMR tube. Samples were anaylysed at 303.1 K on a 600 MHz spectrometer (Bruker Biospin) using an automatic sample changer. Proton data was acquired with noesypr1d sequence, 64 scans, 64k data point, acquisition time of 2.7 s, relaxation delay of 4 s and mixing time of 100 ms. Spectra were processed with zero-filling to 131k, multiplied with an exponential window function, applied a line-broadening of 0.3 Hz, then Fourier transformed, phased and baseline correct using apk and abs function. The final number of samples was n = 2439. 
Further data analysis was performed in Matlab: Chemical shift referencing on the TSP signal at 0.000 ppm.

\subsection{Processing Details}     
\label{sec:methods:experimentaldetails}
The generation of the computer simulated datasets are given in Section \ref{sec:methods:simstudy} and results in Section~\ref{sec:app:simstudy}, the experimental details of the laboratory made synthetic mixtures of Urine Metabolites (A and B) and a description of the Human Urine dataset are given in Section~\ref{sec:methods:datasets}, while the results of the full comparison of all methods for all 10 component model order can be found at Section~\ref{sec:app:allcomponents}. Each dataset was processed as follows:

\begin{enumerate}
    \item Estimate noise variance, $\tilde{\sigma}_\epsilon^2$.
    \item Divide the dataset into interval(s).
    \item Scale the interval data for each sample $\x_n$ by dividing by $\alpha_n$ which balances the noise variance $\tilde{\sigma}_\epsilon^2$ and Frobenius norm $||\x_s||_2^2$ to address that sample concentrations can vary by orders of magnitude.
    \item Setup methods specification.
    \item Apply each method in each interval.
    \item Scale the learned concentrations $\C=[\c_1,\c_2,\ldots,\C_N]$ by multiplying $\c_n$ by $\alpha_n$ to recover the original scale.
    \item Address uniqueness in connection scaling and permutation of the components and shift-invariance. 
\end{enumerate}

In step 1., for each dataset, the signal to noise ratio is estimated by identifying a region without any chemical signals, this region was 9.5 to 10 ppm in each dataset\footnote{The Human Urine datasets had a few samples with visible structure was present, but most samples had
no structure, so the effect of these structures is averaged out over the samples. }. In this region, the variance over ppm for each sample was calculated and then the mean variance across samples used as the noise variance, e.g. $\tilde{\sigma}_\epsilon^2 = \frac{1}{N}\sum_{n=1}^N \mathtt{var}(\x_n^{\text{9.5 to 10 ppm}})$.

In step 2., the intervals were chosen to represent scenarios highlighting different challenges of \textsuperscript{1}H NMR data. While it is possible to apply the developed methods to the full ppm range, this is ill-advised as it is computationally more demanding than dividing the dataset into intervals - due to both size of the data and number of chemicals.

In step 3., within each interval each sample is scaled to account for that fact that compound concentrations can vary by several orders of magnitude between samples. However, naive scaling  can inflate the noise if no or low concentration compounds are present. To avoid inflating the noise, we use a threshold $\beta$ and scale each sample $\x_n$ by $\alpha_n = \sqrt{\mathtt{max}(\beta, ||\x_n||_2^2)}$. For setting $\beta$, we assume a worst case scenario of a sample $\x_n$ that is entirely noise, we then assume this noise is normal distributed with zero mean and an unknown noise variance $\epsilon_{n,t} \sim \DN(\mu,\sigma_\epsilon^2)$. For $T$ draws from this distribution, the largest expected value of the noise can be estimated using approximate order statistics for normal random variables \cite{Royston1982} by
\begin{align}
    \epsilon_{\max} = \mathbb{E}\left[\max([\epsilon_{n,1},\epsilon_{n,2},\ldots,\epsilon_{n,T}])\right] \approx \mu + \sigma_\epsilon \Phi^{-1}\left(\frac{T - \frac{\pi}{8}}{T-\frac{\pi}{4}+1}\right)
\end{align}
where $\Phi^{-1}(\cdot)$ is the inverse cumulative distribution function for the standard normal distribution. We then assume a noise signal, were all values are at maximum $\epsilon_{n,t}  = \epsilon_{\max} \forall_t$ and define $\beta = ||\epsilon_{\max}\cdot T||_2^2$. For each dataset, we let  $\mu=0$ and use the estimated noise variance in step 1.

In step 4., the method specification are setup, here we used a convergence criteria of relative improvement in fit by $10^{-6}$ or maximum of $50.000$ iterations and fit each method with ten repeats. For each repeat, the baseline component was initialized as $\s_0 = \1$ and $\c_{n,0} = \mathtt{min}(\x_n)$ and then $\C$ and $\S$ were initialized from a random uniform distribution and the multiplicative NMF algorithm (from \cite{Gillis2012}) was fitted for five iterations on $\X - \c_0\1^\top$- to ensure proper scaling of the components. Then NMF, CoShiftNMF, or SI-NMF was fitted for 25 iterations without Bayesian automatic relevance determination, the result of which was passed to B-NMF, B-CoShiftNMF, or BSI-NMF and run until convergence or $5.000$ iterations. This was initialization was repeated 10 times at the highest assumed signal-to-noise ratio (50 dB) and the best of these ten initializations was then used fitting the methods. Note, that for B-NMF, B-CoShiftNMF, and BSI-NMF the updating of automatic relevance determination prior $\lambda_d$ started at iteration 5 while shift modeling started at iteration 10. For both CoShiftNMF and SI-NMF, the shifts were initialized by sample specific shifts of the data to maximize the average cross-correlation across all samples. For BSI-NMF, a maximum allowed shift was specified common to all components within a given interval and dataset, this was identified visually to get a rough NMR valid shift constraint.

In step 5., for NMF a specific number of components where chosen and it was run with $D=1,2, \ldots, 10$. For B-NMF, B-CoShiftNMF and BSI-NMF, the method where initialized with $D=10$ components at the highest assumed signal-to-noise ratio (SNR) of 50 dB, after convergence at this SNR level the learned model was used as input to the next highest SNR level 49 dB. The was repeated until SNR was 0 dB. Note, the data was padded with $T\cdot 0.25$ timepoints on the left and right of every sample using zero-order hold.

In step 6., the learned concentrations $\C$ were scaled to match the original data scale. For sample specific shifts, the learned shifts were shifted on the ppm axis, such that the  weighted average shift $\frac{1}{N}\sum_{n=1}^N c_{n,d}\tau_{n,d}\frac{\alpha_n}{\sum_{n'=1}^N \alpha_{n'}}$ was zero. For BSI-NMF this was not done, as the shifts where constrained. 

In step 7., we address the essential uniqueness of NMF, by scaling the components so each spectrum has a maximum value of 1, e.g. we find the scaling $\rho_d =\max(\s_d)$ and then let $\s_d$ be $\s_d/\rho_d$ and $\c_d$ by $\c_d \cdot \rho_d$.  Then we reorder the components, so that the one with the highest contribution is the first component, second highest second, etc., the contribution for component $d$ is calculated as $\frac{\sum_{n,t}c_{n,d}\s_{t,d}}{\sum_{n,t,d'}c_{n,d'}s_{t,d'}}$. When shifts are moved after fitting, then the component spectrum $\s_d$ is redefined by shifting it accordingly.

\subsection{Non-negative Matrix Factorization}
\label{sec:methods:nmf}
Non-negative Matrix Factorization (NMF) is a powerful technique for decomposing complex data sets into interpretable components \cite{Lee1999,paatero1994positive}. In the context of NMR spectroscopy, NMF can be used to estimate overlapping spectra where no sample or compound shift is present. In the chemistry and chemometrics literature, this method also covered under the term multi-variate curve resolution (MCR) \cite{deJuan2021}.

NMF decomposes a non-negative matrix $\X \in \mathbb{R}^{N \times T}$ with $N$ samples and $T$ timepoints into two non-negative matrices, a latent concentrations matrix $\C \in \mathbb{R}^{N \times D}_{\geq 0}$ and a latent spectra matrix $\S\in \mathbb{R}^{T \times D}_{\geq 0}$ with $D$ latent components, such that
\begin{align}
    \X &= \bepsilon + \sum_{d=1}^D \c_d \s_d^\top = \bepsilon + \C \S^\top, \quad \text{s.t. } \C\geq 0, \S\geq 0 \label{eq:nmf}
\end{align}
where $\bepsilon$ is the residual error. Estimating $\C$ and $\S$ is typically done via least squares optimization, $\argmin_{\C,\S} ||\X-\C \S^\top||_F^2$, see \cite{Lee1999,Gillis2012}. From an NMR perspective, the latent components are ideally identical to metabolite signals or chemical compounds, but in practice also includes baseline component(s). The key assumption of NMF is that the latent components ($\C$ and $\S$) are bi-linear and non-negative and the data reconstructed through their product. Importantly, NMF assumes no specific form of the underlying signal and is thus able to capture multiple peaks, multiplets, and lineshapes in one compound if supported by the data. In contrast, methods based on shape assumption - so-called lineshapes - will split the signal from one compound into separate peaks to adheed to the a specific lineshape assumption - moreover any violations to the lineshape assumption will results in artifacts. 

For interpreting the spectra and getting the right concentrations in NMR, a unique solution is essential, as a non-unique solution allows the mixing of components, such that even when $\C$ and $\S$ perfectly describing the data, the true concentrations and spectra are not discernible. When the factorized solution $(\C,\S)$ is not unique, it means the found factor matrices can be rotated by a matrix $\R$ such that the approximation error is the same., e.g. $||\X-\C\S^\top||_F^2=||\X-\C\R\R^{-1}\S^\top||_F^2=||\X-\tilde{\C}\tilde{\S}^\top||_F^2$. In general, NMF does not find unique solutions unless the data spans the non-negative orthant, see Section~\ref{sec:methods:shiftnmf:unique}.

\subsection{Shift-invariant Non-negative Matrix Factorization}
\label{sec:methods:shiftnmf}

Shift-invariant Non-negative matrix factorization (SI-NMF) \cite{Morup2007,morup2008shiftcp} is a modelling framework that extends the capabilities of NMF by allowing sample $n$ and component $d$  specific shifts $\tau_{n,d} \,\forall_{n,d}$ for one of the factors (here $\S$, the spectral component). We propose using SI-NMF to account for \textsuperscript{1}H NMR data as it can resolve both overlapping multiplet signals (due to the bilinarity assumption) and account changes caused by compound specific chemical shifts (due to sample- and component-specific shifts). The SI-NMF model can be formulated for the $n^{th}$ sample as,
\begin{align}
    x_{n,t} &=  \epsilon_{n,t} + \sum_{d=1}^D c_{n,d} s_{t-\tau_{n,d},d} 
    \label{eq:shiftnmf},
\end{align}
such that 
\begin{align}
    \x_{n} &= \bepsilon_n +\c_{n} \S_{\btau_n}^\top \
    ,
\end{align}
where $n=[1,\ldots,N]$ and $\X \in \mathbb{R}^{N\times T}$ is the data with $N$ samples and $T$ timepoints (ppm axis in NMR), $\bepsilon_n$ is the residual error for the $n^{th}$ sample, $\C=[\c_1,\ldots,\c_N]$ are the latent concentrations, $\S$ are the latent spectra, and $\tau_{n,d}$ is the sample and component specific shifts (with $\btau_n=[\tau_{n,1},\ldots,\tau_{n,D}]$). The concept is that $s_{t,d}$ is shifted by $\tau_{n,d}$ indices in the time point axis, e.g. $s_{t-\tau_{n,d},d}$ and we use the short hand notation $\S_{\btau_n}$ to denote that the latent spectra $\S$ are shifted according to ${\btau_n}$. This facilitates that a positive $\tau_{n,d}$ shifts the latent spectra $\s_d$ for sample $n$ to higher ppm values and a negative shift value shifts the spectra to lower ppm values. 
The shifts are calculated in the frequency domain (Fourier transform) for computational efficiency which assume circular shifts. This means that shifting over the right of the axis (index $t=T$) moves the shifted signal to the beginning of the axis ( index $t=1$) and vice versa. In practice, the data can be padded to mitigate the influence of this assumption. We found zero-order hold - first value on left and last value on right end, respectively - works best for NMR data, but zero-padding or min padding may be of use for other application areas. This paper only considers shifts in integers (whole indices in $[1,T]$), but non-integer shifts are possible via gradient based optimization of the shift \cite{morup2008shiftcp}. 

An important distinction is that SI-NMF does not shift or modify the data $\X$ as done by existing methods such as coshift/\textit{i}coshift \cite{Savorani2010-dz}, warping\cite{Tomasi2004,NIELSEN1998}. Instead, it shifts each latent component in each sample to match the data. This allows SI-NMF can be applied directly to the NMR data without any alignment preprocessing to extract metabolite profiles. In practice, it is often advantages to still separate the NMR data into several intervals, as computational complexity increases with more timepoints (ppm) and more latent components - larger intervals have more metabolites and thus more components. Additionally, SI-NMF is still optimized under the Frobenius norm, so large difference in intensity of multiplets - in the same interval - is challenging to resolve. We therefore provide a standardization approach for \textsuperscript{1}H NMR in Section~\ref{sec:methods:experimentaldetails}.

\paragraph{CoShiftNMF}
For a single component, $D=1$, SI-NMF is the similar to CoShift followed by NMF if and only if the optimal reference is used as measured by least squares. However, SI-NMF optimizes the the shifts and factors jointly thus finds the optimal the reference ($\s_{1}$) without a priori knowledge. Importantly, this contribution provides a \emph{reference free approach} to aligning data and is beneficial in cases where the reference is unknown or costly to determine\cite{Vu2013-iy}. Based on this insight, we define CoShiftNMF with an arbitrary number of components by restricting the shifts to be sample but not component specific, e.g. $\tau_{n,d}=\tau_{n,d'} \forall_{d',d}$, leading to the following model,
\begin{align}
    \x_n &= 
    \bepsilon_n + \c_{n} \S_{\tau_n}^\top \label{eq:coshiftnmf},
\end{align}
with $\tau_n$ being a scalar value.

\subsection{Bayesian Shift-invariant Non-negative Matrix Factorization}
\label{sec:methods:bsinmf}
Until now, the number of components has been assumed known. We drop this assumption and cast SI-NMF in a probabilistic setting and infer it using Bayesian inference. This enable the use of so-called automatic relevance determination priors \cite{Mackay1995} which are beneficial in pruning unsupported components and thus facilitate easily interpretable results and efficient model inference. The parameters with highest evidence are identified by optimizing the model using maximum a posterior (MAP) estimation \cite{morup2009automatic} as in conventional NMF modeling of \textsuperscript{1}H NMR spectra as opposed to using computationally more demanding sampling or approximate Bayesian inference procedures \cite{Hinrich2018}. This can be achieved by treating $\C,\ \S,\ \bbeta,\ \blambda$ and $\btau$ as random variables and specifying prior probability distributions on the random variables as also described by the generative process in Figure~\ref{fig:1}, as follows,
\begin{align}
    p(\boldsymbol{\beta}) =& \prod_{n=1}^N\Uni(\beta_n|0,\infty),\\
    p(\blambda) =& \prod_{d=1}^D\Exp(\lambda_d|\eta),\\
    p(\C|\blambda) =& \prod_{d=1}^D\prod_{n=1}^N \Exp(c_{n,d}|\lambda_d),\\
    p(\boldsymbol{\tau}) =& \prod_{d=1}^D\prod_{n=1}^N\Uni(\tau_{n,d}|-\tfrac{T}{2},\tfrac{T}{2}),\\
    p(\S|\blambda) =& \prod_{d=1}^D\prod_{t=1}^T \Exp(s_{t,d}|\lambda_d),\\
     \mathcal{L}\left(\X  | \C, \S, \btau, \bbeta, \blambda, \sigma\right) =& \prod_{n=1}^N \DN\left(\x_n | \bbeta_n + \c_{n} \S_{\btau_{n}}^\top,\  \sigma^2\I_T\right),
\end{align}

where $\Exp(y|\alpha)= \alpha\exp\{-\alpha y\}$ is the exponential distribution,  and $\DN(\y|\boldsymbol{\mu},\sigma^2\I_T)$ is the normal distribution with mean $\bmu$ and known noise variance $\sigma^2$, $\Uni(a,b)$ is the Uniform distribution, and $\mathcal{L}(\X|\cdot)$ is the likelihood under a normal distribution.  The same model formulation can be used for a Bayesian NMF (B-NMF) and Bayesian CoShiftNMF (B-CoShiftNMF) by substituting the appropriate model reconstruction formula and adapting the prior, $p(\btau)$, on the shifts. 

We consider three versions of the specified model; a) The BSI-NMF version with sample- and component-specific shifts $\tau_{n,d}$. b) B-CoShiftNMF with sample shift only $\tau_{n,d} = \tau_{n,d'}\, \forall_{d,d'}$. c) The Bayesian NMF (B-NMF) version without sample shift $\tau_{k,d}=0\, \forall_{k,d}$ (as presented in \cite{Hinrich2018}).

Having specified the likelihood function and prior distributions, the parameters are determined by maximizing the probability or equivalently minimizing the negative log of the probability,
\begin{align}
    \argmax_{\C,\S,\btau,\bbeta,\blambda} &\mathcal{L}\left(\X| \C, \S, \btau, \bbeta, \blambda, \sigma\right)p(\C|\blambda)p(\S|\blambda)p(\blambda)p(\bbeta)p(\btau)\nonumber\\
    &\equiv \argmin_{\C,\S,\btau,\bbeta,\blambda} -\log\left(\mathcal{L}\left(\X | \C, \S, \btau, \bbeta,\blambda, \sigma\right)p(\C|\blambda)p(\S|\blambda)p(\blambda)p(\bbeta)p(\btau)\right)\label{eq:shiftnmf:argmin}
\end{align}
where all random variables are optimized via MAP estimation. In principle, the noise variance $\sigma^2$ can also be treated as a random variable and be estimated via MAP, but this leads to underestimation of $\sigma$ as the model overfits to the noise \cite{morup2009automatic}. If the noise variance is unknown - as is commonly the case - this can be addressed by; 1) Fixing $\sigma^2=||\X||_F^2((1+10^{SNR/10})*(nnz(\X)))^{-1}$ at a specific signal-to-noise ratio (SNR) in dB as suggested in \cite{morup2009automatic}. 2) Providing an estimate of the noise variance - for instance by considering the empirical noise variances in an interval (ppm) without chemical signals (as described in Section~\ref{sec:methods:experimentaldetails}). 

We propose using a modification of 1) as we consider a range of SNRs (0 to 50 dB in steps of 1 dB), starting at the highest assumed signal level (50dB) and fitting the first model. Once converged to a local optima, that model is then used as a starting point for the next highest assumed SNR (here 49 dB). This process is repeated until it terminates at the lowest assumed SNR (0 dB). This gives a regularization path which shows the noise assumptions that give rise to different number of components. Ideally, the model should remain expressive at lower SNRs but with lower complexity. Note, that due to the sparsity of NMR compounds, we found the Frobenius norm definition of the noise variance to be insufficient, as it over penalizes initially (at 50 dB or even 100 dB) and is unable to fully prune at low dB (-50 or 0 dB) prunes to zero components. Instead, we define $\sigma^2$ through the infinity norm, e.g. $\sigma^2=||\X||_\infty^2(1+10^{SNR/10}))^{-1}$. The associated objective function and update rules of BSI-NMF are given in Appendix~\ref{sup:sec:objfunandupdates}.

\subsubsection{Uniqueness of BSI-NMF}
\label{sec:methods:shiftnmf:unique}
Model uniqueness is essential in order to be able to correctly uncover the underlying metabolites in \textsuperscript{1}H NMR data. The uniqueness of non-negative matrix factorization has been carefully studied already in the original work of positive matrix factorization in 
\cite{paatero1994positive} considering conditions of so-called p-rotatability (denoting positive rotatability) in which solutions can be modified by a linear operator and its inverse, i.e. $\mathbf{X}\approx \mathbf{CS}^\top=(\mathbf{CP})(\mathbf{P^{-1}S^\top})$ such that the resulting modified factorization remained positive. It was here observed that if the components had structured zero entries, such rotability reduce to simple scaling and permutation (i.e, what is also defined as essential uniqueness). The uniqueness of NMF has been further mathematically formalized in 
\cite{donoho2003does,laurberg2008theorems,huang2013nmf} where it has been established that the data need to suitably span the positive orthant in order for the NMF model to be unique. When this is not the case and uniqueness by the standard NMF cannot be achieved, it has been argued in \cite{laurberg2007affine,laurberg2008theorems} that affine transformation corresponding in the context of \textsuperscript{1}H NMR modeling to removing baseline effects (i.e., sample wise constant offsets) can potentially correct the data characterized by the NMF model to suitably span the positive and thereby enhance uniqueness. However, even baseline correction does not guarantee the uniqueness of solutions. In circumstances where uniqueness cannot be achieved, a common strategy is to regularize the solution space to promote sparse representations of the extracted components, typically based on $\ell_1$ regularization, see also \cite{laurberg2008theorems,huang2013nmf} for a discussion of sparsity constraints and how it can promote model uniqueness. The BSI-NMF include translation invariance by the learned baseline parameter $\boldsymbol{\beta}$ and impose sparse regularization by the exponential priors with length scales $\boldsymbol{\lambda}$. Unfortunately, these existing strategies does not guarantee correct recovery of the underlying metabolites if chemical shift changes occurs despite promoting unique representations. 

In \cite{harshman2003shifted} the shift-invariant matrix factorization procedure was proposed and uniqueness properties of this decomposition discussed. It was here observed that sample level shifts (as used in the context of NMR by the \textit{i}coshift procedure) does not guarantee uniqueness of the decomposition, whereas shifts also at the component level was argued to enable unique representations. When analyzing the shifted factor analysis in its frequency representation it was observed in \cite{morup2007shifted} that whereas shifts can improve upon uniqueness it does not guarantee in itself unique representations. We also observe this lack of uniqueness for the BSI-NMF when no pure compounds are present in Urine Metabolite A and B, see Appendix~\ref{sec:app:waternopure}, where the true spectra and concentrations are linearly intermixed.

Importantly, component wise shifts in the context of the shift-invariant NMF model can substantially enhance the model uniqueness when the standard NMF model fails \cite{Morup2007} as the component shifts and the induced diversity strengthens the models ability to span the positive orthant. This was experimentally also observed in \cite{Morup2007}. 

We illustrate the enhanced uniqueness of shift-invariant NMF considering the following very minimal problem where two latent components with four spectral values is observed across three samples forming a $3 \times 4$ data matrix $\mathbf{X}$ given by
\begin{align}
\mathbf{X}&=\mathbf{CS}=\left[
\begin{array}{cc}
1 & 0 \\
0 & 1\\
0.5 & 0.5\end{array}
\right]\left[
\begin{array}{cccc}
0 &1 & 0 & 0\\
0 & 1 & 1 & 0
\end{array}
\right]=
\left[
\begin{array}{cccc}
0& 1& 0 & 0\\
0 & 1 & 1 & 0\\
0 &  1 & 0.5 & 0
\end{array}
\right].
\end{align}
Notably, this problem is not unique when applying the standard NMF. As such, the following is a also valid two component solution
\begin{align}
\mathbf{CS}=\left[
\begin{array}{cc}
 1& 0  \\
 1& 1\\
 1&  0.5
 \end{array}
\right]\left[
\begin{array}{cccc}
0 & 1 & 0 & 0\\
0 &  0 & 1 & 0 
\end{array}
\right]=
\left[
\begin{array}{cccc}
0 & 1 & 0 & 0\\
0 & 1 & 1 & 0\\
0 & 1 & 0.5 & 0
\end{array}
\right],
\end{align}
thus reconstructing also perfectly the dataset.

However, let us now assume for the last sample that the first component is shifted to the left by one index  (i.e., $\boldsymbol{\tau}=\left[\begin{array}{cc} 0 &0\\ 0 &0\\ -1 &0\end{array}\right]$) such that the observed data matrix becomes
\begin{align}
\tilde{\boldsymbol{X}}=
\left[
\begin{array}{cccc}
 0 & 1 &0 & 0 \\
0 & 1 & 1 & 0\\
0.5 & 0.5 & 0.5 & 0
\end{array}
\right].
\end{align}
The standard NMF solution cannot perfectly account for these data unless using three components producing 
solution where for instance $\mathbf{C}$ can be defined as the data matrix itself and $\mathbf{S}$ the identity matrix, i.e., $\mathbf{C}=\tilde{\mathbf{X}}$ and $\mathbf{S}=\mathbf{I}$ providing a valid but incorrect solution. Notably, The BSI-NMF model solution will be able to correctly recover the true two component solution uniquely, as it can account for the shift in the first component of the last sample. Importantly, this shift reveals the underlying structure of the two latent components, as the latent components have to include a component profile with only one non-zero element to produce the first observed sample. To produce a spectrum with three consecutive non-zero entries in the third sample, this implies that the second latent spectrum must have at least two non-zero elements. However, to construct the second observed sample, we know that this second component can maximally have two non-zero elements. Due to the non-negativity constraint these non-zero elements have to be consecutive, otherwise the second sample would not be possible to construct with zeros in the first and last entries. The two values further have to be identical otherwise it would not be possible to construct the third sample with its values observed. Consequently, the profile structure is revealed as having two non-zero elements of equal value that are next to each other. Had the shift in the third sample instead been one index to the right, i.e.,  $\boldsymbol{\tau}=\left[\begin{array}{cc} 0 &0\\ 0 &0\\ 1 &0\end{array}\right]$ this would not suffice to uncover the two latent components as it would still overlap with the second component and not result in a sufficient span of the positive orthant by the shift. As such, the observed data matrix can be trivially decomposed by an incorrect two component NMF solution 
\begin{align}
\hat{\boldsymbol{X}}=
\left[
\begin{array}{cccc}
 0 & 1 &0 & 0 \\
0 & 1 & 1 & 0\\
0 & 0.5 & 1 & 0
\end{array}
\right]=\left[
\begin{array}{cc}
 1& 0  \\
 1& 1\\
 0.5& 1
 \end{array}
\right]\left[
\begin{array}{cccc}
0 & 1 & 0 & 0\\
0 &  0 & 1 & 0 
\end{array}
\right].
\end{align}
As a result, uniqueness requires the presence of sufficient shifts to reveal the underlying components.

Consequently, the BSI-NMF model for sufficient presence of shifts is essential unique, i.e., unique up to permutation, scaling, and shifting each component by a constant. These redundancies can trivially be resolved by centering the shifts, ordering components by importance and scaling the extracted latent spectra to have maximum amplitude of 1.

\subsection{Comparison methods: \textit{i}coshift with NMF and Peak fitting}
\label{sec:methods:comparision}
The proposed BSI-NMF method is compared to two existing methodologies for estimating the concentration of each metabolite, namely citric acid, dimethylamine, creatinine, and creatine. First, the methodologies are introduced and our implementation of them discussed, then they are applied to all datasets and the results presented in Section~\ref{sec:app:comparision}. The main text includes only a summary of this comparison in Section~\ref{sec:results:comparison} and Table~\ref{tab:comparison}.

Importantly, our implementations have taken great care of optimizing the comparison methodologies to the two cases considered, namely 2.5 to 2.75 ppm and 3.03 to 3.08 ppm, see Section~\ref{sec:results:citric} and \ref{sec:results:creatine}, respectively. Our implementations in MATLAB are available as part of the supplementary code.

The results of BSI-NMF to two existing methodologies for estimating concentrations, peak fitting  and interval co-shifting with NMF (\textit{i}coshift). Peak fitting fits a peak shape function (here a Lorentzian peak) to each peak in the spectrum and peaks are then grouped afterwards.

\subsubsection{\textit{i}coshift with Non-negative Matrix Factorization}
This methods consists of a two-step procedure, first the samples are aligned and then the aligned samples are factorized using non-negative matrix factorization (or equivalently multi-variate curve resolution). \textit{i}coshift \cite{Savorani2010-dz} is an extension of CoShift to align multiple intervals - either manually defined start and end point or automatically based on a number of desired intervals - to a user defined reference spectrum. Then NMF decomposes the aligned dataset into a latent concentration matrix and a latent score matrix, i.e. $\X_{aligned}  = \sum_{d=1}^D\c_d\s_d^\top+\epsilon = \C\S^\top+\epsilon$, for a given number of latent components. 

In this work, we use the \textit{i}coshift \cite{Savorani2010-dz} implementation from University of Copenhagen Chemometrics Group\footnote{\url{https://ucphchemometrics.com/icoshift-an-ultra-rapid-and-versatile-tool-for-the-alignment-of-spectral-datasets/}} with the reference spectra being the maximum value at each ppm over all samples. For the NMF step, we use the proposed BSI-NMF framework, but without no shifts allowed and no automatic relevance determination - equivalent to hierarchical alternating least squares NMF with a baseline component.

\textit{i}coshift was applied to the citric acid and dimethylamine interval (2.5 to 2.75 ppm) for the Urine Metabolites A and B and the Human Urine dataset. While it is possible to apply it to the creatinine-creatine interval (3.03 to 3.08 ppm) for Urine Metabolites A and B, it is not possible in general, as creatinine and creatine are not baseline separated and this overlap prevents identifying sub-intervals that can perfectly align each peak. This is a known limitation of \textit{i}coshift with NMF.

For the Urine Metabolites A and B, we manually defined three intervals separating the dimethylamine in one interval and the doublets of citric acid in two separate intervals. For each interval, we then fit a one component $D=1$ NMF model and then numerically integrated the component using the trapz-algorithm (in MATLAB R2025b). For citric acid, we added the integrated value of the two doublet intervals to estimate its concentration. 

For the Human Urine dataset, we first automatically aligned two intervals and then on this aligned data, we further sub-aligned the dimethylamine peak. Then the three intervals containing dimethylamine, one doublet of citric acid, and another doublet of citric acid where fit with a one component NMF and analysed as above. While fast, this alignment in some cases caused the dimethylamine peak to be shifted into the region of the citric acid doublet providing suboptimal results. Therefore, we manually aligned three hundred spectra of the dataset (in batches with clear separation) to get perfectly \textit{i}coshift aligned data.

\subsubsection{Peak fitting}
Peak fitting  is an approach where peaks are assumed to have a parameterized shape given by a so-called \emph{peak shape function} where the typical choices are Gaussian, Lorentzian, Voigt or Pseduo-Voigt. The choice of peak shape function need to closely match the data or it will introduce modeling artifacts. In this work, we have chosen to use the Lorentzian peak shape function, namely,
\begin{equation*}
    L(x|x_0, w, A) = A \cdot \frac{(0.5 w)^2}{(x-x_0)^2 + (0.5 w)^2}
\end{equation*}
where $x$ is a position on the ppm axis, $x_0$ is the center of the peak, $w$ is the width of the peak at full-width-half-maximum, and $A$ is the amplitude of the peak. To get the estimated spectra, the function is evaluated each ppm value. The area of the peak is found through numerical integration using the integral function in MATLAB R2025b. 

Peak fitting must be applied each peak in the spectrum for every sample and afterwards the peaks have to be grouped within an across samples. There are many was to approach fitting multiple peaks through non-linear optimization methods and the found solution is highly influenced by the starting guess and the allowed range of the parameters. Therefore, we implemented two slight different versions of peak fitting, one tailored at separating citric acid and dimethylamine and another at separating creatinine and creatine.

For the citric acid and dimethylamine, the peaks are well separated in an individual samples and far larger than other contaminating peaks. Therefore, the implement approach takes the maximum value of the NMR spectrum and fits a peak at that position with very little slack allowed. Then this peak is subtracted from the original data and the procedure is run again until five peaks have been found. Afterwards, the dimethylamine peak is easily identified and the remaining peaks belong to citric acid. 

For creatinine and creatine, the peaks are not well separated and the implementation jointly fits to peaks to the spectrum. Here, initialization of the parameters of the two peak shape functions is highly influential. Building on domain knowledge, we know that creatinine is the higher intensity peak and is always at a higher ppm value and vice versa for creatine. Therefore, we find the top of the creatinine and assume a monotonic function with maximum at that point, then track the function towards lower ppm values, then the next top is set as the maximum of the creatine peak. This initialization along with strict confidence regions for the parameters allow a good characterization of both creatinine and creatine.

\ifarxiv
\else
\backmatter
\fi
\newpage

\section*{Supplementary information}
This article contains the following supplementary information:

\begin{enumerate}
    \item BSI-NMF objective function and update rules (Appendix~\ref{sup:sec:objfunandupdates}).
    \item Simulation Study Results (Appendix~\ref{sec:app:simstudy}).
    \item Comparison Study: BSI-NMF, ICoShift with NMF, and Peakfitting (Appendix~\ref{sec:app:comparision}).
    \item Analysis of Urine Metabolite A and B without pure samples (Appendix~\ref{sec:app:waternopure}).
    \item Optimal number of components as selected by BSI-NMF, but for all methods (Appendix~\ref{sec:app:rightcomponents}).
    \item The best solution found for each model and each choice of component (Appendix~\ref{sec:app:allcomponents}).
    \item Full dataset - 0 to 10 ppm - for the two artificial datasets, Urine Metabolites A (Appendix~\ref{sec:app:allcomp:urineA}) and B (Appendix~\ref{sec:app:allcomp:urineB}).
     \item Source code for B-NMF, B-CoShiftNMF, and BSI-NMF 
     as well as code for comparison of BSI-NMF to ICoShift with NMF and peak fitting in MATLAB (Appendix~\ref{sec:app:sourcecode}). 
    \item Data availability statement (Appendix~\ref{sec:app:dataavailability}).

\end{enumerate}

\begin{appendices}

\section{BSI-NMF objective function and update rules}
\label{sup:sec:objfunandupdates}
\subsection{BSI-NMF objective function}
The update rules, presented in Section~\ref{sec:updaterules}, are guaranteed to monotonically decrease the objective function and reach a local optimum. To assess convergence, we explicitly calculate the objective function, which is,
\begin{align}
    -\log&\left(\mathcal{L}\left(\X| \C, \S, \btau, \bbeta, \blambda, \sigma\right) P(\blambda, \C, \S,\btau,\bbeta)\right) = -\log\prod_{n=1}^N\DN\left(\x_n |\beta_n+ \c_n\S_{\btau_n}^\top,\  \sigma^2\I_T\right) \nonumber\\
    &- \log\prod_{d=1}^D\Exp(\lambda_d|\eta) \nonumber\\
    &-  \log \prod_{d=1}^D\prod_{n=1}^N \Exp(c_{n,d}|\lambda_d)  \nonumber\\
    &- \log \prod_{d=1}^D\prod_{t=1}^T \Exp(s_{t,d}|\lambda_d)\nonumber\\
    &-\log \prod_{d=1}^D\prod_{n=1}^N\Uni(\tau_{n,d}|-\tfrac{T}{2},\tfrac{T}{2})\nonumber\\
    &- \log \prod_{n=1}^N\Uni(\beta_n|0,\infty),\\
    =& -\frac{N\cdot T}{2}\log(2 \pi \sigma^2) +\dfrac{\sum_{n=1}^N||\x_n - \sum_{d=1}^D \c_n \S_{\btau_n}^\top||_{F}^2}{2\sigma^2}\nonumber\\
    &-N \sum_{d=1}^D \log(\lambda_d) + \sum_{n=1}^N \blambda \c_n^\top \nonumber\\
    &-T \sum_{d=1}^D \log(\lambda_d) + \sum_{t=1}^t \blambda \s_t^\top\nonumber\\
    &- D\log(\eta) + \eta\sum_{d=1}^D\lambda_d \nonumber\\
    &-N \log(0) + N\cdot D \log(T).
\end{align}
Note, the $\log(0)=-\infty$ comes from the Uniform prior on the baseline $\beta_n$ and can be ignored when calculating the objective function.

For convergence, we calculate the objective function $\omega$ at iteration $i$ and $i+1$ and assess the relative change, $\frac{\omega_{i}-\omega_{i+1}}{\omega_i}$, until it is below the convergence threshold.

\paragraph{BSI-NMF with different regularization}
In the above, we have focused on BSI-NMF with exponential distributions as priors for the latent concentration $\C$ and latent spectra $\S$. This is similar to considering a regularized SI-NMF with $\ell_1$-norm penalization (also known as LASSO penalization) on each component - in the maximum likelihood setting. Another possible regularization is to consider the $\ell_2$-norm penalization (also known as Ridge penalization) on each component. For BSI-NMF this can be done by changing the prior distribution on $\C$ and $\S$ to be  normal distributions truncated at zero and infinity, e.g., 

\begin{align}
    p(\C|\blambda) =& \prod_{d=1}^D\prod_{n=1}^N \mathcal{N}_{[0,\infty]}(c_{n,d}|0,\lambda_d) = \sum_{d=1}^D\sum_{n=1}^N\frac{1}{Z^c_{n,d}} \frac{\sqrt{\lambda_d}}{\sqrt{2\pi}} \exp\left\{-\frac{\lambda_d c_{n,d}^2}{2}\right\}\\
    p(\S|\blambda) =& \prod_{d=1}^D\prod_{t=1}^T \mathcal{N}_{[0,\infty]}(s_{t,d}|0,\lambda_d) = \sum_{d=1}^D\sum_{t=1}^T\frac{1}{Z^s_{t,d}} \frac{\sqrt{\lambda_d}}{\sqrt{2\pi}} \exp\left\{-\frac{\lambda_d s_{t,d}^2}{2}\right\}
\end{align}
where $Z_{i,j}^k=1-\Phi(k_{i,j}\sqrt{\lambda_j})$ is a normalization constant with $\Phi(z)=\frac{1}{2}(1+\mathtt{erf}(z/\sqrt{2})$ being the cumulative distribution function for a standard Gaussian distribution and $\mathtt{erf}$ the error function. For a truncated Gaussian on $[0,\infty]$ then $Z=1/2$. In the context of Bayesian NMF, this was explored in \cite{Hinrich2018}.

Similarly, BSI-NMF can trivially be transformed into BSI-MF by changing the prior on $\C$ and $\S$ to follow normal distribution or Laplace distribution for $\ell_2$ and $\ell_1$ regularization, respectively. 

\subsection{BSI-NMF update rules}
\label{sec:updaterules}
Optimizing BSI-NMF is a non-convex problem, but each independent set of random variables can be split into a convex problem given a fixed value of all other random variables. The resulting subproblem is a convex problem and closed form update rules are derived. Here, we consider only maximum a posteriori estimation (MAP) of the parameters of the posterior distributions. Once all update rules are established, these can be followed in an alternating pattern to arrive at a local optima of the optimization, e.g. eq. \eqref{eq:shiftnmf:argmin}. 

For computational efficiency, the \textsuperscript{1}H NMR dataset $\X \in \mathbb{R}^{N \times T}$ is once more Fourier transformed resulting in $\tilde{\X} \in \mathbb{C}^{N \times F}$ were $f=1,2,\ldots, F$ with $F=\lfloor T/2\rfloor$. This is done as the shift in the the NMR frequency domain (the ppm axis) are convolutions, but shifts in the data frequency domain (Fourier transformed domain) are multiplications. So $s_{t-\tau_{n,d},d}$ becomes $\tilde{s}_{f,d} \exp\{-\tau_{n,d}\phi_f\}$ with $\phi_f = -2\pi i \frac{f-1}{T}$. This can be written for all frequencies as  $\tilde{\s}_{d} \exp\{-\tau_{n,d}\bphi\}$ where $\bphi = [\phi_1, \phi_2,\ldots, \phi_F]$, the results can then be moved back to the time domain by an inverse Fourier transform. 
By performing calculations in the Fourier domain it is assumed that shifts are circular, i.e., shifting across the right interval limit causes values to enter at the left interval limit and vice versa. To mitigate the effect of this circular assumption, the data for each interval is padded by $25\%$ on each size with a zero-order hold, i.e., copying the first and last observed values respectively.

\paragraph{Updating spectra}
\label{sec:update:S}
For estimation of $\S\in \mathbb{R}_{\geq 0}^{T \times D}$ with $T$ points on the ppm axis and $D$ components, we consider one column at a time $\s_d$ with all other columns fixed, $\s_{d}' \forall_{d'\neq d}$, which in case of MLE is called column-wise updating or hierarchical alternating least squares \cite{Gillis2012}. For $\s_d$ the MAP estimate is,

\begin{align}
    \s_d =& \mathtt{max}\left(\0,\left(-\lambda_d + \sigma^{-2} \mathtt{ifft}\left(
    \left(\c_d \circ \exp\{\btau_d\bphi\} \circ \tilde{\X}\right)^\top\1_N - \g_d \right) \right)^\top \left(\sigma^{-2}\c_d^\top\c_d\right)^{-1}\right),\\
    \g_d =& \sum_{d'\neq d} \left(\tilde{\s}_{d'} \circ \exp\left\lbrace\bphi^\top (\tau_d+\tau_{d'})^\top\right\rbrace (\c_d \circ \c_{d'})\right)\1_N
\end{align}
where $\g_d$ is the contribution from the fixed components and $^\top$ is the non-conjugated transpose. The $\mathtt{max}()$ operator is used for non-negativity. Sufficient statistics based on $\bphi, \tau $ and $\C$ can be precomputed for all $d$ and then multiple update steps can be done for each $d$ without increasing computational complexity. In practice, we update until the relative difference between the current iteration $i$ and last iteration $i-1$ is below the convergence threshold, e.g.  $||\S^{(i-1)}-\S^{(i)}||_F^2 / ||\S^{(i-1)}||_F^2 < 10^{-6} $. 

\paragraph{Updating concentrations}
\label{sec:update:C}
For estimation of $\C\in \mathbb{R}_{\geq 0}^{N \times D}$, the update is also done column-wise $\c_d$, as follows. 

\begin{align}
    \c_d =& \mathtt{max}\left(\0,\left(\sigma^{-2} \s_d^\top\s_d\right)^{-1}\left(\sigma^{-2}\left((\X\circ\s_{\btau_d,d}^\top)\1_T -\h_d\right)-\lambda_d\right)\right),\\
     h_d =& \sum_{d'\neq d} \c_{d'}\circ(\s_{\btau_{d'},d'}\circ\s_{\btau_d,d})^\top \1_T
\end{align}
where $\g_d$ is the contribution from the fixed components and  $\s_{\btau_d,d}$ is $\s_d$ but shifted for each sample $\tau_d = [\tau_{1,d},\ldots,\tau_{N,d}]$ and concatenated into a $T \times N$ matrix
Again the max-operator used for non-negativity. Note, $\s_d^\top\s_d$ is not shifted by $\btau_d$ as shifting does not change the inner product for the same component. Similar to the updates for $\S$, multiple updates can be made cheap by precomputing some of the terms. 


\paragraph{Updating automatic relevance determination}
\label{sec:update:lambda}
The relevance or length scale of each component was estimated based on ARD, the MAP update for $\blambda$ is
\begin{align}
    \lambda_d =& \frac{(N+T)}{\sum_{n=1}^N c_{n,d} + \sum_{t=1}^T s_{t,d} + \eta}.
\end{align}
The shifts do not play a role in this update and it is this identical to Bayesian NMF with exponential distributed factors and automatic relevance priors based on Gamma distributions, see \cite{Hinrich2018}.

An issue with ARD is that it can be slow to prune irrelevant components, therefore many existing approaches add a threshold and remove components once they are below it. In our work, no threshold was applied and a component $d$ was removed when its contribution to the solution was zero, i.e. either  $\sum_{n=1}^N \c_{n,d}=0$, or $\sum_{t=1}^T s_{t,d}=0$.

\paragraph{Updating shifts}
\label{sec:update:tau}
For updating the shift parameter, we follow the approach suggested in \cite{morup2008shiftcp}, but updating each shift $\tau_{n,d}$ also allows for updating $\c_{n,d}$ in $O(T)$ complexity, so MAP estimation of $\c_{n,d}$ is updated as well. The procedure is as follows,

First calculate the residual error in the Fourier domain, $\tilde{\E} = \tilde{\X}-\sum_{d=1}^D \c_d \tilde{\s}_d \circ \exp\{\btau_d \bphi\}$. Then for each component $d$, remove its contribution from the residual error $\tilde{\E}_{+d} = \tilde{\E} + \c_d \tilde{\s}_d \circ \exp\{\btau_d \bphi\}$. Now calculate the cross-correlation matrix, $\tilde{\K}=\tilde{\E}_{+d} \circ \tilde{\s_d}^H$ and move it back to the time domain by $\K = \mathtt{ifft}(\K)$. For each sample $n$, find the index $t_{\max}$ of the maximum  of $\k_n$ if $\C\geq 0$ or maximum of $\k_n$  if $\C$ is unconstrained. Once $t_{\max}$ has been found, the shift is updated,
\begin{align}
    \tau_{n,d} = (t_{\max} - T) -1
\end{align}

 Note, that it is also possible to constrain the maximum allowed shift per component by masking out $\k_n$ elements that are outside the allowed shifts. This is particular relevant for NMR where many singlets have identical shape which allows the shift-invariant property of BSINMF to use the same singlet component to describe singlets of different compounds across samples. In practice, it is easy to get rough estimates of the shift range of compounds and constrain the shifts accordingly. Alternatively, one can resolve too shifted components afterwards by post-processing the estimated shifts and splitting them.

After the shift is updated, each element of $\c_d$ can now be updated via MAP by,
\begin{align}
    c_{n,d} = \mathtt{max}\left(0,\left(\sigma^{-2}\s_d^\top\s_d\right)^{-1}\left(\k_{n,t_{\max}} - \lambda_d\right)\right)
\end{align}

The residual error is then updated by $\tilde{\E} = \tilde{\E}_{+d} - \c_d \tilde{\s_d} \circ \mathtt{exp}\{\btau_d \f\}$ and the process is repeated for a different $d$ until all shifts have been updated once.

For CoShiftNMF, the update is simpler, we first calculate the residual error without the baseline component $\bbeta$ (if any), then residual error $\E$ is element-wise multiplied by the reconstructed data $\M=\sum_{d=1}^D \c_d \tilde{\s_d} \circ \mathtt{exp}\{\btau_d \f\}$, e.g. $\K=\E \circ \M$. Then we find the index $t_{\max}$ for the maximum in $\k_n$  and update the shifts $\btau=[\tau_1,\tau_2,\ldots,\tau_N]^\top$ as,
\begin{align}
    \tau_n= (t_{\max}-T)-1
\end{align}

\paragraph{Permutation of components}
\label{sec:permutecomponents}
Solving the (B)SI-NMF problem is an non-convex optimization problem, but a local minima can be found via alternating sequential optimization of the individual parameters. This approach is sensitive to initial values of the parameters $\btau, \C,$ and $\S$ and can get stuck in local minima. To mitigate the risk of local minima, we apply the following permutation of components check to see if a signal modelled by component $\alpha$ might be better modelled by component $\beta$. Let $\mathcal{H}$ be an operator handling how the shift is applied, then the procedure is,

\begin{enumerate}
    \item[1)] Calculate the residual error $\bepsilon=\x_n-\c_n\mathcal{H}(\S,\btau_n)^\top$. 
    \item[2)] Calculate the residual without component $\alpha$, e.g. $\bepsilon_{\alpha}=\x_n-\sum_{d\neq \alpha}c_{n,d}\mathcal{H}(\s_d,\tau_{n,d})^\top$. 
    \item[3)] Fit component $\beta$, its concentration $\hat{c}_\beta$ and associate shift $\hat{\tau}_\beta$ on $\bepsilon_{\alpha}$ and calculate the residual error, e.g. $\bepsilon_{\alpha\rightarrow\beta}=\x_n-\hat{c}_{\beta}\mathcal{H}(\s_\beta,\hat{\tau}_{\beta})^\top-\sum_{d\neq \alpha}c_{n,d}\mathcal{H}(\s_d, \tau_{n,d})^\top $. 
    \item[4)] Calculate the residual error using the updated component $\beta$ replacing $\alpha$ and the original component $\beta$ removed, e.g.,$\bepsilon_{\alpha,\beta} = \x_n-\hat{c}_\beta\mathcal{H}(\s_\beta,\hat{\tau}_{\beta})^\top-\sum_{d\neq \alpha,\beta}c_{s,d}\mathcal{H}(\s_d, \tau_{n,d})^\top$. 
    \item[5)] Fit component $\alpha$ and its associate concentation $\hat{c}_\alpha$ and associated  shift parameter $\hat{\tau}_\alpha$ to $\bepsilon_{\\\alpha,\\\beta}$ and calculate the residual error based on component $\beta$ and $\alpha$ now swapped, e.g. $\bepsilon_{\alpha\leftarrow\beta}=\x_n-\hat{c}_{\alpha}\mathcal{H}(\s_\alpha, \hat{\tau}_{\alpha})^\top-
\hat{c}_{\beta}\mathcal{H}(\s_\beta, \hat{\tau}_{\beta})^\top
    -
    \sum_{d\neq \alpha,\beta}c_{n,d}\mathcal{H}(\s_d, \tau_{n,d})^\top $.
\end{enumerate}

Now, to assess if components should be permute, we consider if $\bepsilon_{\alpha\leftarrow\beta}$ is less than $\bepsilon$. If it is, we update $\c_n$ and $\tau_n$ accordingly to keep the solution found by $\bepsilon_{\alpha\leftarrow\beta}$ and $\bepsilon_{\alpha\rightarrow\beta}$. Importantly, this permutation scheme strictly decreases the error and does not affect the convergence guarantee of the alternating optimization procedure. In principle, this trick is applicable to any factorization model, but we only consider it for the (B)SI-NMF model.

Considering all possible permutations has complexity O$(N\cdot D^4 \cdot T)$ and represents a significant cost to the optimization. Therefore, we amortize this cost by only checking permutations every D iterations and only for the $N/D$ samples with the highest absolute shift values for each component. This results in an amortized cost of O$(N\cdot D^2 \cdot T)$ per iterations which is the same complexity as the updates for $\S$. 

\section{Simulation Study Result}
\label{sec:app:simstudy}
\begin{figure}[tbp]
    \centering
        \subfloat[No shift, No Overlap, SNR 50dB]{
            \includegraphics[width=0.49\linewidth]{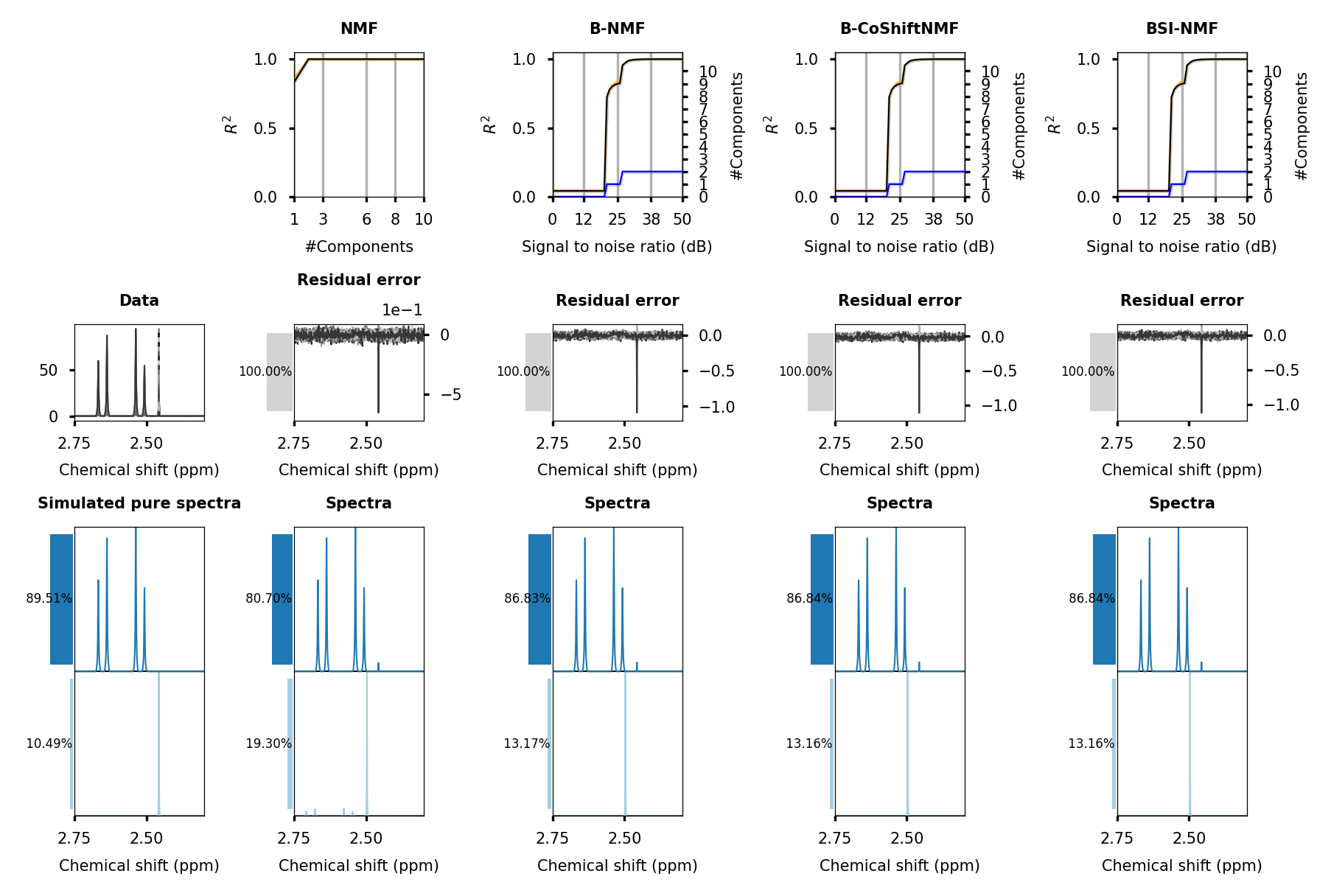}    
        }
        \subfloat[No shift, SNR 50dB]{
            \includegraphics[width=0.4\linewidth]{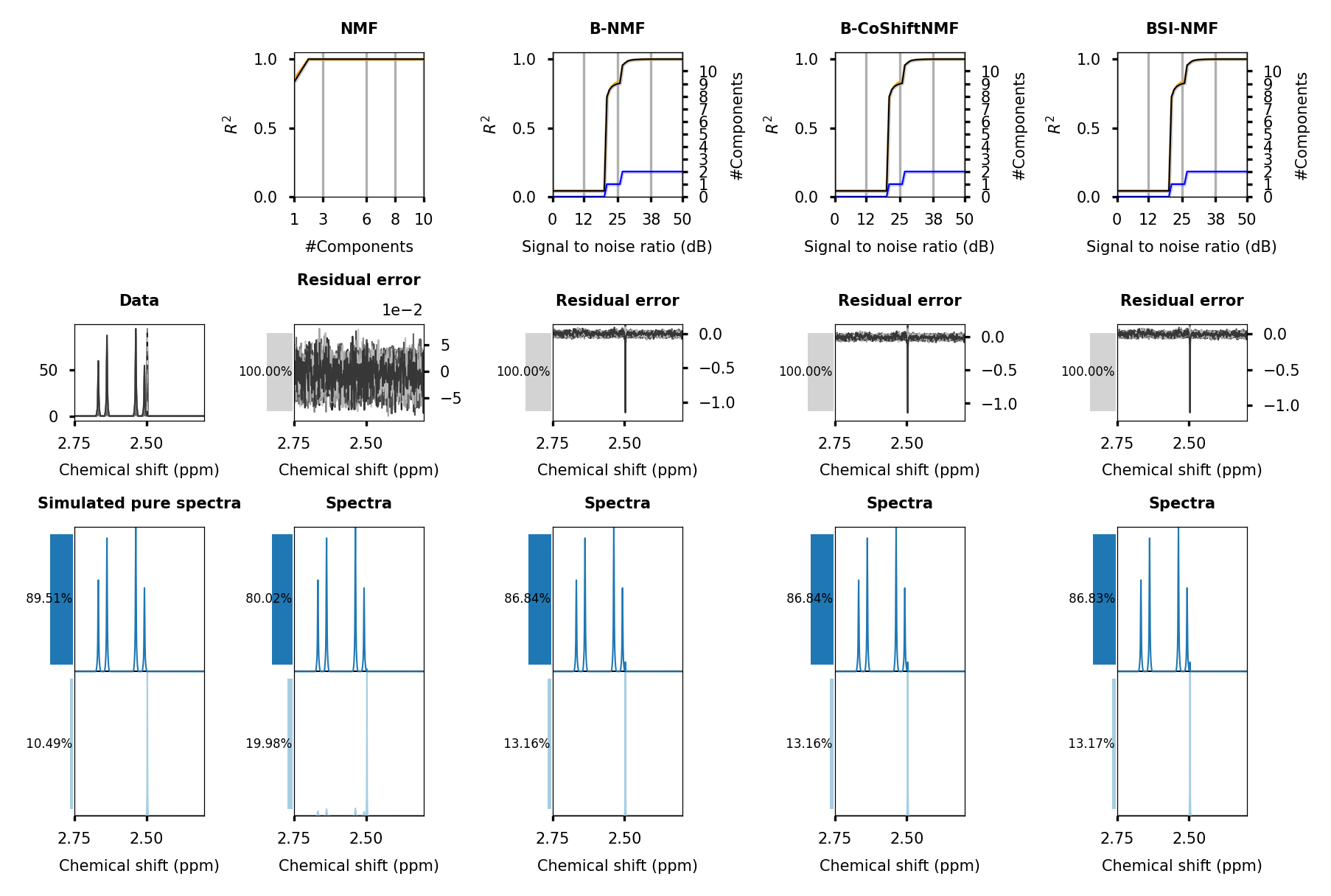}    
        }\\
        \subfloat[Sample shift only, SNR 50dB]{
            \includegraphics[width=0.65\linewidth]{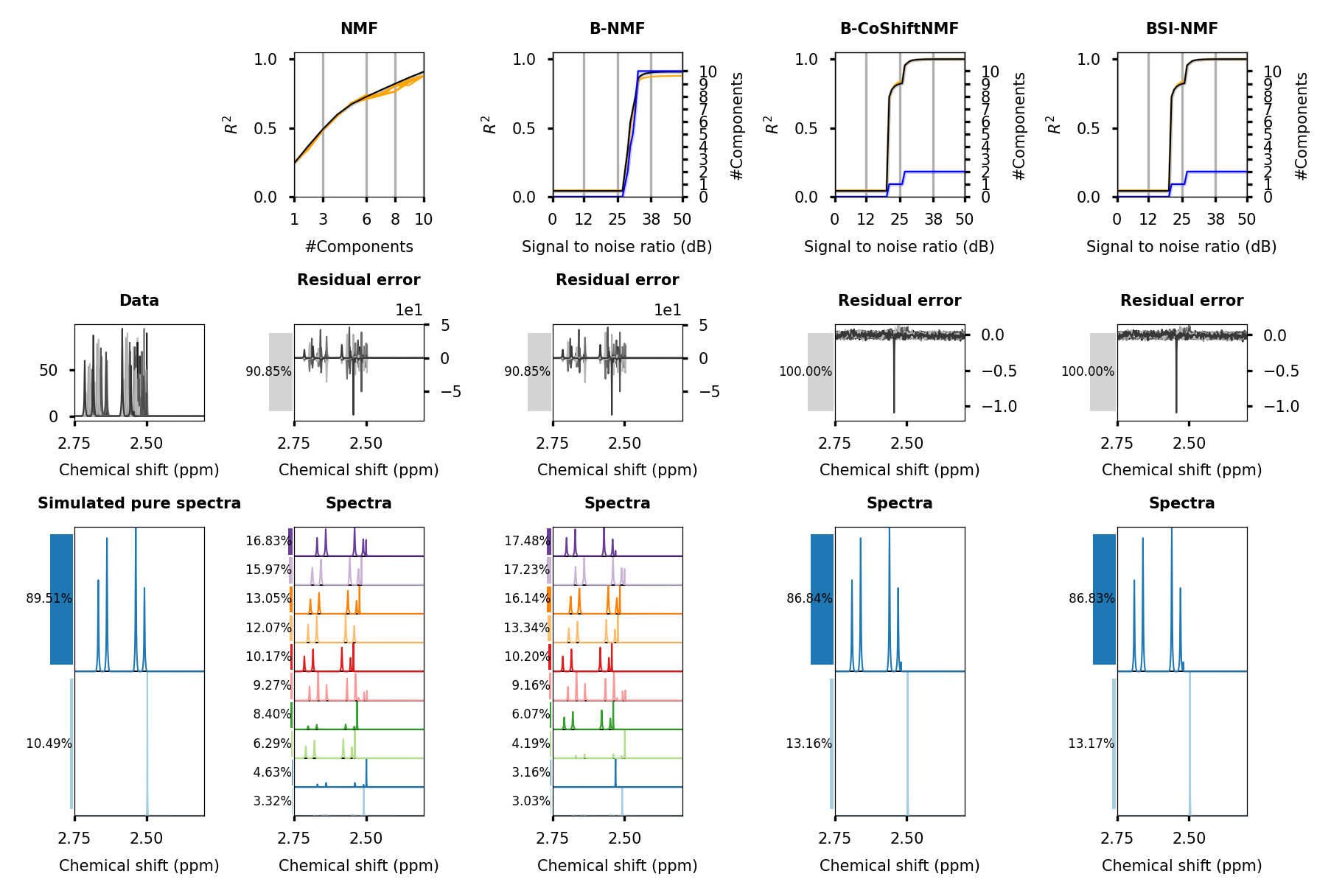}    
        }\\
        \subfloat[Metabolite specific shift, SNR 50dB]{
            \includegraphics[width=0.65\linewidth]{illustrations/figure-2-SimulatedMetaboliteShifts-2.3,2.75-comp-best.png}    
        }
    \caption{Computer simulated data under four conditions (a) No shifts and no overlapping spectra, (b) No shifts, (c) spectra jointly shifted for each sample, and (d) spectra individually shifted in each sample. In (a) and (b) all methods identify the correct number of components, but have non-unique solutions. In (c), only correcting for shift identifies the correct number of components, but joint shifts still results a non-unique solution. In (d), only SiNMF identifies the correct number of components and also provides a unique (up to scaling, permutation, and shifting) solution.}
    \label{fig:app:simulated}
\end{figure}

The simulated study considers four scenarios for generating the data, the detailed simulation settings are described in Section~\ref{sec:methods:simstudy}. All results are shown in Figure~\ref{fig:app:simulated} (a) to (d).

In the first scenario, the two simulated spectra have zero spectral overlap, e.g. are baseline separated and no shift. 
The results are given in Figure~\ref{fig:app:simulated} and show all methods correctly identifying the number of simulated spectra, but they all suffer from  non-unique solutions (see Section~\ref{sec:methods:shiftnmf:unique} for uniqueness discussion) and thus do not recover the true underlying spectra or concentrations. 

In scenario two, Figure~\ref{fig:app:simulated}(b), the two simulated spectra have significant spectral overlap and but still no shifts. Again, all methods are able to identify the right number of components, but are hampered by non-unique solutions.

In scenario three, Figure~\ref{fig:app:simulated}(c), the two simulated spectra have significant spectral overlap and a sample specific shift - e.g. both components move on the ppm axis by the same amount within a sample (global shift). This shift violates the assumption of NMF and B-NMF and neither are  able to correctly identify the number of components and by extension unable to recover the two simulated spectra. Both B-CoShiftNMF and BSI-NMF are able to account for the sample specific shifts and clearly identifies the right number of components. However, the recovered spectra still suffers from non-uniqueness, as correcting for sample specific shift is equivalent to shifting each data sample and then performing NMF (or B-NMF), for details see Section~\ref{sec:methods:shiftnmf:unique}. 

In scenario four, Figure~\ref{fig:app:simulated}(d), the two simulated spectra have significant spectral overlap and metabolite/component specific shifts - e.g. each component shifts individually on the ppm axis within a sample. The assumptions of NMF, B-NMF, and B-CoShiftNMF are now invalid and they all fail to recover the right number of components. Interestingly, the spectra for B-CoShiftNMF show how it finds the main component (citric acid) in the data and correctly aligns it across samples, but it is unable to recover the other component. In contrast, BSI-NMF is able to identify the number of components but also \emph{uniquely} identify the spectra as sufficient metabolite/component shifts make the BSI-NMF model unique, see Section~\ref{sec:methods:shiftnmf:unique}.

In summary, the proposed BSI-NMF is able to unique recover metabolites if they have sufficient individual shifts within each sample. If shifts are samples specific or there are no shifts, then BSI-NMF learns this and performs similar to B-CoShiftNMF and B-NMF/NMF, respectively, and suffers from their known issues regarding uniqueness as well. 

Note, that these simulations had no samples with pure spectra, as including pure spectra for all metabolites trivially solves the uniqueness issue, as the data then spans the non-negative orthant.

\FloatBarrier

\section{Comparison Study: BSI-NMF, \textit{i}coshift with NMF, Peak Fitting}
\label{sec:app:comparision}

The comparison of BSI-NMF versus ICoShift with NMF and Peak fitting is summarized in Table~\ref{tab:comparison}. The comparison methods are described in Section~\ref{sec:methods:comparision} and here the detailed results of the comparison is provided.

\begin{figure}[tbp]
    \centering
        \subfloat[Urine Metabolites A - Citric Acid]{
            \includegraphics[width=0.49\linewidth]{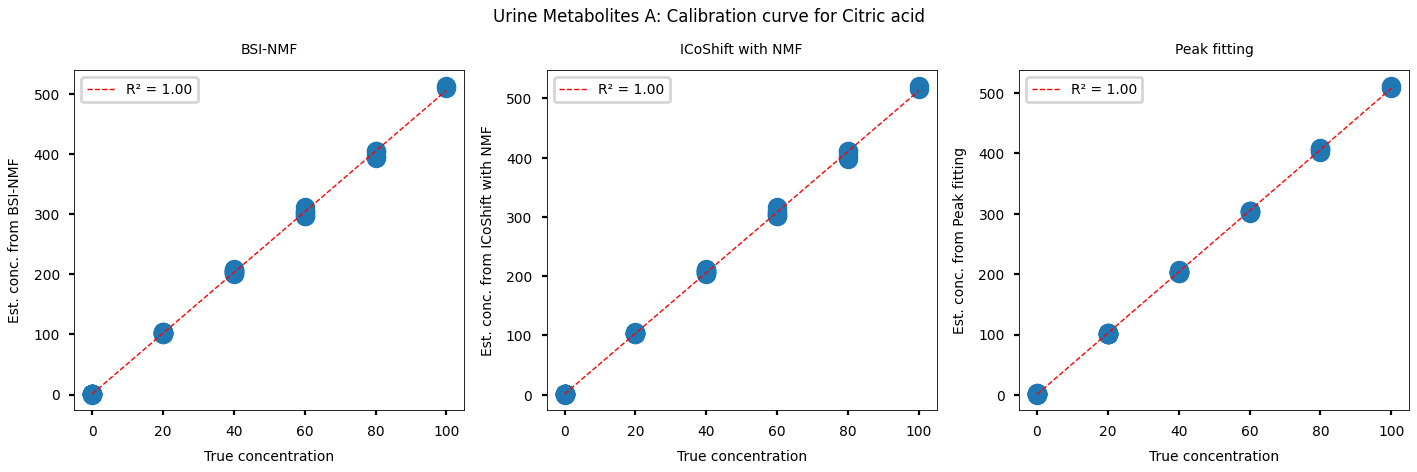}    
        }
        \subfloat[Urine Metabolites A - DMA]{
            \includegraphics[width=0.49\linewidth]{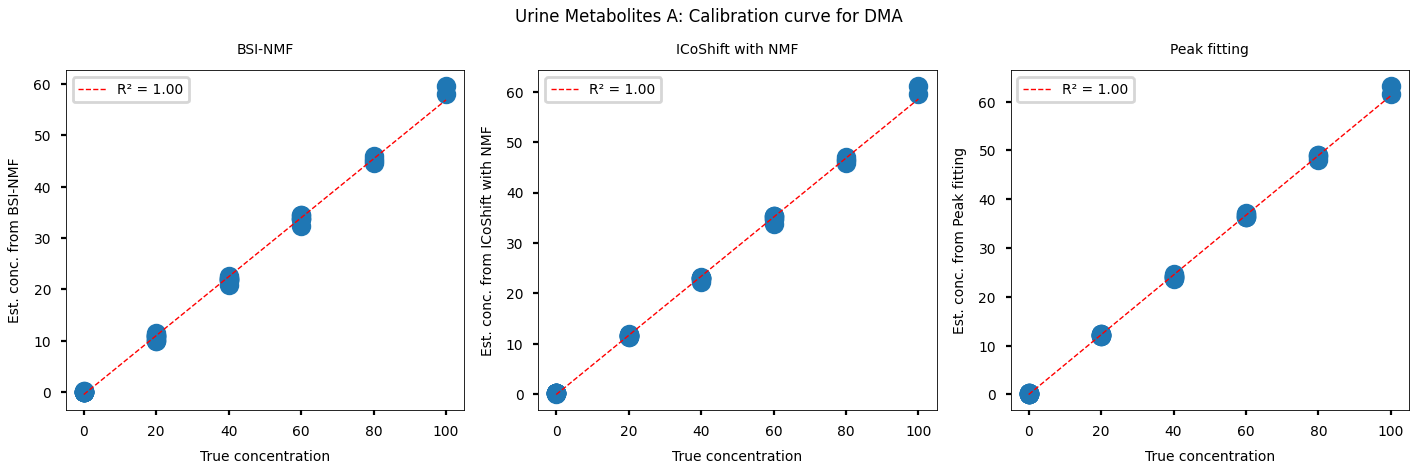} }\\
        
        \subfloat[Urine Metabolites B - Citric Acid]{
            \includegraphics[width=0.49\linewidth]{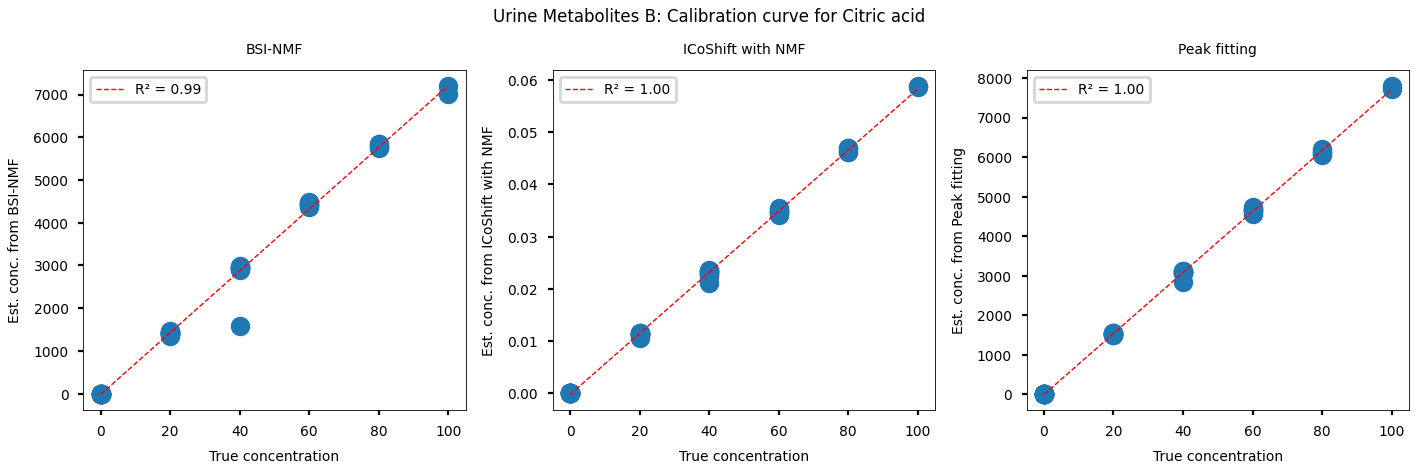} 
        }
        \subfloat[Urine Metabolites B - DMA]{
            \includegraphics[width=0.49\linewidth]{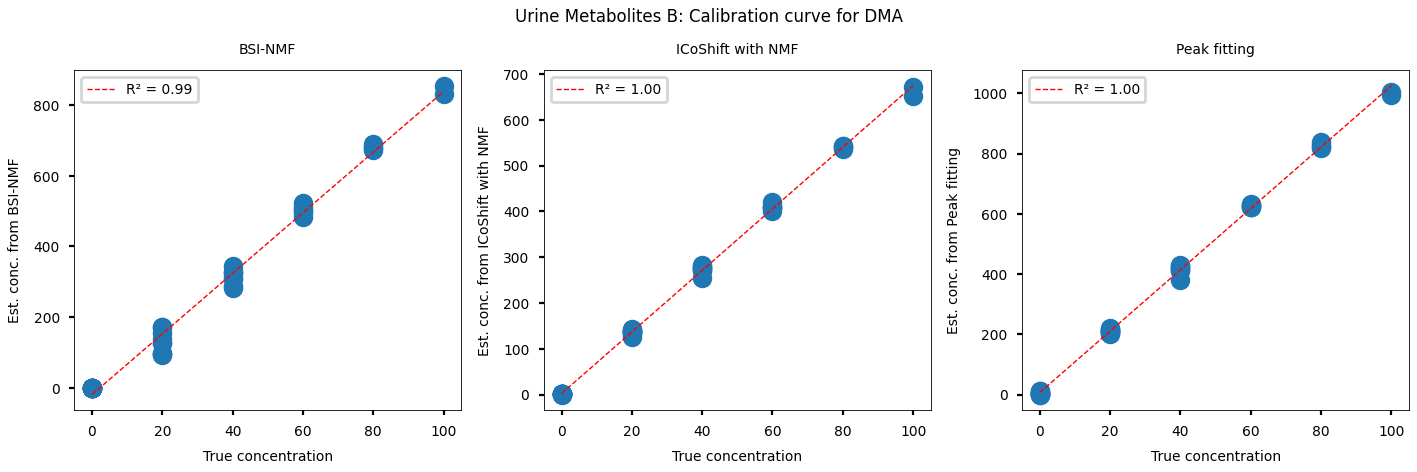} }\\

        \subfloat[Human Urine - Citric Acid]{
            \includegraphics[width=0.49\linewidth]{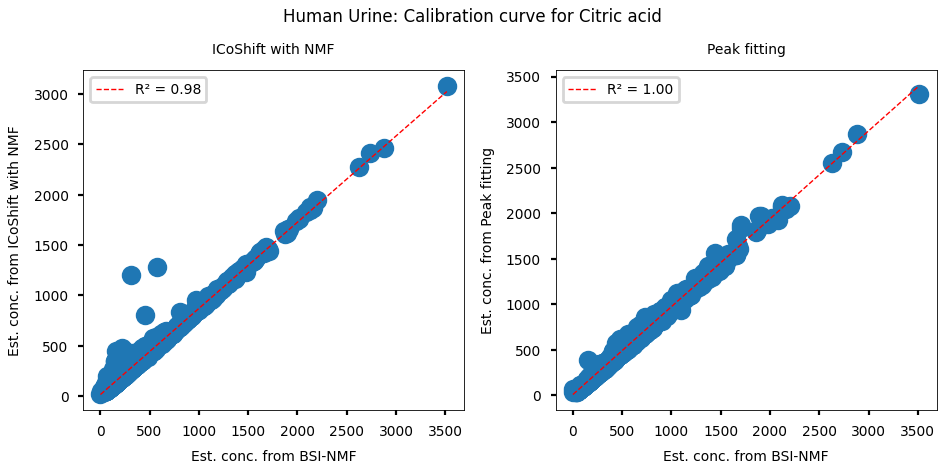}    
        }
        \subfloat[Human Urine - DMA]{
            \includegraphics[width=0.49\linewidth]{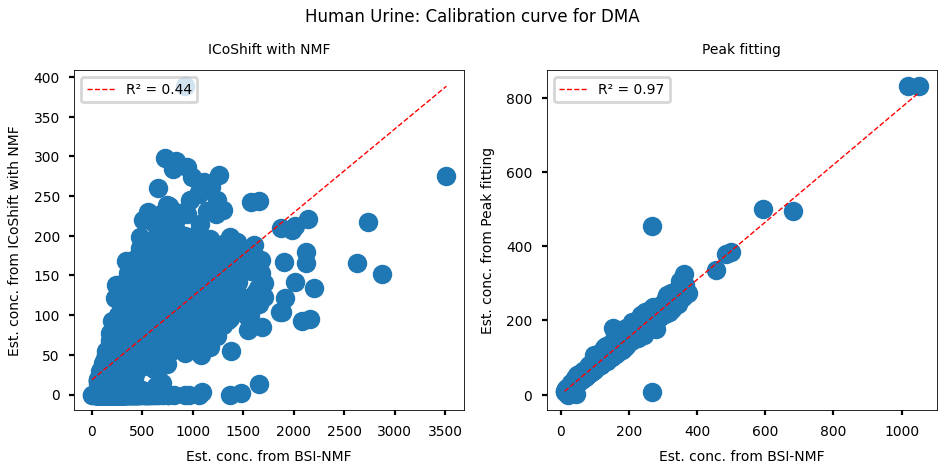} }\\

    \subfloat[Human Urine - DMA (300 Manually verified samples)]{
        \includegraphics[width=0.49\linewidth]{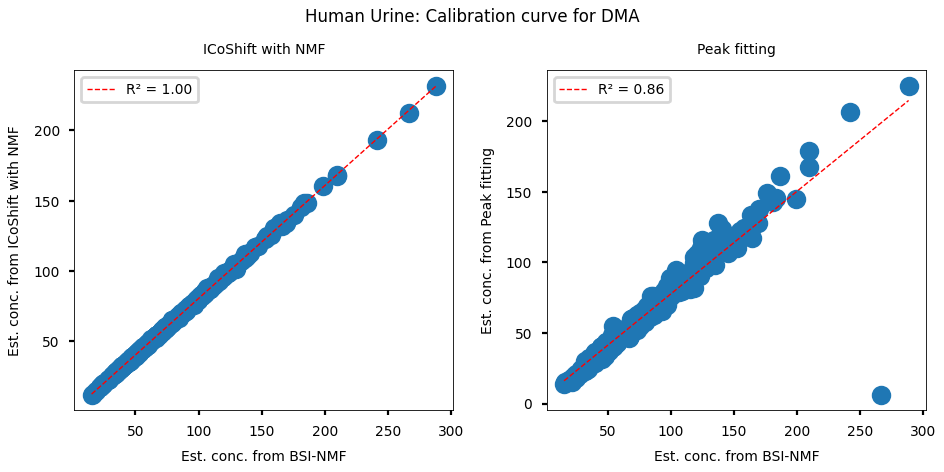} }
    \caption{Interval 2.5 to 2.75 ppm}
    \label{fig:app:comparison:citric}
\end{figure}

\begin{figure}[tbp]
    \centering
        \subfloat[Urine Metabolites A - Creatinine]{
            \includegraphics[width=0.49\linewidth]{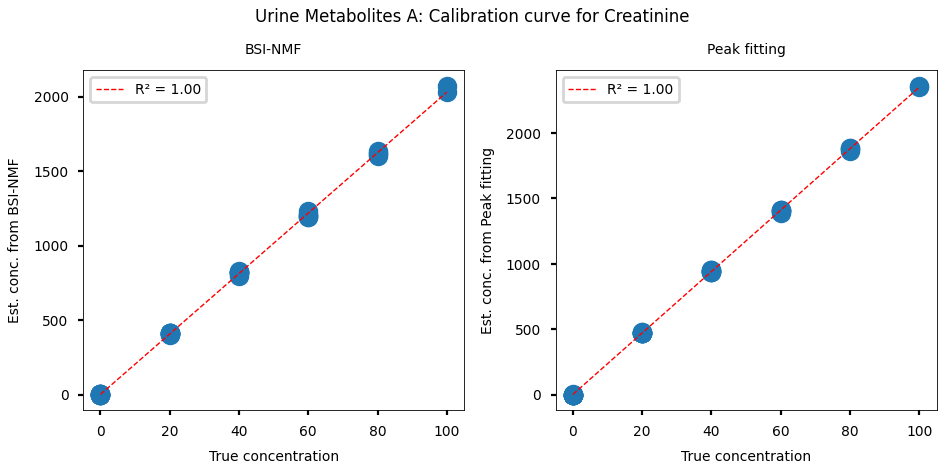}    
        }
        \subfloat[Urine Metabolites A - Creatine]{
            \includegraphics[width=0.49\linewidth]{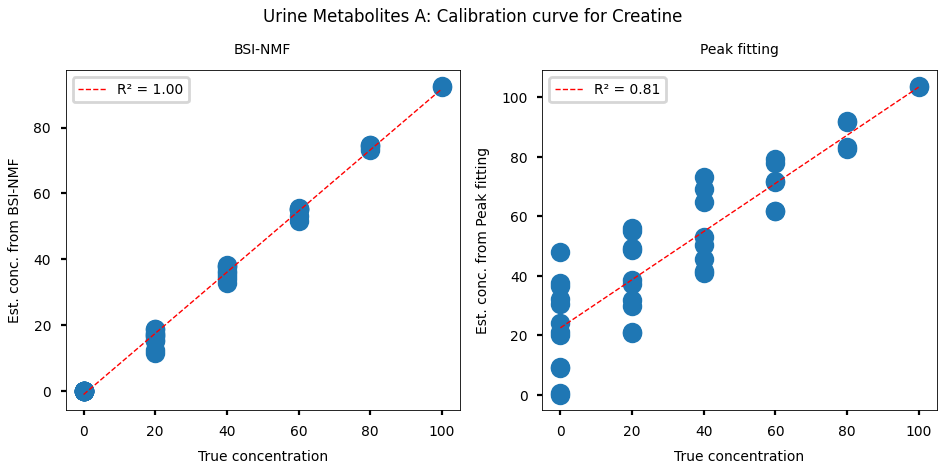}
            }\\
        \subfloat[Urine Metabolites B - Creatinine]{
            \includegraphics[width=0.49\linewidth]{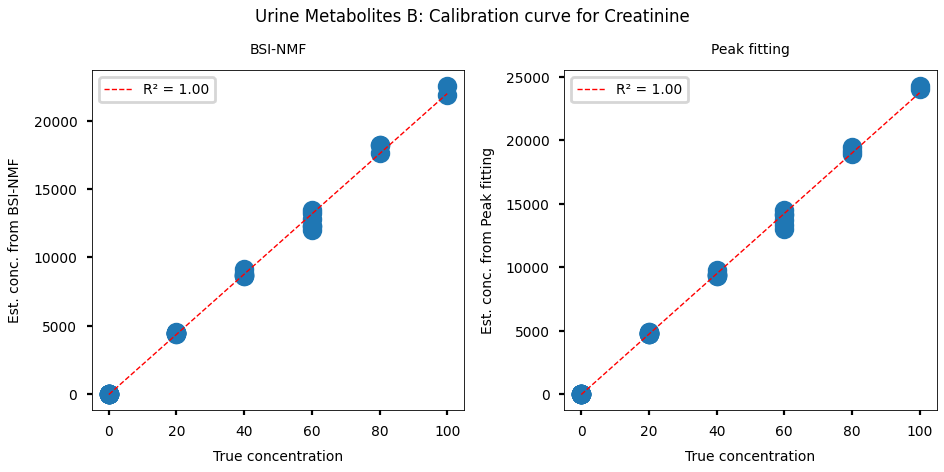}
        }
        \subfloat[Urine Metabolites B - Creatine]{
            \includegraphics[width=0.49\linewidth]{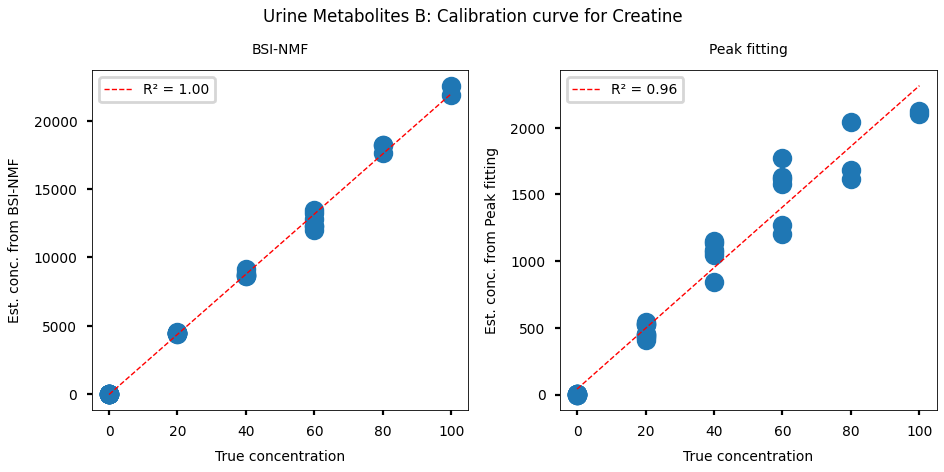} }\\

        \subfloat[Human Urine - Creatinine]{
            \includegraphics[width=0.33\linewidth]{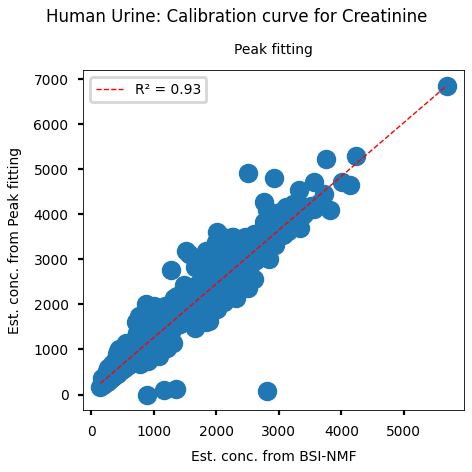}
        }
        \subfloat[Human Urine - Creatine]{
            \includegraphics[width=0.33\linewidth]{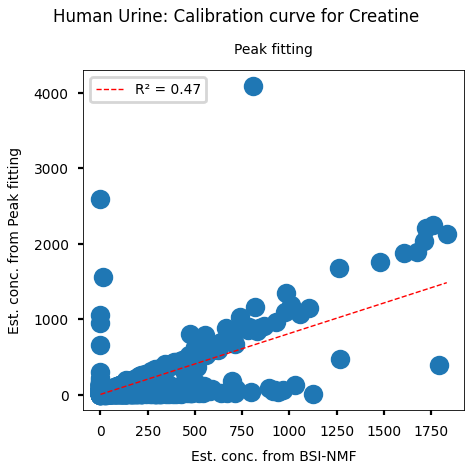} 
            }
        \subfloat[Human Urine - Creatine (100 Manually verified samples)]{
        \includegraphics[width=0.33\linewidth]{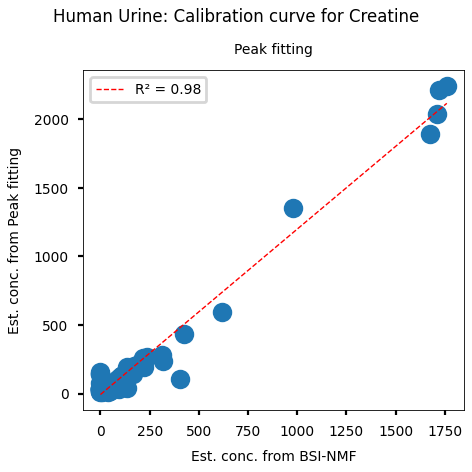} }
    \caption{Interval 3.03 to 3.08 ppm}
    \label{fig:app:comparison:creatinine}
\end{figure}

For each metabolite (citric acid, dimethylamine, creatinine, and creatine) and each dataset (Urine Metabolites A and B, and Human Urine) a correlation curve is plotted, e.g. Figure~\ref{fig:app:comparison:citric} and \ref{fig:app:comparison:creatinine} for interval 2.5 to 2.75 ppm (citric acid and DMA) and interval 3.03 to 3.08 ppm (creatinine and creatine), respectively.

\begin{table}[tbp]
    \centering
    \fontsize{5pt}{5pt}\selectfont
    \begin{tabular}{l|ccc|ccc|cc}    
    \hline \hline
        & \multicolumn{3}{c}{Urine Metabolites A} & \multicolumn{3}{c}{Urine Metabolites B} & \multicolumn{2}{c}{Human Urine}\\
        & BSI-NMF & ICoShift &  Peak Fit.  & BSI-NMF & ICoShift &  Peak Fit. & ICoShift &  Peak Fit. \\
        \hline
         Citric acid & 1 & 1 & 1 & 0.99 & 1 & 1 & 0.98 & 1 \\
         DMA  & 1& 1 & 1 & 0.99& 1 & 1 & 0.44 & 0.97 \\
         DMA (man.) &   &   &   &   &   &   &   1 & 0.86\\
         \hline
         Creatinine & 1 & N/A &  1  & 1 & N/A & 1 & N/A & 0.93 \\
         Creatine  &  1 & N/A & 0.81 & 1 & N/A & 0.96 & N/A & 0.47\\
         Creatine (man.) &  & & & & & & & 0.98 \\
         \hline \hline
    \end{tabular}
    \caption{Coefficient of regression $R^2$ for the estimated concentrations against the reference method. For Urine Metabolites A and B, the reference is the known concentrations, while for Human Urine, the reference is the estimated concentration by BSI-NMF. The N/A indicates scenarios where ICoShift was not applicable. The manually checked subset of samples (man.) was only done for DMA and Creatine in the Human Urine dataset.}
    \label{tab:comparison}
\end{table}

For the Urine Metabolites A and B dataset, the true concentrations (x-axis) are known and the estimated concentrations (area of the peak) are compared against it (y-axis). For the Human Urine dataset, the true concentrations are unknown and we compare the estimated concentrations of BSI-NMF (x-axis) to the other methods (y-axis). For each calibration curve, the $R^2$ for a linear fit is reported - which is also collected in Table~\ref{tab:comparison} in the main text.

For citric acid and dimethylamine (Figure~\ref{fig:app:comparison:citric}), all methods perfectly correlates with the true concentrations in Urine Metabolites A and B. For the human urine dataset, citric acid is almost equivalently recovered by BSI-NMF, ICoShift with NMF, and Peak fitting. The slight difference with ICoShift with NMF is due to ICoShift alignment failing as in some samples the dimethylamine peak is shifted into a doublet of citric acid. This is a limitation of aligning the entire dataset at once, as described in Section~\ref{sec:methods:comparision}. This also results in dimethylamine having a poor coefficient of variation $R^2=0.44$ for BSI-NMF vs ICoShift with NMF, see Figure~\ref{fig:app:comparison:citric}(f). In contrast, this issue is not present for peak fitting and the correlation is much higher $R^2=0.97$. For completeness, we manually examined three-hundred samples and ICoShift aligned them in batches (one to twenty samples at a time) and this results in perfect correlation between BSI-NMF and ICoShift with NMF, which is also the expected results. On this subset, the correlation between BSI-NMF and peak fitting is lower.

For creatinine and creatine (Figure~\ref{fig:app:comparison:creatinine}), the creatinine peak is perfectly recovered by both BSI-NMF and Peak fitting in the Urine Metabolites A and B datasets. Note, ICoShift with NMF was not applied, see Section~\ref{sec:methods:comparision}). In contrast, creatine is only perfectly recovered by BSI-NMF ($R^2=1$) with Peak fitting having good but more varied performance, i.e. $R^2=0.81$ and $R^2=0.96$ for Urine Metabolites A and B, respectively. 

For the Human Urine dataset, creatinine is similarly recovered by both methods and the comparative performance is mostly affected by a modest variation and a few outliers. For creatine, the performance is much worse $R^2=0.47$ for the entire dataset, but if we manually examine (via visual inspection) and select a hundred samples where peak fitting has successfully captured the creatine peak, then there is almost perfect correspondence between BSI-NMF and Peak fitting, Figure~\ref{fig:app:comparison:creatinine}(g). 

In general, verification of the performance on the Human Urine dataset is difficult, as we do not have the true concentrations of any of the metabolites. We sought to remedy this by manual examination of good samples for ICoShift (citric acid and dimethylamine) and Peak fitting (creatine) which shows that we great care and optimization of these two comparison methods then they can perform equal to BSI-NMF. The advantage of BSI-NMF is that is automatically handles shifting peaks and identifying the number of metabolites - the comparison methods had the unfair advantage of knowing the number of metabolites before analysis.

\section{Analysis of Urine Metabolites A and B without pure samples}
\label{sec:app:waternopure}
In the main text, the artificially create Urine Metabolites datasets (A and B) was analysed with pure samples (100\% one stock solution), as pure samples ensures that the data spans the non-negative orthant and thus NMF, CoShiftNMF, ShiftNMF, and their Bayesian counterparts will find unique solutions. For a discussion of uniquenes see Section~\ref{sec:methods:shiftnmf:unique}. 

\begin{figure}[tbp]
    \subfloat[Urine Metabolites A without pure samples]{
    \centering     
    \includegraphics[width=0.49\linewidth]{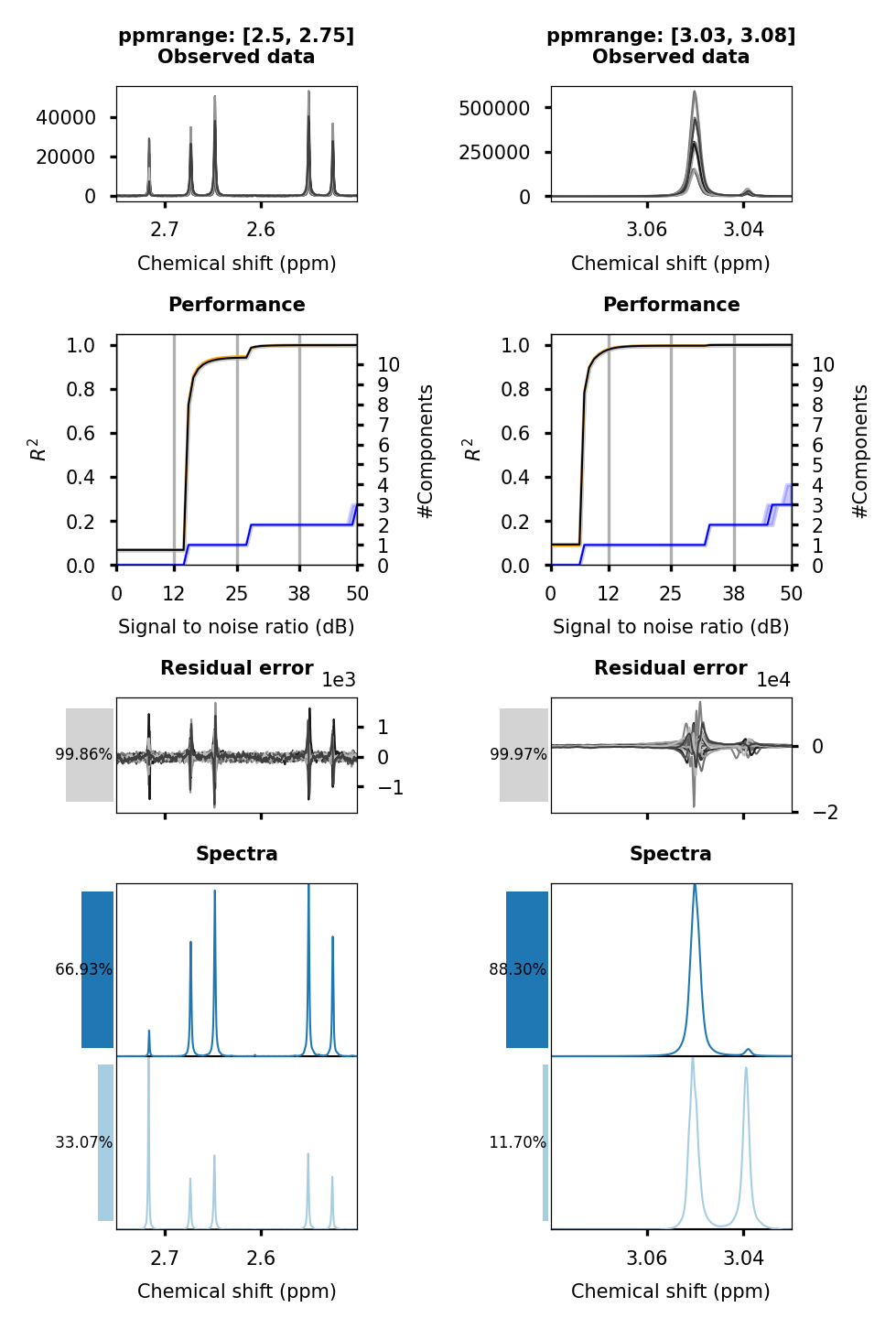} 
    \label{fig:app:nopuresamples:labA}
    }
    \subfloat[Urine Metabolites B without pure samples]{
    \centering
        \includegraphics[width=0.49\linewidth]{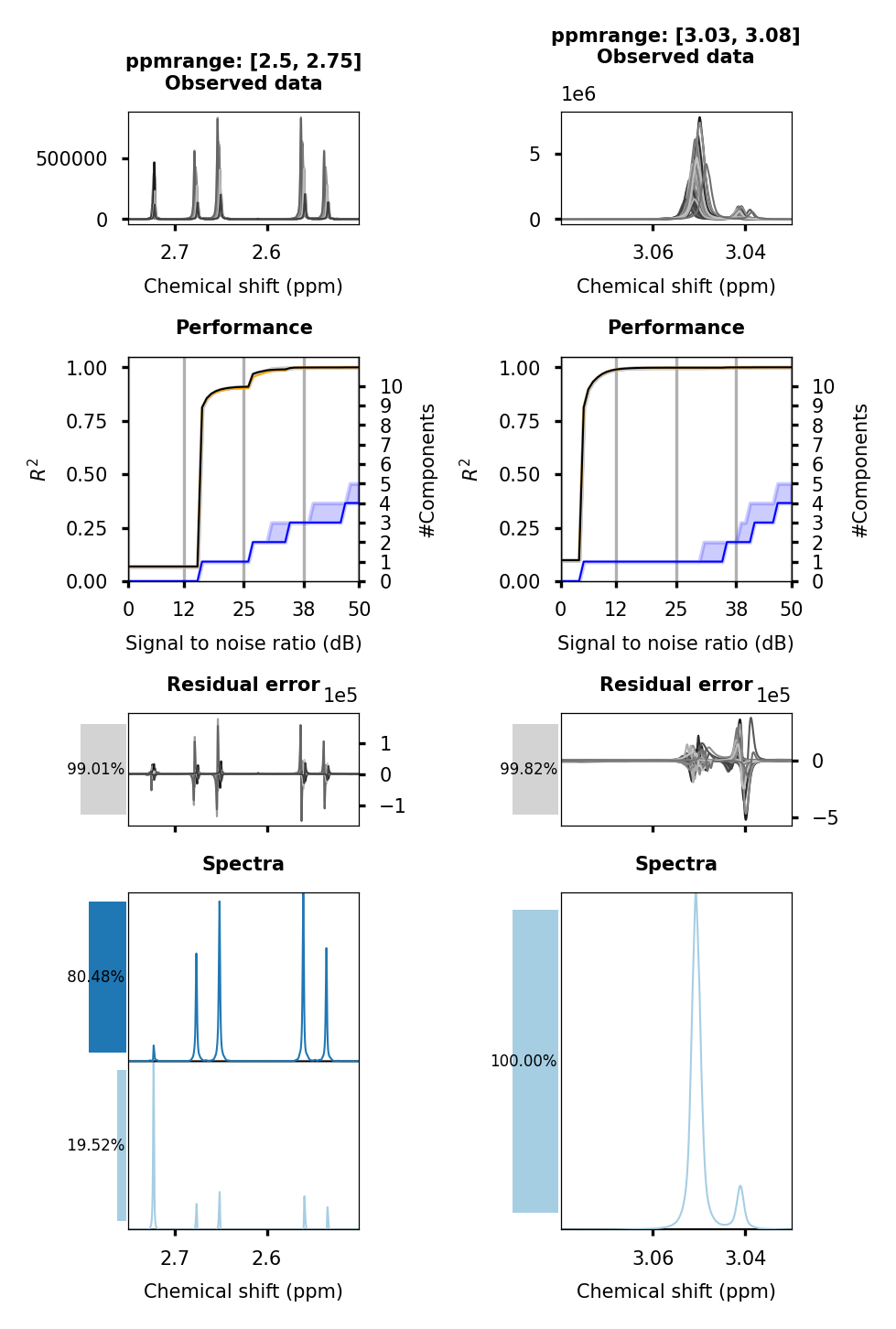}    
        \label{fig:app:nopuresamples:labB}
    }
    \caption{Urine Metabolites A and B without pure samples. }
    \label{fig:app:nopuresamples}
\end{figure}

For illustrative purposes, we analysed Urine Metabolites A and B without pure samples using the same processing steps, see Section~\ref{sec:methods:experimentaldetails}. The results are shown in Figure~\ref{fig:app:nopuresamples}. The performance curves in terms of $R^2$ and number of components are similar to when pure samples are included, but the estimated latent spectra and by extension concentration are not unique. This non-uniqueness shows up as peaks that are duplicated across components and thus intermix the signal. 

The shifts or lack thereof in Urine Metabolites A, Figure~\ref{fig:app:nopuresamples:labA}, means that it is essentially an NMF problem and the data does not fulfill the uniqueness conditions for this method.  More surprising is that citric acid and dimethylamine are not separated in Urine Metabolites B, Figure~\ref{fig:app:nopuresamples:labB}, as the data has slight shifts. Here, we hypothesis that this lack of uniqueness is do to the sample concentration and shift having perfect co-variation due to the experimental design. For creatinine and creatine, they are in the same stock solution and is thus most efficiently described as one component.

The main takeaway from this experiment is that pure samples or metabolites that are only present  (in the selected interval) for some samples is always desirable to include to improve the possibility of having BSI-NMF provide unique solutions. 

\FloatBarrier

\section{Optimal number of components as selected by BSI-NMF, but for all methods}
 \label{sec:app:rightcomponents}
In this section, we compare the number of components estimated by BSI-NMF and how that same number of components looks for the NMF, B-NMF, and B-CoShiftNMF. The plots are shown for the Urine Metabolites A and B datasets and the Human Urine dataset for each interval, i.e. 2.5 to 2.75 ppm, 3.03 to 3.08 ppm, and 5.66 to 5.987 ppm.

For each plot, the measured \textsuperscript{1}H NMR samples are shown to the left (Data). The results of applying NMF, B-NMF, B-CoShiftNMF, and BSI-NMF are shown in column 2 through 5, respectively. In the first row, the performance across the chosen number of components (for NMF) and across the chosen SNR range 50 to 0 dB (for B-NMF, B-CoShiftNMF, and BSI-NMF) is shown as coefficient of regression $R^2$ with (black line) and without (dotted orange) the baseline component. The blue line indicate the number of components identified by the automatic relevance determination at each SNR. The shaded gray and blue area denotes the range of values across repeats - if the shaded area is not visible the variation is negligible. Similarly, if the orange dotted line is not visible, the contribution of the baseline offset is negligible. In the second row, the residual error for each model is shown and a bar with the percent explained variance is shown, note the y-axis have individual scales. In the third row, the spectra for the best model at the specified number of components are shown and their percentage contribution to the reconstruction of the data as bars.

\begin{figure}[tbp]
    \centering
    \subfloat[Human Urine - 2.5 to 2.75 ppm]{
            \includegraphics[width=0.6\linewidth]{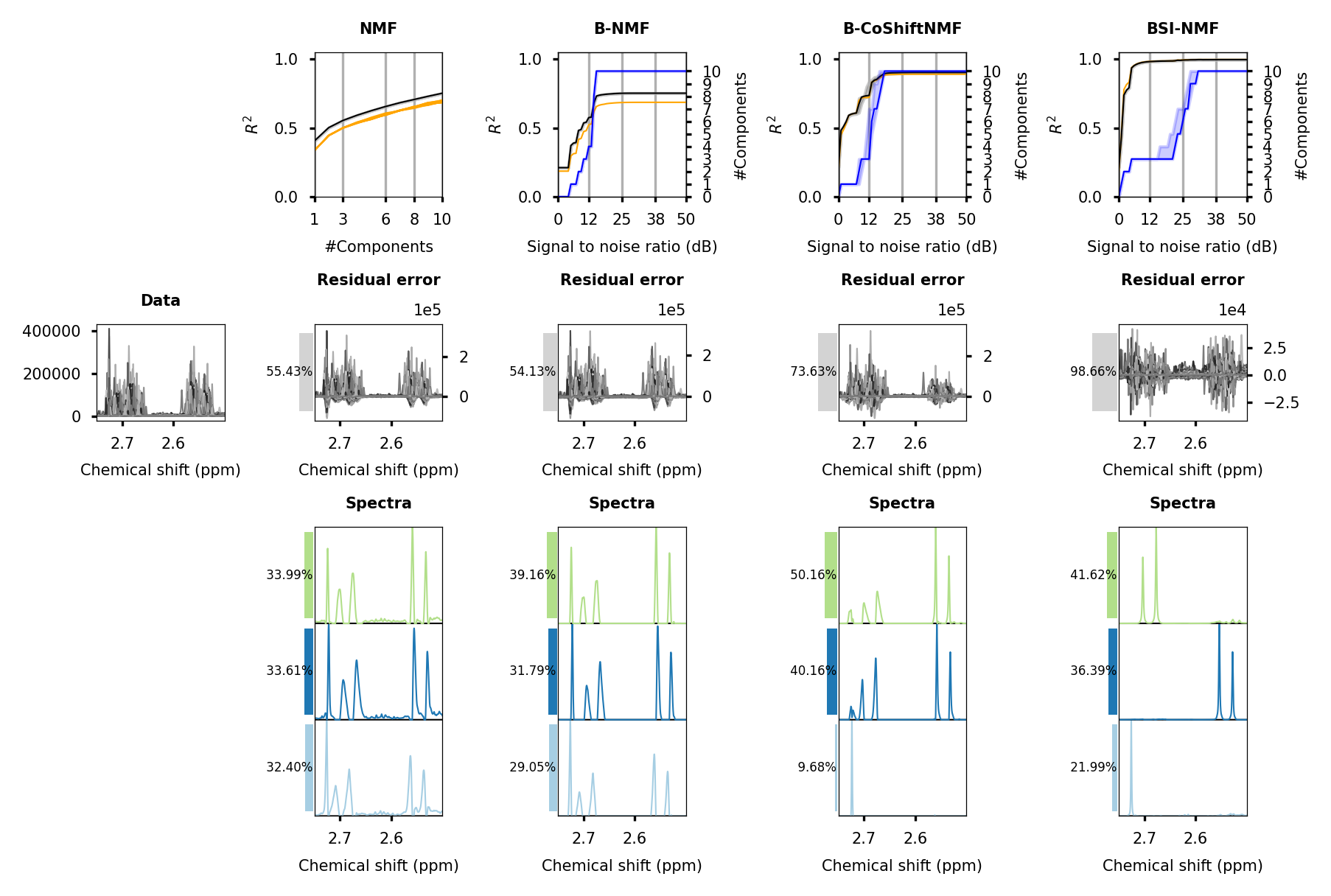}
            \label{fig:app:rightcomponents:human:citric}
        }\\
        \subfloat[Human Urine - 3.03 to 3.08 ppm]{
        \includegraphics[width=0.6\linewidth]{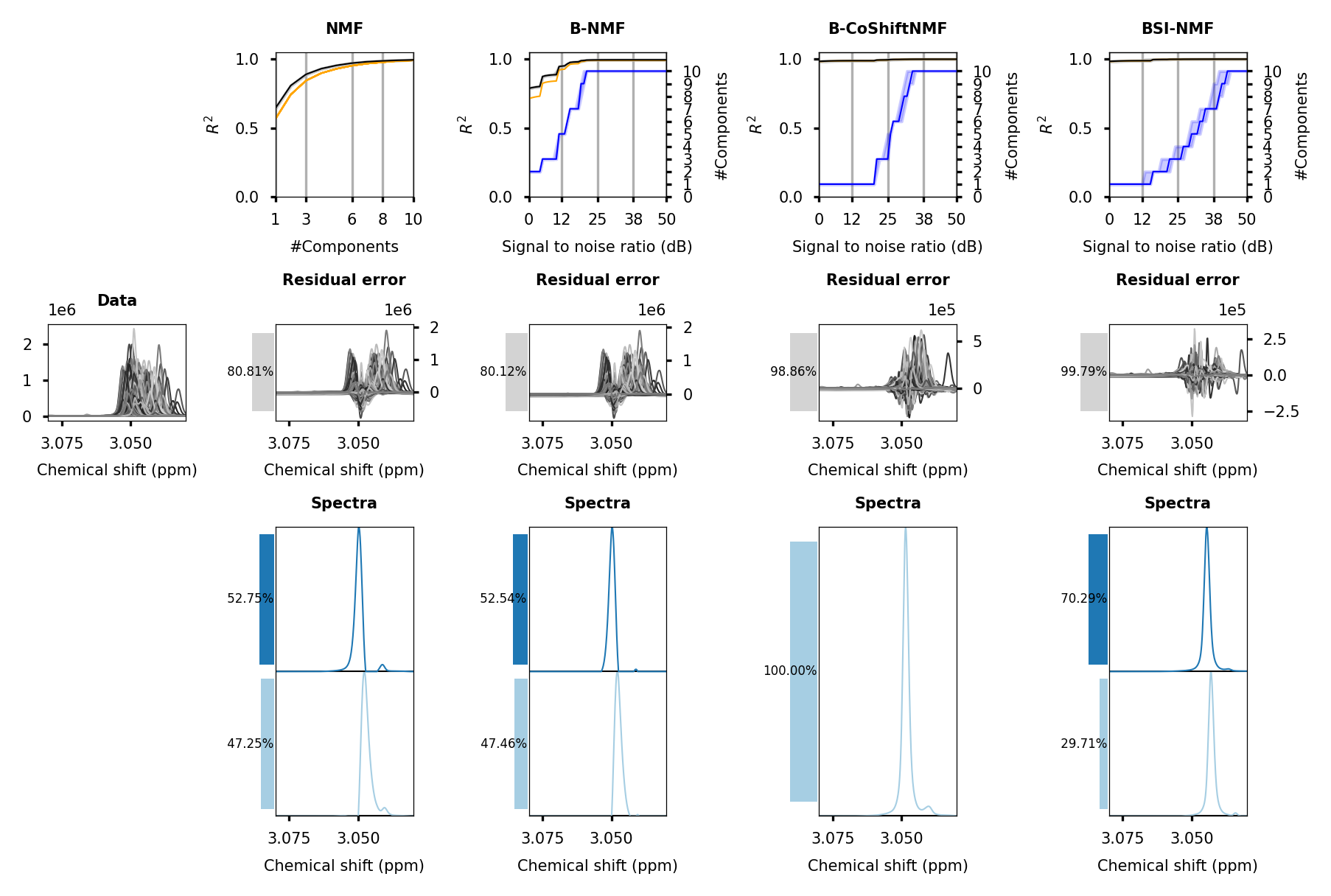}
        \label{fig:app:rightcomponents:human:creatine}
        }\\
        \subfloat[Human Urine - 5.66 to 5.987 ppm]{
            \includegraphics[width=0.6\linewidth]{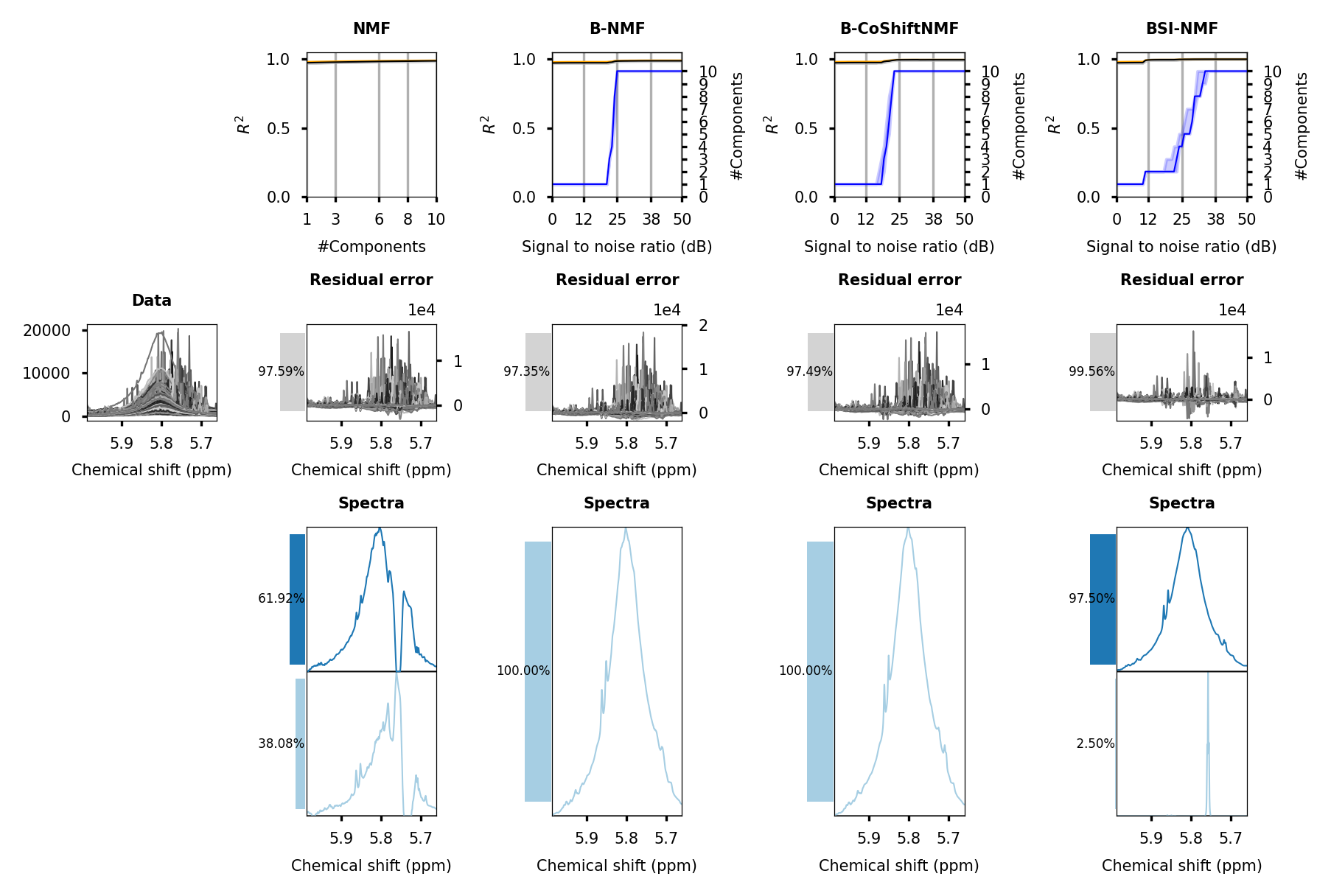}
            \label{fig:app:rightcomponents:human:urea}
        }
    \caption{Here the number of components identified by BSI-NMF are plotted for all methods for all three intervals.}
    \label{fig:app:rightcomponents:human}
\end{figure}

\begin{figure}[tbp]
    \centering
        \subfloat[Urine Metabolites A (w. pure samples) - 2.5 to 2.75 ppm]{
            \includegraphics[width=0.8\linewidth]{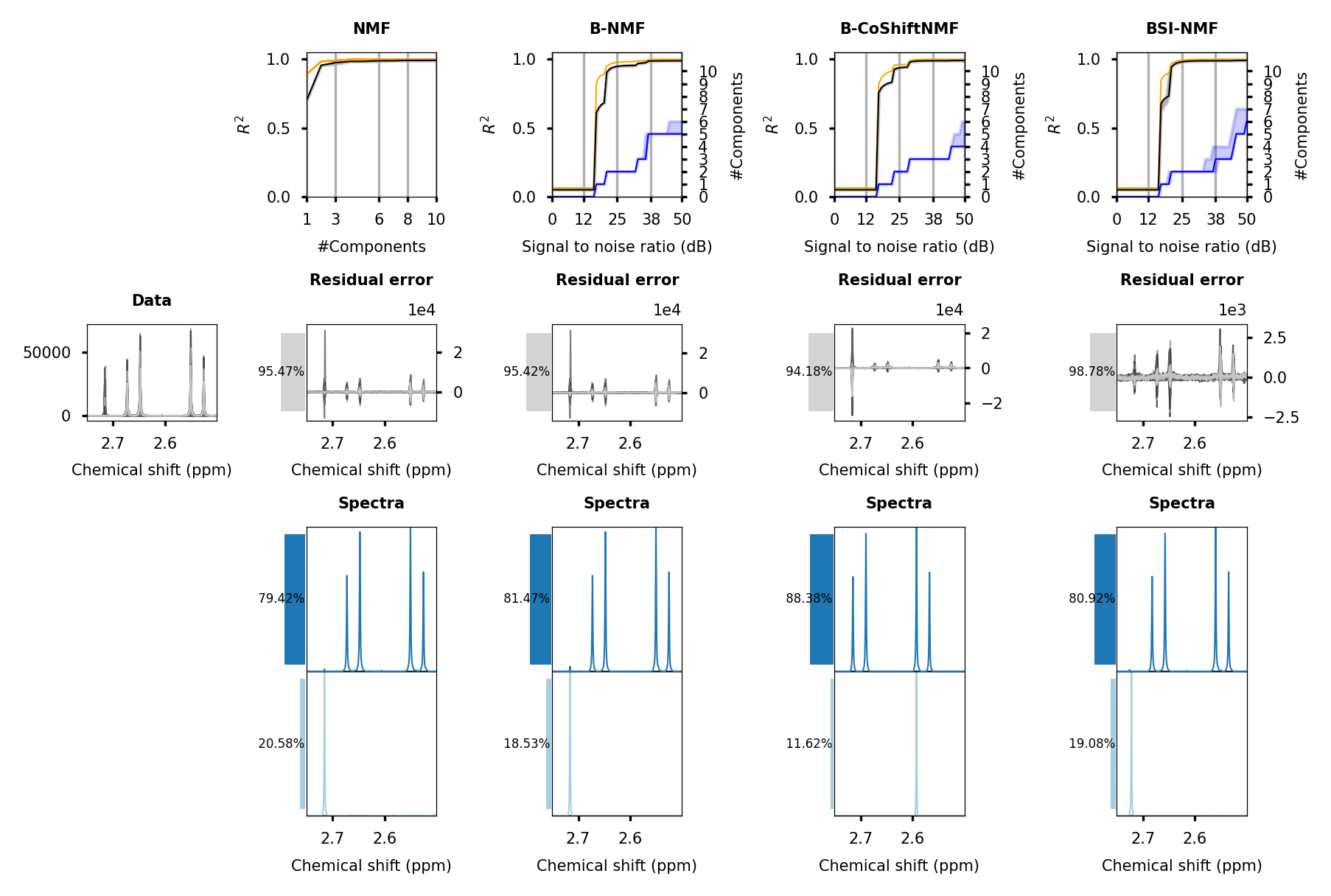}
            \label{fig:app:rightcomponents:urineA:citric}
        }

        \subfloat[Urine Metabolites A (w. pure samples) - 3.03 to 3.08 ppm]{
            \includegraphics[width=0.8\linewidth]{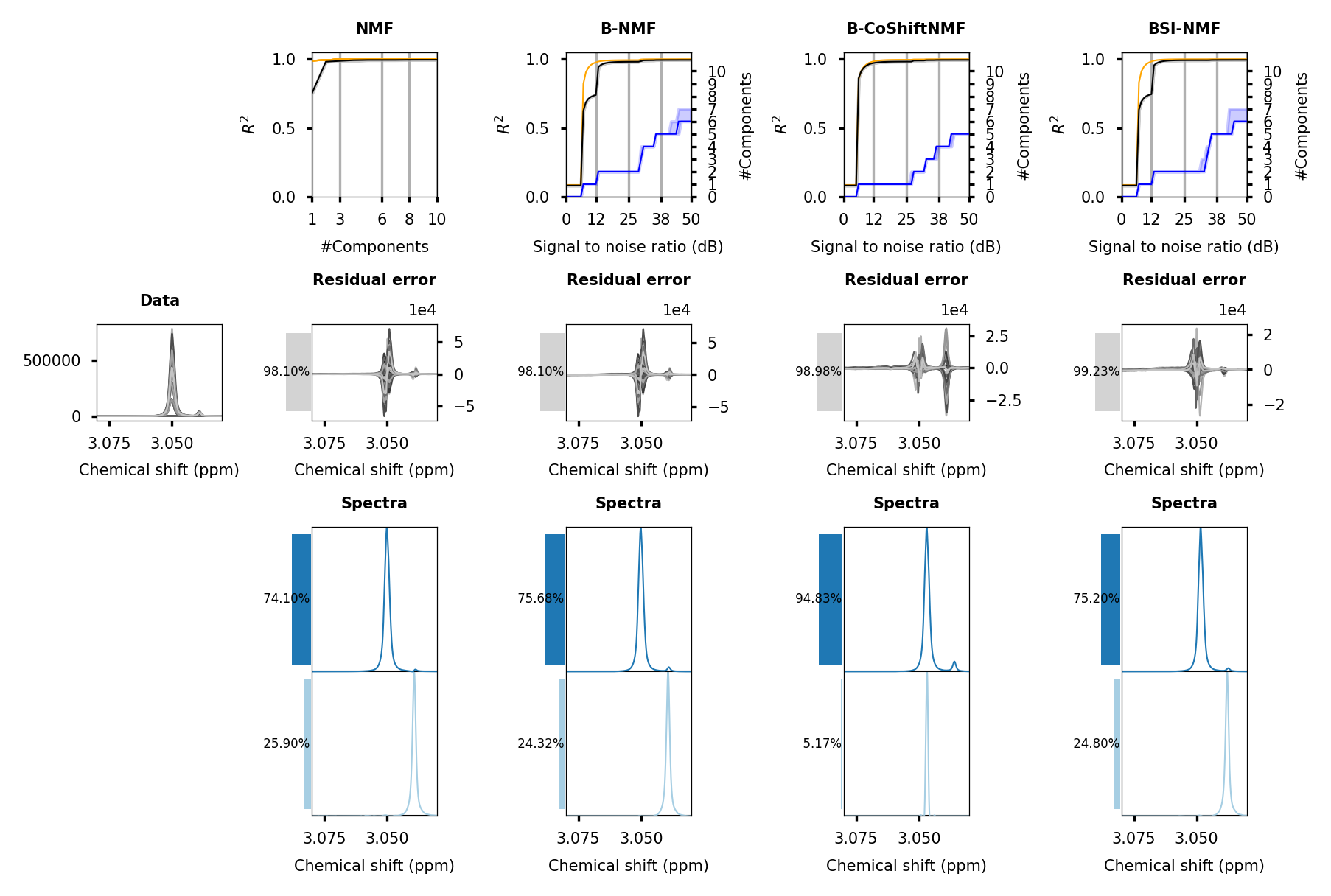}
            \label{fig:app:rightcomponents:creatine}
        }
    \caption{Urine Metabolites A: Here the number of components identified by BSI-NMF are plotted for all methods for all three intervals for the dataset both without pure samples (left column) and with pure samples (right column).}
    \label{fig:app:rightcomponents:urineA}
\end{figure}

\begin{figure}[tbp]
    \centering
        \subfloat[Urine Metabolites B (w. pure samples) - 2.5 to 2.75 ppm]{
            \includegraphics[width=0.8\linewidth]{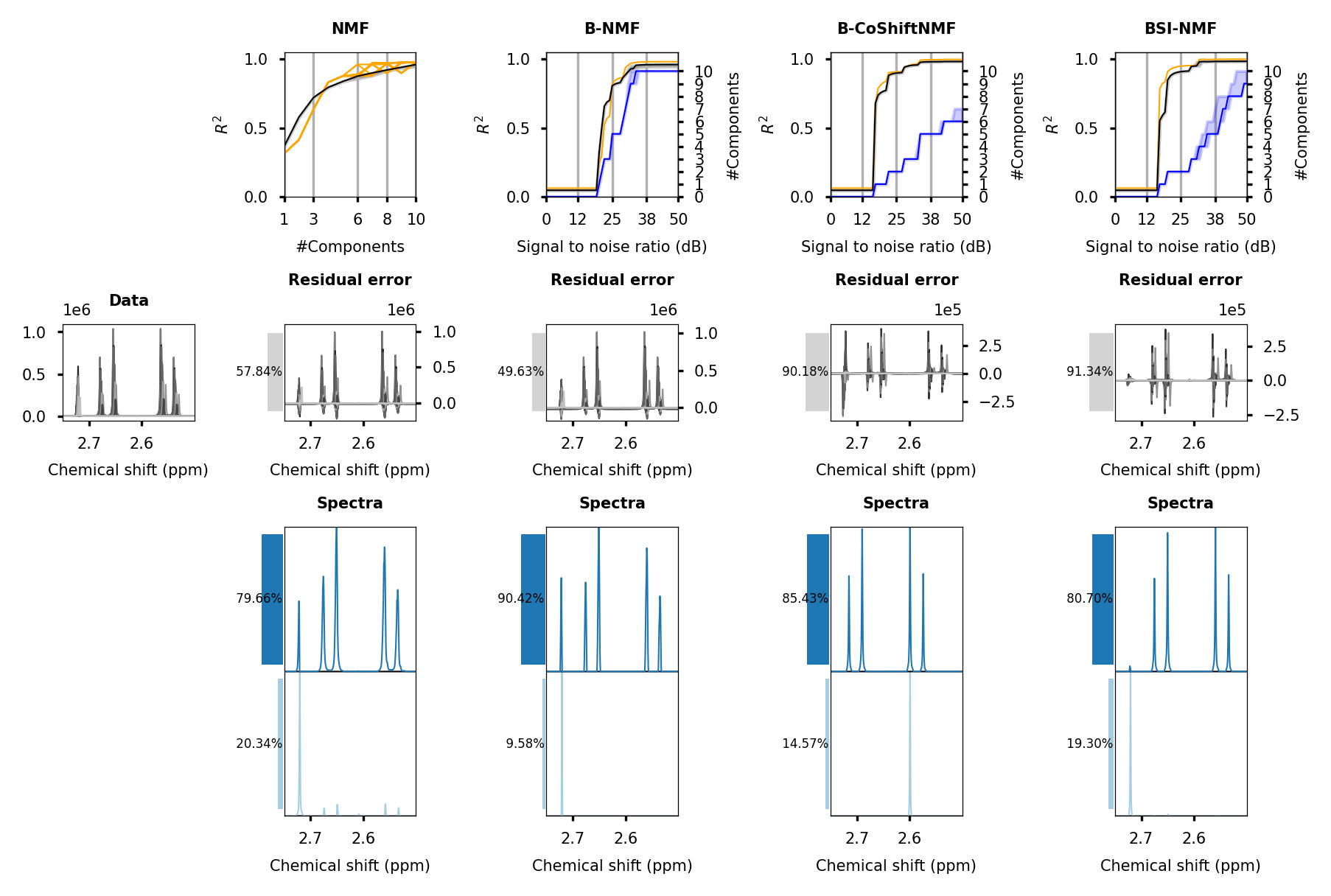}
            \label{fig:app:rightcomponents:urineB:citric}
        }

        \subfloat[Urine Metabolites B (w. pure samples) - 3.03 to 3.08 ppm]{
            \includegraphics[width=0.8\linewidth]{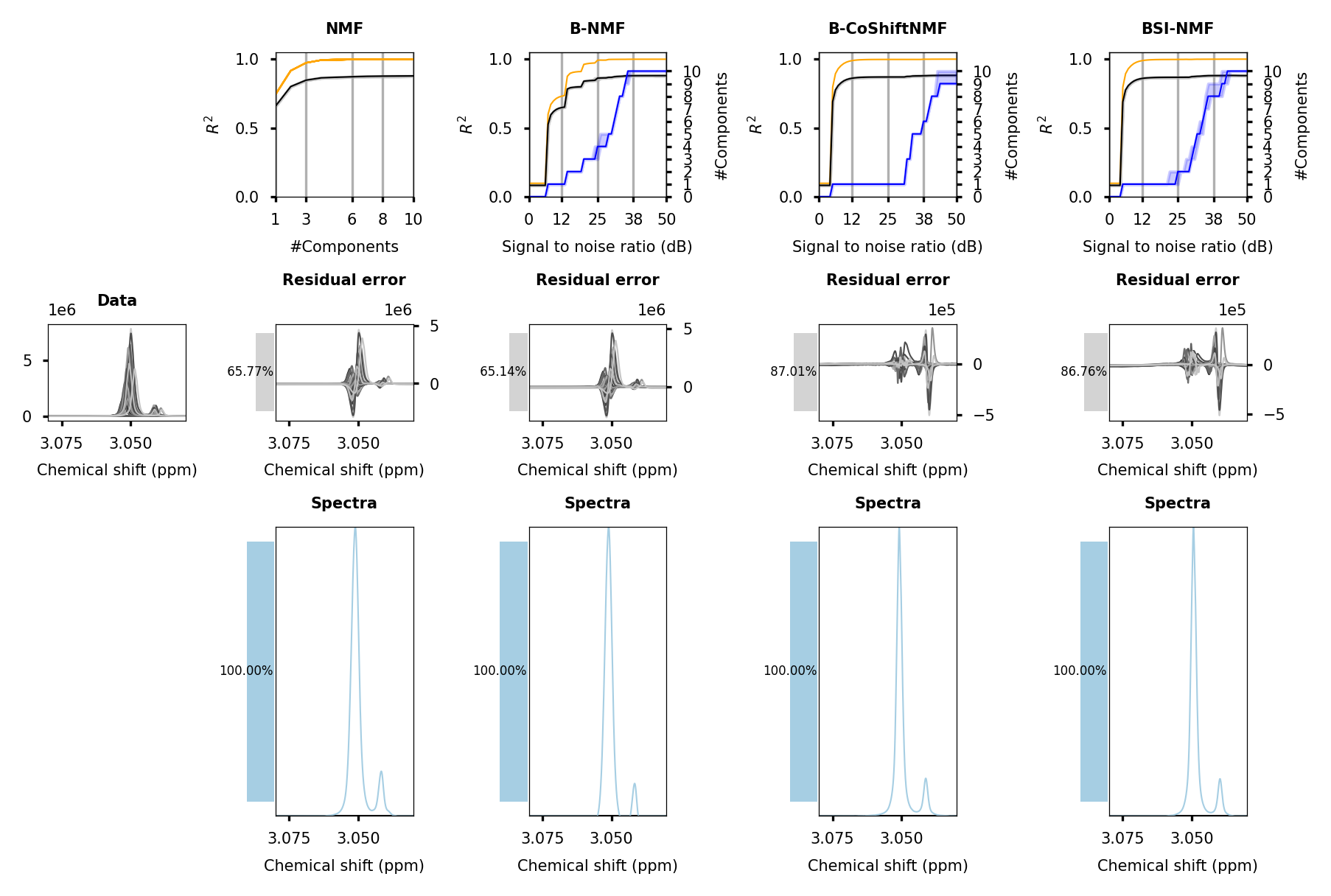}
            \label{fig:app:rightcomponents:urineB:creatine}
        }
    \caption{Urine Metabolites B: Here the number of components identified by BSI-NMF are plotted for all methods for all intervals for the dataset both without pure samples (left column) and with pure samples (right column).}
    \label{fig:app:rightcomponents:urineB}
\end{figure}

The results fot the Human Urine dataset is shown in Figure~\ref{fig:app:rightcomponents:human} while Urine Metabolites A and B datasets are shown in Figure~\ref{fig:app:rightcomponents:urineA} and \ref{fig:app:rightcomponents:urineB},  respectively.


\FloatBarrier

\section{The best solution found for each model and each choice of component}
\label{sec:app:allcomponents}

This section contains detailed information on each of the tested methods (NMF, B-NMF, B-CoShiftNMF, BSI-NMF) for each case and each dataset. The resulting 19 figures are shown in this section. With the Human Urine, Urine Metabolite B, Urine Metabolite A, shown in Section~\ref{sec:app:allcomp:human}, \ref{sec:app:allcomp:urineB}, and \ref{sec:app:allcomp:urineA}, respectively. For all datasets, we show the 2.5 to 2.75 ppm and 3.03 to 3.08 ppm case and for Human Urine we also show 5.66 to 5.987 ppm.

Each figure contains, in the first row, the measured \textsuperscript{1}H NMR data at the original scale. In the second row, the measured \textsuperscript{1}H NMR data but scaled by the $\alpha_n$, as presented in Section~\ref{sec:methods:experimentaldetails}. 

The third row gives the performance for each method where performance is across the components $d=1,2,\ldots, 10$ for NMF and across the assumed signal-to-noise ratios (SNRs) from 50 to 0 dB for B-NMF, B-CoShiftNMF, and BSI-NMF. The left y-axis is the coefficient of regression $R^2$ with (black line) and without (dotted orange) the baseline component. The right y-axis - the blue line - indicate the number of components at each SNR. The shaded gray and blue area denotes the range of values across ten repeats - if the shaded area is not visible the variation is negligible. Similarly, if the orange dotted line is not visible, the baseline offset is negligible.

For each number of components (NoC), the residual error (in the original scale) and each component ($d=1,2,\ldots, D$) is plotted. For B-NMF, B-CoShiftNMF, and BSI-NMF, some NoCs are empty as no model with that number of components is supported by the data (e.g. it is pruned by the automatic relevance determination). The residual errors have individual scales on the y-axis and to the left is a bar with the percent explained variance (in the original scale) and to the right the percent explained variance in the scaled data. The latent spectra for best model are shown and their percentage contribution to the reconstruction of the data as bars.

\subsection{Human Urine}
\label{sec:app:allcomp:human}
\begin{figure}[tbp]
    \centering
        \includegraphics[width=0.55\linewidth]{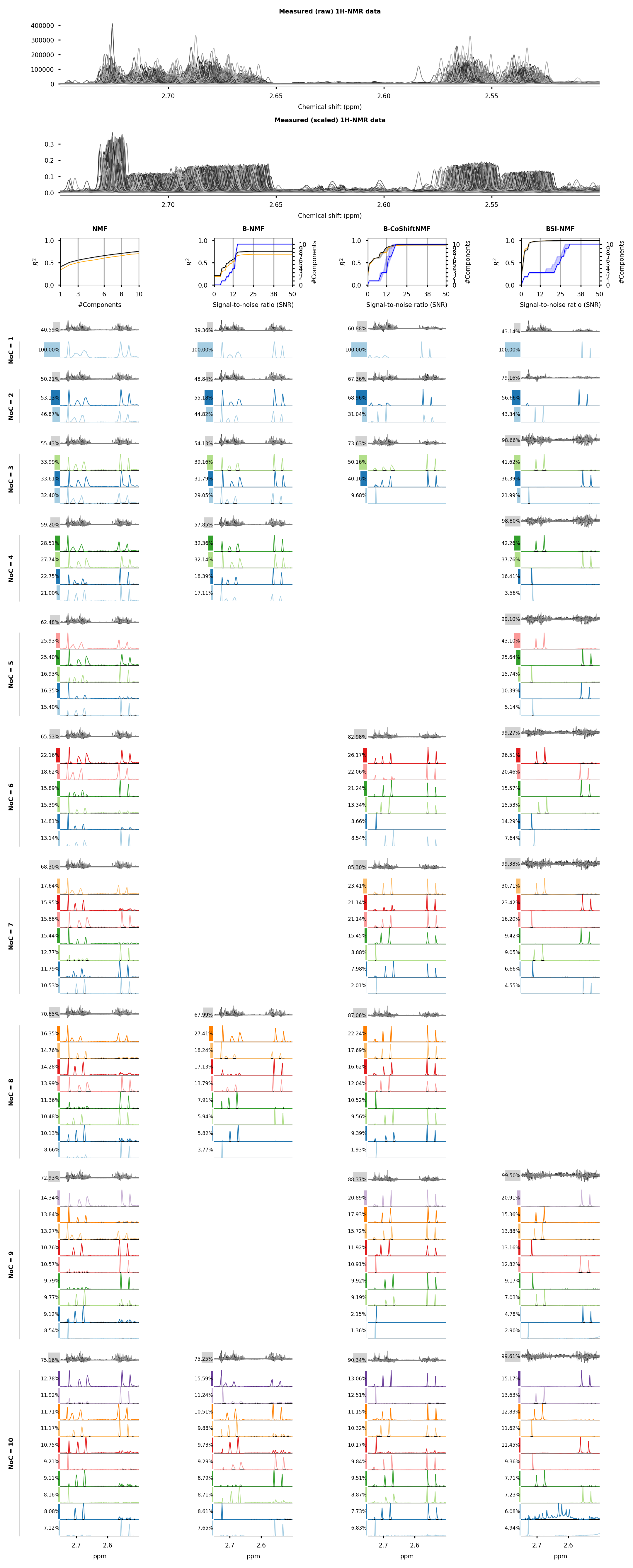}
    \caption{Human Urine (2.5 to 2.75 ppm): The measured and scaled data, as well as the performance and the the best model at each number of components (NoC) for all methods. See Section~\ref{sec:app:allcomponents} for detailed description.}
    \label{fig:app:allcomp:human:citric}
\end{figure}

For the citric acid and dimethylamine interval (2.5 to 2.75 ppm), Figure~\ref{fig:app:allcomp:human:citric}, clearly shows that NMF, B-NMF, and B-CoShiftNMF fail to completely capture citric acid (split as two doublets) and DMA as the individual shifts cannot be handles. BSI-NMF determines the right number of components and identifies both the dimethylamine peak and two doublets belonging to citric acid. We validated that the splitting of the two citric acid doublets into separate components is a consequence of the data not the model. This can easily be seen visually by investigating the scaled data (second row) in Figure~\ref{fig:app:allcomp:human:citric} as the doublet at lower ppm has less shifts in its peaks than the doublet at higher ppm. 

\begin{figure}[tbp]
    \centering
        \includegraphics[width=0.55\linewidth]{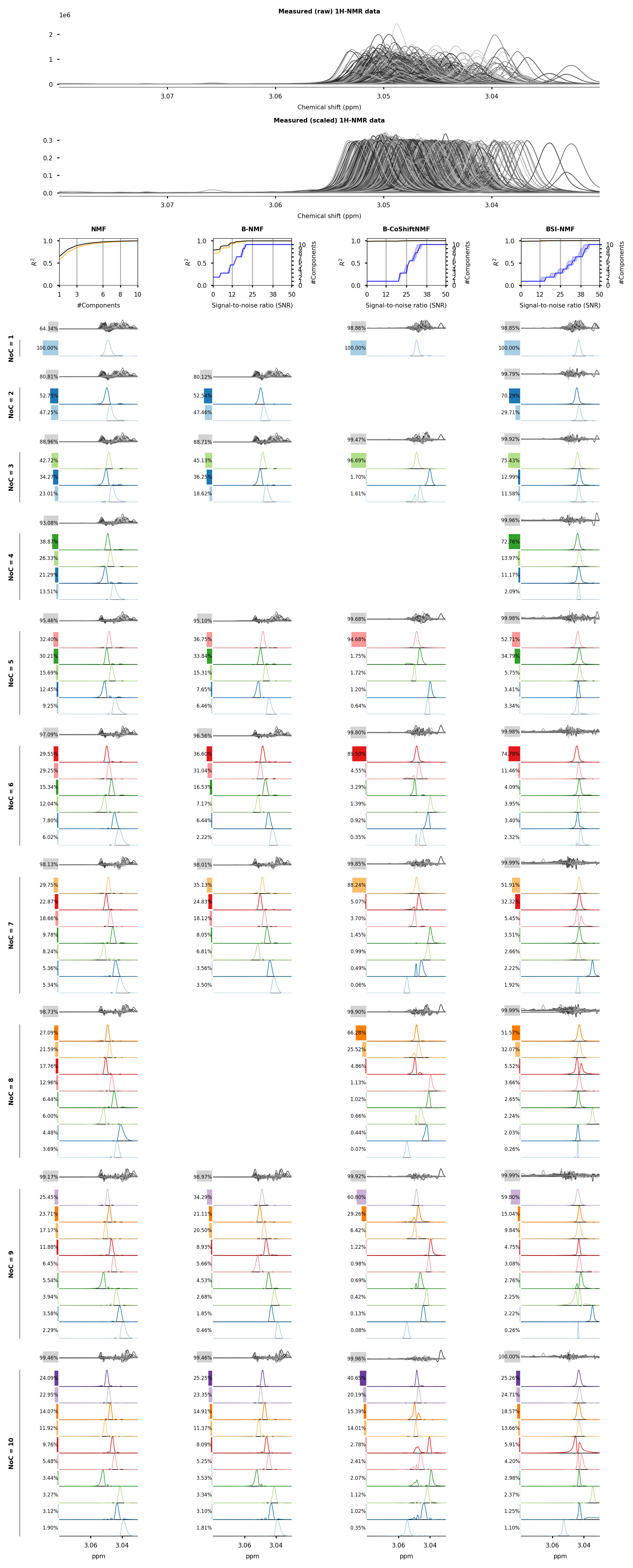}
    \caption{Human Urine (3.03 to 3.08 ppm): The measured and scaled data, as well as the performance and the the best model at each number of components (NoC) for all methods. See Section~\ref{sec:app:allcomponents} for detailed description.}
    \label{fig:app:allcomp:human:creatine}
\end{figure}

For the creatinine and creatine interval (3.03 to 3.08 ppm), Figure \ref{fig:app:allcomp:human:creatine}, the NMF, B-NMF, and B-CoShiftNMF does not identify the correct number of components and does not recover both creatinine and creatine. For B-CoShiftNMF, the creatinine peak is recovered, as it is the main peak and B-CoShiftNMF can handle a single sample shift, but it does not recover creatine. BSI-NMF recovers both creatinine and creatine, but is affected by the shift ambiguity of having two singlets that have the same shape. Consequently, in a few samples the component used to describe the creatinine peak is the creatine components and vice versa. This can be handled by specifying a post-hoc procedure were creatinine is always the larger or at higher ppm. This is also why the explained variance $R^2$ is a poor measure for distinguishing between peaks and non-peaks when the magnitude and shape is almost identical. 

\begin{figure}[tbp]
    \centering
        \includegraphics[width=0.55\linewidth]{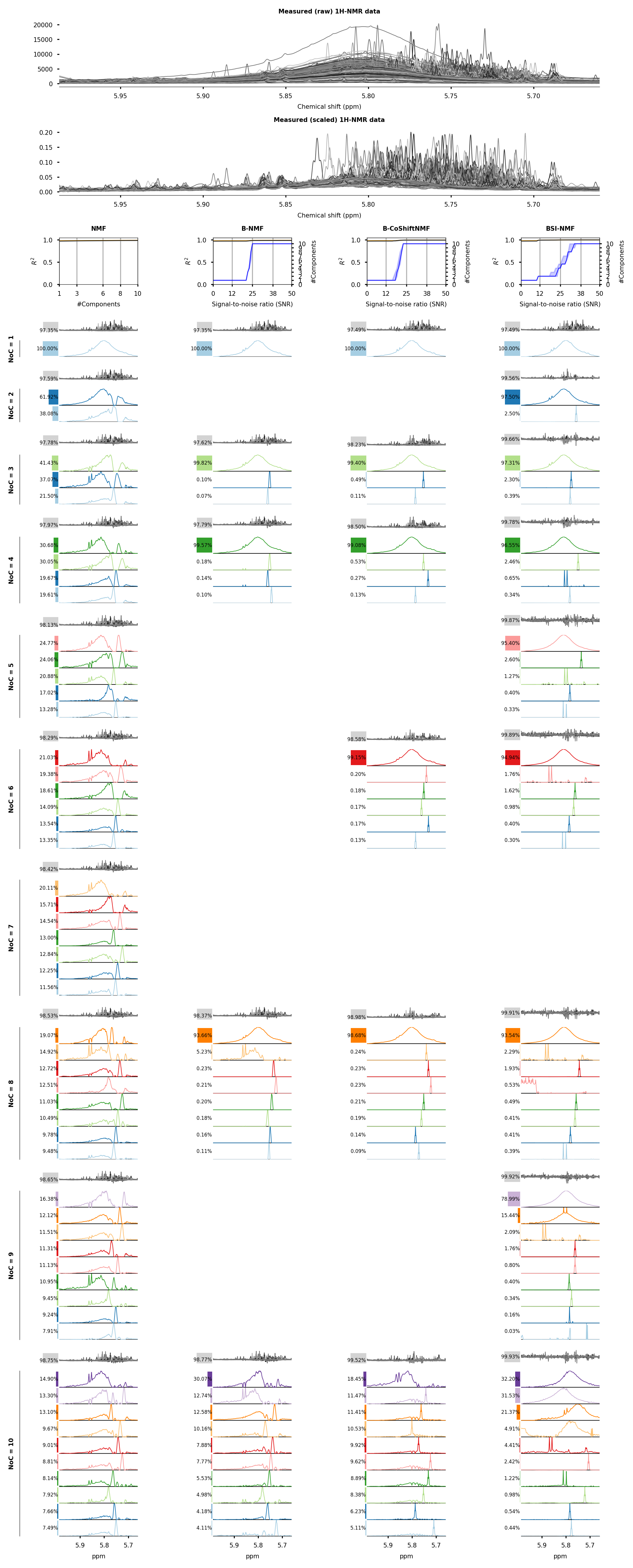}
    \caption{Human Urine (5.66 to 5.987 ppm): The measured and scaled data, as well as the performance and the the best model at each number of components (NoC) for all methods. See Section~\ref{sec:app:allcomponents} for detailed description.}
    \label{fig:app:allcomp:human:urea}
\end{figure}

For the urea interval (5.66 to 5.987 ppm), Figure~\ref{fig:app:allcomp:human:urea}, all methods are able to capture the urea peak with some partially co-varying artifacts and doing so describes most of the variation in the data. Both NMF and B-NMF fail in describing the minor metabolites, but interestingly they do so in different ways. NMF splits the urea component into chucks as more components are added, in contrast, B-NMF keeps the urea component and describes additional variance by duplicating the minor metabolite component at different ppm position - until the automatic relevance determination is essentially (SNR=50, NoC=10) turned of. B-CoShiftNMF gets slightly better performance (0.14\% more variance explained) than B-NMF by allowing the urea component to shift, but otherwise performs similarly. It is difficult to say how many metabolites this interval contains, BSI-NMF points to two components ($D=2$) based on the $R^2$ value being high and the number of components plateauing. However, multiple possible values of $D$ is plausible, for instance $D=5$ which separates the smaller peaks that partially co-vary with the urea peak, but a better measure than $R^2$ is needed to clearly distinguish between the different solutions.

\FloatBarrier

\subsection{Urine Metabolites A}
\label{sec:app:allcomp:urineA}
\begin{figure}[tbp]
    \centering
        \includegraphics[width=0.55\linewidth]{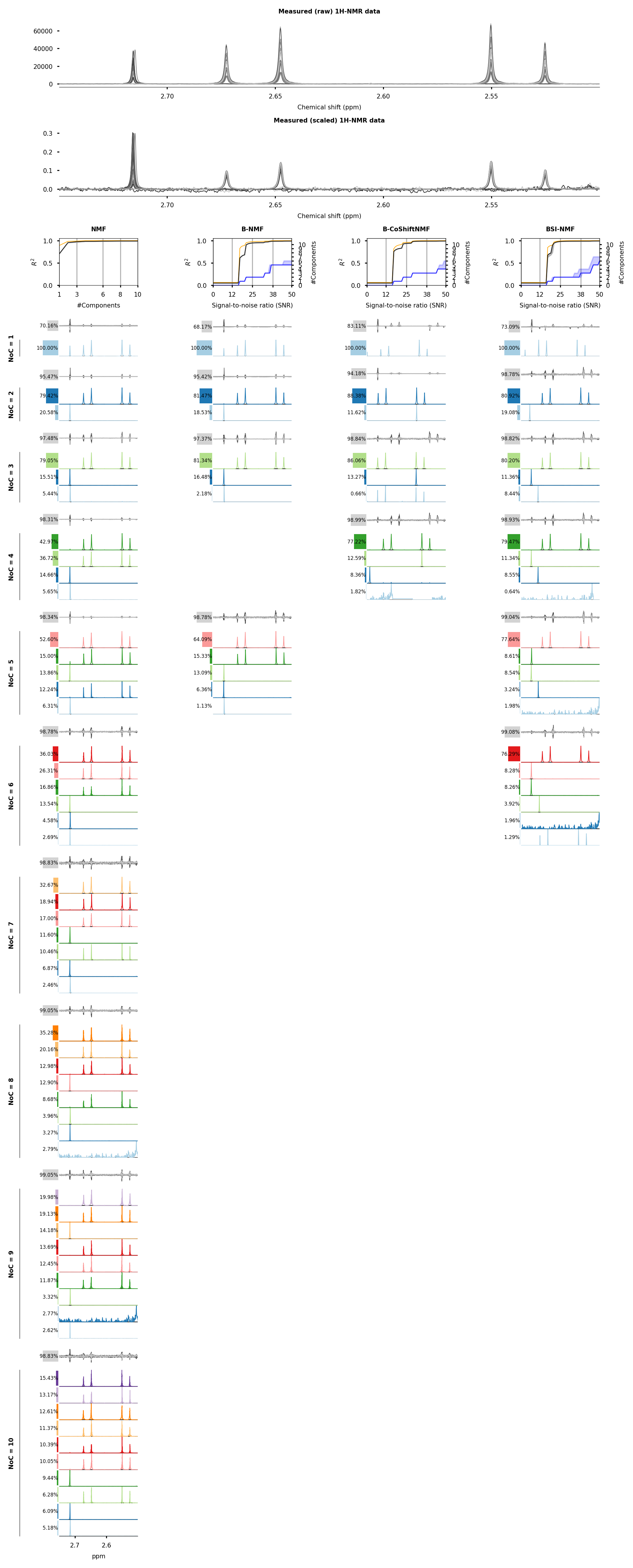}
    \caption{Urine Metabolites A with pure samples, $N=42$, (2.5 to 2.75 ppm): The measured and scaled data, as well as the performance and the the best model at each number of components (NoC) for all methods. See Section~\ref{sec:app:allcomponents} for detailed description.}
    \label{fig:app:allcomp:urineA:citric}
\end{figure}

For the citric acid and dimethylamine interval (2.5 to 2.75 pm), Figure~\ref{fig:app:allcomp:urineA:citric}, the NMF, B-NMF, and BSI-NMF methods point to a two component solution ($D=2$) - based on the $R^2$ and explained variance (percentages given in the light grey bar). BSI-NMF a little more of the data variance $\sim 3\% $ . The non-shifted methods (NMF and B-NMF) works well because the data has little to no shift, so failure to model it is not a significant downside. In addition, since samples with pure metabolites (e.g. only one stock solution) are present, the uniqueness criteria based on spanning the non-negative orthant is fulfilled. Interestingly, B-CoShiftNMF points to a three components solution ($D=3$) and mixes the spectra a bit. This is hypothesized to be due to CoShift causing a slight misalignment of the metabolites by introducing sample shift.

\begin{figure}[tbp]
    \centering
    \includegraphics[width=0.55\linewidth]{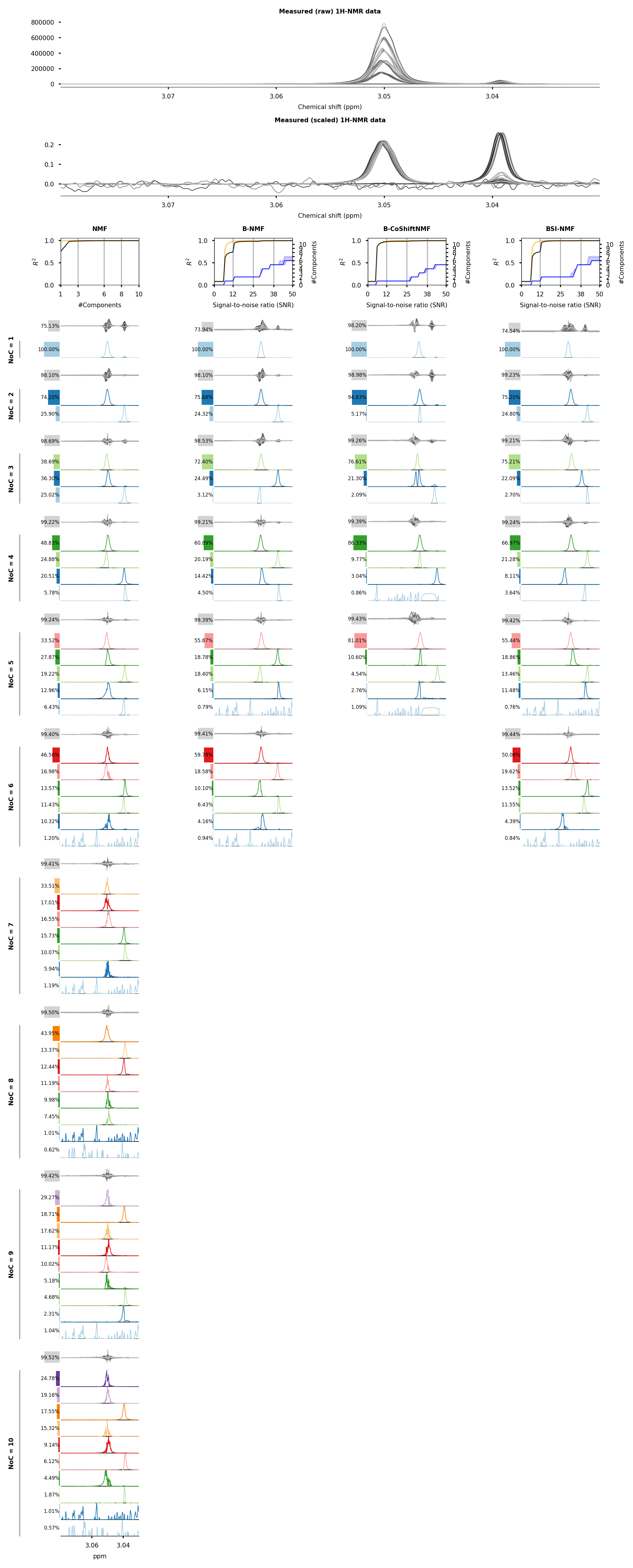}
    \caption{Urine Metabolites A with pure samples, $N=42$, (3.03 to 3.08 ppm): The measured and scaled data, as well as the performance and the the best model at each number of components (NoC) for all methods. See Section~\ref{sec:app:allcomponents} for detailed description.}
    \label{fig:app:allcomp:urineA:creatine}
\end{figure}

For the creatinine and creatine interval (3.03 to 3.08 ppm),  Figure~\ref{fig:app:allcomp:urineA:creatine}, the results and conclusions are identical. The slight shift present in the data is sought accounted for by B-CoShiftNMF which then causes a slight misalignment that the method is unable to correct. In contrast, all other methods recover the two spectra.

\FloatBarrier

\subsection{Urine Metabolites B}
\label{sec:app:allcomp:urineB}
\begin{figure}[tbp]
    \centering
        \includegraphics[width=0.55\linewidth]{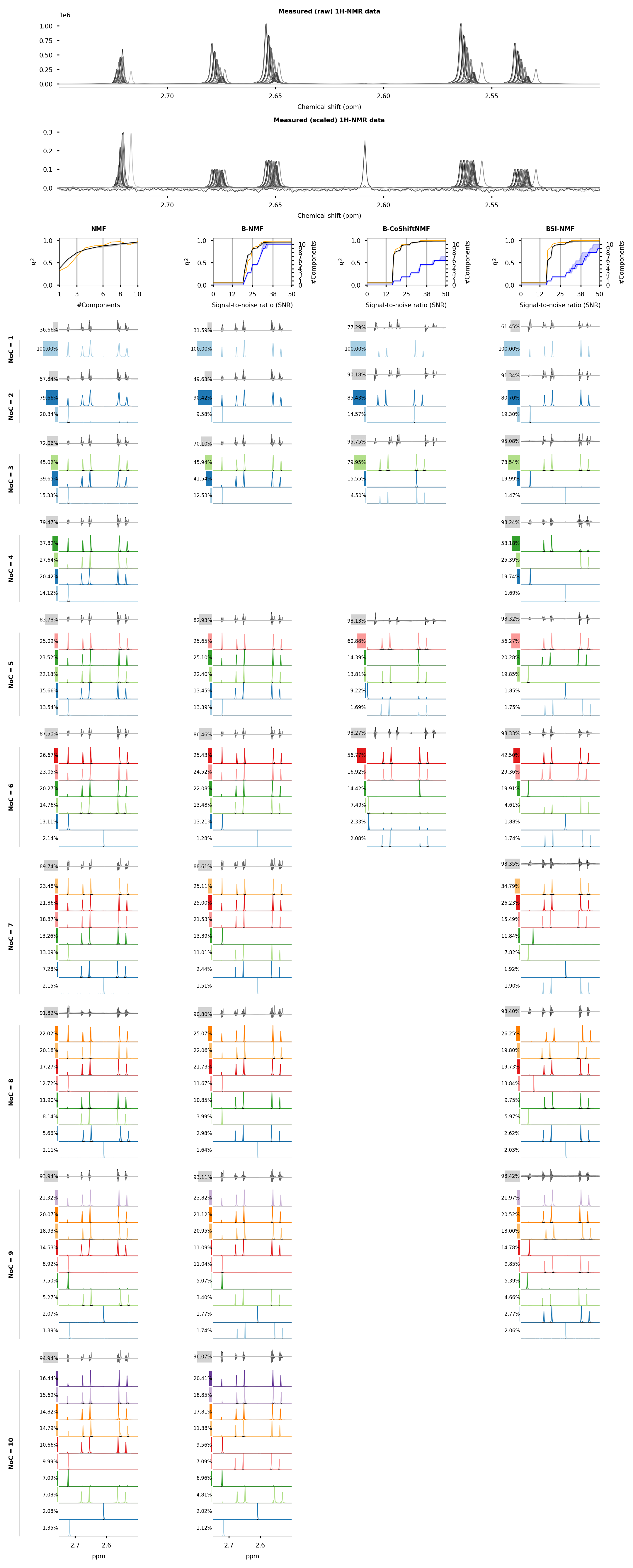}
    \caption{Urine Metabolites B with pure samples, $N=42$, (2.5 to 2.75 ppm): The measured and scaled data, as well as the performance and the the best model at each number of components (NoC) for all methods. See Section~\ref{sec:app:allcomponents} for detailed description.}
    \label{fig:app:allcomp:urineB:citric}
\end{figure}

\begin{figure}[tbp]
    \centering
        \includegraphics[width=0.55\linewidth]{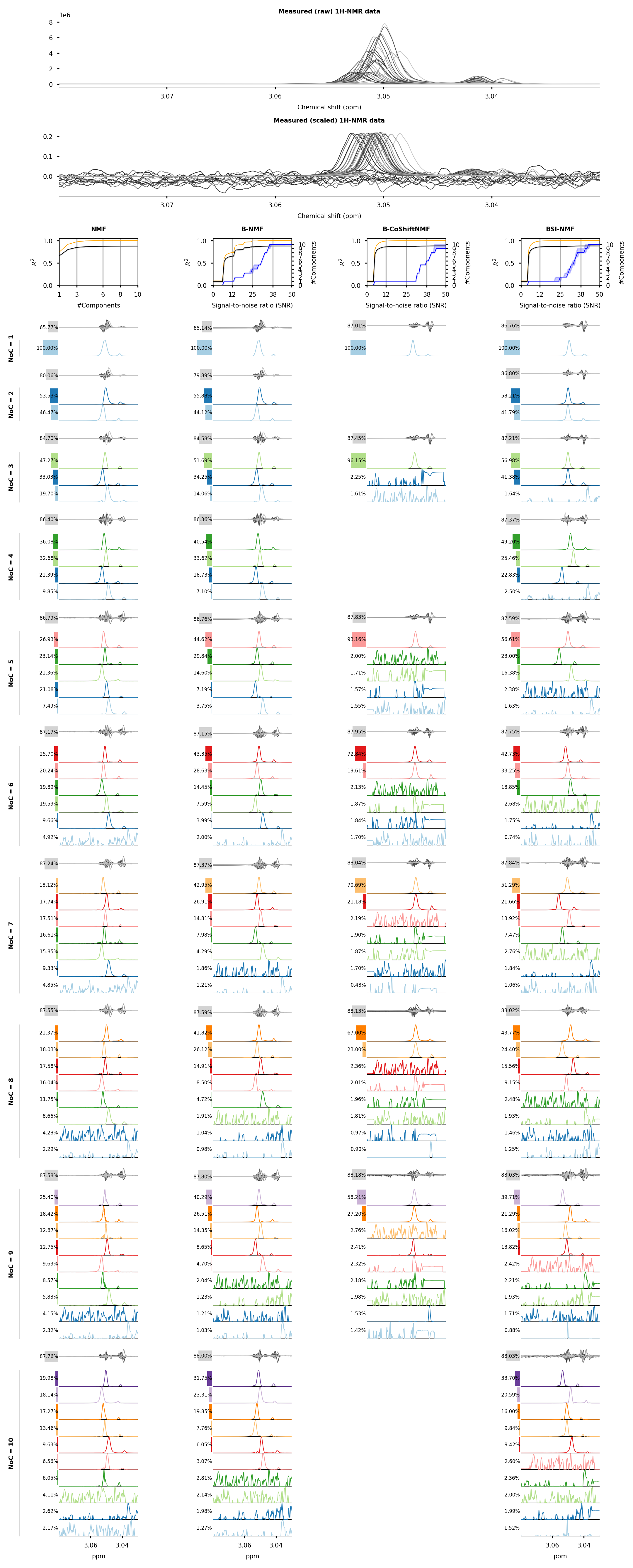}
        
    \caption{Urine Metabolites B with pure samples, $N=42$, (3.03 to 3.08 ppm): The measured and scaled data, as well as the performance and the the best model at each number of components (NoC) for all methods. See Section~\ref{sec:app:allcomponents} for detailed description.}
    \label{fig:app:allcomp:urineB:creatine}
\end{figure}

In contrast to Urine Metabolites A, the Urine Metabolites B dataset has slight shifts of the metabolites. This causes NMF and B-NMF to fail to recover the underlying spectra in both the citric acid and dimethylamine interval (2.5 to 2.75 pm), Figure~\ref{fig:app:allcomp:urineB:citric}, and the creatinine and creatine interval (3.03 to 3.08 ppm),  Figure \ref{fig:app:allcomp:urineB:creatine}.  In contrast, BSI-NMF accounts for the shifts and identify the two component solution ($D=2$) as being optimal and recovers the  spectra of citric acid and dimethylamine. For the creatinine and creatine interval, both B-CoShiftNMF and BSI-NMF correctly points to a one component solution ($D=1$)  as the two metabolites were in the same stock solution and can therefore not be separated by assuming bi-linearity. 

There is a caveat for the citric acid and dimethylamine interval, as it includes samples that should have zero signal (they are from the third stock solution) but have a small contamination. Since there are no other signals in these samples they contaminat is scaled up by the standardization scheme (see Online Methods, Section ~\ref{sec:methods:experimentaldetails}) are blown up. This contaminant is highly visible in the scaled \textsuperscript{1}H NMR data in Figure~\ref{fig:app:allcomp:urineB:citric}.

\FloatBarrier

\section{Source code}
\label{sec:app:sourcecode}
Source code for B-NMF, B-CoShiftNMF, and BSI-NMF 
     as well as code for comparison of BSI-NMF to ICoShift with NMF and peak fitting in MATLAB are available for download from
\url{https://github.com/JesperLH/bsi-nmf}.

\section{Data availability}
\label{sec:app:dataavailability}
The data used for the experimentation can be shared upon request by contacting the authors.

\end{appendices}

\FloatBarrier

\bibliographystyle{unsrt}  
\bibliography{main_references}

\end{document}